\documentclass[10pt,aps,showpacs,nofootinbib,prd,aps,epsf,floats,
               amsmath,amssymb,amsfonts,axodraw,twocolumn]{revtex4}
\usepackage{amsmath, amssymb}
\newcommand{\mathsym}[1]{{}}

\usepackage{graphicx}
\usepackage{amsmath}
\usepackage{amssymb}
\usepackage{bm}
\newcommand{\rem}[1]{}
\newsavebox{\PSLASH}
 \sbox{\PSLASH}{$p$\hspace{-1.8mm}/}
 
\renewcommand{\theequation}{\thesection.\arabic{equation}}
\newcounter{saveeqn}
\newcommand{\add}{\addtocounter{equation}{1}}
\newcommand{\alpheqn}{\setcounter{saveeqn}{\value{equation}}%
\setcounter{equation}{0}%
\renewcommand{\theequation}{\mbox{\thesection.\arabic{saveeqn}{\alph{equation}}}}}
\newcommand{\reseteqn}{\setcounter{equation}{\value{saveeqn}}%
\renewcommand{\theequation}{\thesection.\arabic{equation}}}

 \newsavebox{\notrightarrow}
 \sbox{\notrightarrow}{$\to$\hspace{-4mm}/}
 
 \newsavebox{\PARTIALSLASH}
 \sbox{\PARTIALSLASH}{$\partial$\hspace{-1.6mm}/}
 
 \newsavebox{\ASLASH}
 \sbox{\ASLASH}{$A$\hspace{-2.1mm}/}
 
 \newsavebox{\KSLASH}
 \sbox{\KSLASH}{$k$\hspace{-1.8mm}/}
 
 \newsavebox{\LSLASH}
 \sbox{\LSLASH}{$\ell$\hspace{-1.8mm}/}
 
 \newsavebox{\QSLASH}
 \sbox{\QSLASH}{$q$\hspace{-1.8mm}/}
 
 \newsavebox{\DSLASH}
 \sbox{\DSLASH}{$D$\hspace{-2.2mm}/}
 
 \newsavebox{\DbfSLASH}
 \sbox{\DbfSLASH}{${\mathbf D}$\hspace{-2.8mm}/}
 
 \newsavebox{\DELVECRIGHT}
 \sbox{\DELVECRIGHT}{$\stackrel{\rightarrow}{\partial}$}
 
 \newcommand{\blue}{\IfColor{\textCadetBlue}{}}
\newcommand{\black}{\IfColor{\textBlack}{}}
\newcommand{\red}{\IfColor{\textRed}{}}
\newcommand{\green}{\IfColor{\textOliveGreen}{}}
\newcommand{\lila}{\IfColor{\textRedViolet}{}}

\begin{document}
\title{Anomalous magnetic moment of hot quarks, inverse magnetic catalysis and\\
reentrance of chiral symmetry broken phase}
\author{Sh. Fayazbakhsh$^{a}$}\email{shfayazbakhsh@ipm.ir}
\author{N. Sadooghi$^{b}$}\email{sadooghi@physics.sharif.ir}
\affiliation{ $^{a}$Institute for Research in Fundamental Sciences
(IPM), School of Particles and Accelerators, P.O. Box 19395-5531,
Tehran-Iran\\
$^{b}$Department of Physics, Sharif University of Technology, P.O.
Box 11155-9161, Tehran-Iran}
\begin{abstract}
\par\noindent
The effect of anomalous magnetic moment of quarks on thermodynamic properties of the chiral condensate is studied, using of a two-flavor Nambu--Jona-Lasinio model at finite temperature $T$, chemical potential $\mu$, and in the presence of a uniform magnetic field $eB$. To this purpose, the Schwinger linear-in-$B$ ansatz for the quark anomalous magnetic moment in term of the nonperturbative Bohr magneton is considered. In a two-dimensional flavor space, it leads to the correction $\hat{T}_{Sch}=\hat{\kappa}\hat{Q} eB$ in the energy dispersion relation of quarks. Here, $\hat{Q}$ is the quark charge matrix. We consider three different sets for $\hat{\kappa}$, and numerically determine the dependence of the constituent quark mass on $T,\mu$ and $eB$ for fixed $\hat{\kappa}$. By exploring the complete phase portrait of this model in $T$-$\mu$, $\mu$-$eB$, and $T$-$eB$ phase spaces for various fixed $eB$, $T$, $\mu$ and $\hat{\kappa}$, we observe that inverse magnetic catalysis occurs for large enough $\hat{\kappa}$. Moreover, in the regime of weak magnetic fields, the phenomenon of reentrance of chiral symmetry broken and restored phases occurs for $T, \mu$ and $eB$ dependent $\hat{\kappa}$.
\end{abstract}
 \maketitle
\section{Introduction}\label{sec1}
\par\noindent
The effects of uniform magnetic fields on relativistic quark matter have been studied extensively in recent years (for an overview, see \cite{magnetic}). The main interest arises from the relevance of these effects on the physics of neutron stars, as well as on the dynamics of the quark-gluon plasma created in relativistic heavy ion collisions. It is known, that magnetars, a certain class of neutron stars, exhibit magnetic fields up to $10^{12}-10^{13}$ Gau\ss~on their surface, and $10^{18}-10^{20}$ Gau\ss~in their inner core \cite{neutronstar}. In \cite{duncan1992}, the dynamo effect during the first few seconds after the star's gravitational collapse is made responsible for the generation of these unusual magnetic fields.
Even larger magnetic fields are believed to be created in early stages of non-central heavy ion collisions. Depending on the collision energies and impact parameters, the strength of these magnetic fields are estimated to be of the order $eB\sim 0.03$ GeV$^{2}$ at RHIC and $eB\sim 0.3$ GeV$^{2}$ at LHC \cite{skokov2009}.\footnote{Here, $eB=1$ GeV$^{2}$ corresponds to $B\sim 1.7\times 10^{20}$ Gau\ss.} The mechanism for the creation of strong magnetic fields in heavy ion collisions is originally discussed in \cite{mclerran2007} (see e.g. \cite{tuchin2013} for a recent review on this topic).
\par
A uniform and spatially constant magnetic field breaks the Lorentz invariance of the physical system, and consequently induces anisotropies in the parallel and perpendicular directions with respect to the direction of the background magnetic field. The anisotropies include those in the neutrino emission and magnetic winds in the astrophysics of magnetars \cite{duncan1992}, or anisotropies arising in the refraction indices \cite{fayazbakhsh3}, and decay constants of mesons in hot and magnetized quark matter \cite{fayazbakhsh4, andersen2012}. The latter may be relevant for the physics of heavy ion collisions. Recently, the pressure anisotropies for a gas of protons and a gas of neutrons are studied in \cite{strickland2012}. It is, in particular, demonstrated that the inclusion of the anomalous magnetic moment (AMM) of protons and neutrons increases the level of anisotropies in both cases.
Pressure anisotropies, arising from uniform and spatially fixed magnetic fields, are supposed to have significant effects on the elliptic flow $v_{2}$ in heavy ion collisions \cite{endrodi2014}.
\par
In the present paper, we apply the method used in \cite{strickland2012}, to study the effect of various $(T,\mu,eB)$ dependent and independent AMM of quarks on the phase diagram of a two-flavor magnetized Nambu--Jona-Lasinio (NJL) model at finite temperature $T$ and chemical potential $\mu$. We will explore the phenomenon of magnetic catalysis  (MC)\footnote{The phenomenon of MC is introduced originally in \cite{klimenko1992, miransky1995}, and is extensively discussed in the literature (for a recent review, see \cite{shovkovy2012}).}, and inverse magnetic catalysis (IMC)\footnote{To the best of our knowledge, the term ``inverse magnetic catalysis'' is used for the first time in \cite{rebhan2011}, but, the phenomenon of IMC had been previously observed, e.g., in \cite{inagaki2003}.}, as well as the reentrance of chiral symmetry broken and
restored phases in the phase space of our model. The phenomenon of reentrance is already observed in the condensed matter physics of superfluidity \cite{reentrance-cond} and liquid crystals \cite{reentrance-crystal}, as well as in the astrophysics of neutron stars \cite{reentrance-stars}.
\par
In our previous works \cite{fayazbakhsh1,fayazbakhsh2}, we have already studied in detail the complete $T$-$\mu$, $\mu$-$eB$, and $T$--$eB$ phase portraits of the magnetized two-flavor NJL model, without inclusion of the quark AMM. We mainly worked in the supercritical regime of the NJL coupling, i.e., the coupling constant of the NJL model was chosen in such a way that the system exhibited chiral symmetry breaking even for vanishing magnetic fields. In this case, the magnetic field only enhances the production of bound states; they consist of mesons in chiral symmetry broken phase ($\chi$SB) and diquarks in color symmetry broken phase. In this sense, in \cite{fayazbakhsh2}, we focused on the phenomenon of MC of chiral and color symmetry breaking, and reported about various interesting phenomena, such as de Haas-van Alphen (dHvA) oscillations \cite{alphen1930}, that lead to reentrant chiral and color symmetry broken phases, mainly in the regime of weak magnetic fields. The phenomenon of MC is discussed intensively in many contexts \cite{shovkovy2012}.  At zero temperature, it arises from a dimensional reduction from $D$ to $D-2$ dimensions in the lowest Landau level (LLL). This is believed to be responsible for the aforementioned enhancement of bound state formation. As a consequence, in a massless theory, whose original Lagrangian density satisfies the chiral symmetry, a finite mass is dynamically created, which then breaks the chiral symmetry of the theory. Finite temperature and baryonic chemical potential compete with external magnetic fields in this regard, and, except in the regime of very strong magnetic fields, suppress the formation of mesonic bound states (see \cite{fayazbakhsh2} for more details). In our model, this regime is characterized by a threshold magnetic field $eB_{t}\sim 0.5$ GeV$^{2}$, above which the dynamics of the fermions is solely dominated by LLL. In this regime, the constituent quark mass monotonically increases with increasing $eB$ as a consequence of MC. In the regime $eB<0.5$ GeV$^{2}$, however, the $eB$ dependence of the condensate exhibits dHvA oscillations, which occur whenever Landau levels pass the quark Fermi level.\footnote{De Haas-van Alphen oscillations are studied in \cite{ebert1998, shovkovy2007}, and recently in \cite{simonov2014}.} Thus, because of these oscillations in the regime of $eB<eB_{t}$, at a fixed temperature and for both vanishing and nonvanishing chemical potential, the chiral condensate exhibits two different behaviors: In some regimes, it increases with increasing $eB$, this is related to the phenomenon of MC. But, in some other regimes, it decreases with increasing $eB$. This is related to the phenomenon of IMC.
\par
In \cite{fayazbakhsh2}, we have also studied the effect of external magnetic fields on $T$--$\mu$, $T$--$eB$ and $\mu$--$eB$ phase portraits of magnetized two-flavor NJL model. We have shown that, as a consequence of the background magnetic fields, second order chiral phase transitions turn into first order ones, and, in the regime above $eB_{t}\sim 0.5$ GeV$^{2}$, for fixed $\mu$ and $T$, the critical temperature $T_{c}$ and chemical potential $\mu_{c}$ of $\chi$SB, monotonically increase with increasing $eB$. This has been supposed to be an indication of the phenomenon of MC, which is mainly a LLL dominated effect. On the other hand, it has been shown, that in the regime of weak magnetic fields, the aforementioned dHvA oscillations, lead again to two different  behaviors of $T_{c}$ and $\mu_{c}$ as a function of $eB$: In some regimes, $T_{c}$ and $\mu_{c}$ increase with increasing $eB$, because of MC. In some other regimes, however, as a result of IMC, $T_{c}$ and $\mu_{c}$ decrease with increasing $eB$. The phenomenon of reentrance of chiral (color) symmetry broken phases at certain fixed $(T,\mu,eB)$ is also believed to be a consequence of dHvA oscillations in the weak field regime \cite{fayazbakhsh2}.
\par
Several other groups also investigate the effects of external magnetic fields on the phase diagram of hot QCD. These includes the groups working on lattice QCD at finite $T$, zero $\mu$ and nonvanishing $eB$ \cite{polikarpov2008, bali2011, ilgenfritz2013, ellia2013, endrodi2013, bali2014}, or investigating the QCD phase diagram by making use of functional renormalization group method \cite{pawlowski2012, andersen2013,kamikado2013}. Many other groups elaborate different QCD-like models at finite $(T,\mu,eB)$, that exhibit $\chi$SB. The latter consists of, e.g., NJL-model \cite{fukushima2012, farias2014}, Polyakov-linear-$\sigma$ model \cite{mizher2010}, Polyakov-NJL model \cite{ruggieri2010, providencia2013, providencia2014}, Polyakov-Quark-Meson model \cite{fraga2013, andersen2014}, NJL model including axial chemical potential \cite{ruggieri2011,huang2013}, Ginzburg-Landau model \cite{ruggieri2014}, charged scalar model with spontaneous $\chi$SB \cite{ayala2014}, NJL model with dynamical AMM generation \cite{ferrer2014}.
Let us notice that the interest on this subject grows up after the still not fully comprehended lattice results by Bali et al. in \cite{bali2011}. They reported an unprecedented decrease of $T_{c}$ as a function of $eB$ within an ab initio lattice QCD simulation, and declared this as a signature of IMC in this framework. The physical explanation of the phenomenon of IMC is, in particular, still under investigation: In \cite{endrodi2013,providencia2013}, the interaction of the magnetic field with the sea quarks leading to a backreaction of the Polyakov loop, and in \cite{fukushima2012} the magnetic inhibition because of neutral meson fluctuations are made responsible for IMC. In \cite{huang2013}, it is explained how IMC is induced by sphalerons. Recently, in \cite{farias2014, ayala2014, providencia2014, ferrer2014}, the running of the corresponding coupling constants to the considered models, and therefore their $eB$ dependence are taken into account, and it is shown how the lattice results on the $eB$ dependence of $T_{c}$ in \cite{bali2011} can be successfully reproduced.
\par
In a parallel development, the competition between mass and $eB$ contributions to QCD pressure is considered up to two-loop radiative corrections in \cite{blaizot2012}. It is argued that the deviation of the Land\'e $g$-factor from $2$, which is of about $g-2\sim 10^{-3}$, may produce sizable corrections to the QCD pressure. This affects the energy of the LLL by turning the mass into $m_{\mbox{\tiny{eff}}}^{2}=m^{2}+(g-2)eB$. For the relevant $eB>0.01$ GeV$^{2}$, the correction will be in the MeV range. It may thus compete with $m$, and cannot be ignored in quantitative studies \cite{blaizot2012}.
The effect of dynamically generated AMM of fermions on the phase diagram of a magnetized one-flavor NJL model is recently investigated in \cite{ferrer2013}\footnote{See also \cite{martinez2013, martinez2014-1} for similar studies.}. Here, inspired by the results presented in \cite{ferrer2009}, a nonperturbative mechanism for the generation of the quark AMM is introduced. To this purpose, a one-flavor NJL Lagrangian density including an appropriate tensor channel is used. It is in particular shown, that in the subcritical regime of the NJL coupling, where the phenomenon of MC is solely responsible for $\chi$SB, two independent condensates emerge in the LLL dominant regime. They correspond to the dynamical mass and the AMM of quarks. The fact that the dynamical generation of AMM is mainly a LLL effect suggests that the Schwinger linear-in-$B$ ansatz \cite{schwinger1948} for the quark AMM is not appropriate for massless fermions. This is argued to be in sharp contrast to theories with massive fermions, where a the Schwinger ansatz is allowed in the limit of weak magnetic fields \cite{ioffe1984}. Moreover, in \cite{ferrer2013}, the critical temperature of the second order phase transition of $\chi$SB is determined in a LLL approximation, and it is shown that the generation of the quark AMM increases $T_{c}$ as a function of $eB$, and therefore it cannot be responsible for the phenomenon of IMC.
In \cite{ferrer2014}, however, following the aforementioned proposal of $eB$ dependent  coupling constant, the effect of running coupling constant of the one-flavor NJL model on $T_{c}$ is studied. It is shown, that in the regime of strong magnetic fields, characterized by $eB\gg \Lambda_{\mbox{\tiny{QCD}}}^{2}$, the coupling constant of the model becomes anisotropic, and only the coupling parallel to the direction of the background magnetic field receives contributions from quarks in a LLL approximation. Interestingly, as a result of a certain antiscreening arising from the these quarks, this coupling decreases with increasing $eB$. The dependence of $T_{c}$ on this specific coupling leads to the desired phenomenon of IMC in the strong magnetic field limit.
\par
In the present paper, motivated by the above intriguing developments, we will consider the quark AMM in the one-loop effective potential of a magnetized two-flavor NJL model at finite $T$ and $\mu$. In contrast to \cite{martinez2013,ferrer2013, martinez2014-1}, and similar to the method used in \cite{strickland2012}, we will induce the quark AMM by an additional term $\hat{a}\sigma_{\mu\nu}F^{\mu\nu}$ in the original Lagrangian of the model, including massive quarks. The coefficient $\hat{a}$, proportional to the Bohr magneton $\mu_{B}\equiv \frac{e}{2m}$, will then be identified with the quark AMM. Here, $\mu_{B}$ depends on the constituent quark mass $m\equiv m_{0}+\sigma_{0}$, where $m_{0}$ is the current quark mass and $\sigma_{0}$ is the chiral condensate. This is also in contrast to \cite{martinez2014-1}, where the Bohr magneton is inversely proportional to current (bare) quark mass $m_{0}$.
The aforementioned additional term $\hat{a}\sigma_{\mu\nu}F^{\mu\nu}$ in the Lagrangian density of the NJL model leads to the energy dispersion
\begin{eqnarray}\label{Z1}
\hspace{-0.5cm}E^{(p,s)}_{q_{f}}=\sqrt{p_{3}^{2}+[(2p|q_{f}eB|+m^{2})^{1/2}-sT_{f}]^{2}},
\end{eqnarray}
for up and down quarks in the presence of a constant magnetic field. Here, $T_{f}=\kappa_{f}q_{f}eB$ with $\kappa_{f}\equiv \frac{\alpha_{f}}{2m}$, $p$ labels the Landau levels, $s=\pm 1$ states for the spin of quarks, $f=u,d$ labels the up ($u$) and down ($d$) flavors, and $q_{f}=2/3,-1/3$ is the charge of up and down quarks. Moreover, $\alpha_{f}$ is related to the deviation of the Land\'e $g$-factor from $2$. The above additional term $T_{f}$ is equivalent to the linear-in-$B$ ansatz by Schwinger $T_{f}^{sch.}=\alpha_{f}q_{f}\mu_{B}B$ for $T_{f}$, with the nonperturbative (effective) Bohr magneton \cite{ferrer2009} $\mu_{B}=\frac{e}{2m}$. Let us notice, that in the one-loop level, $\alpha_{f}$ is proportional to the electromagnetic fine structure constant $\alpha_{e}=\frac{1}{137}$. Perturbatively, it receives radiative corrections from the vertex function of quarks and the background photon field. In the framework of constituent quark model \cite{AMM1998}, it appears in the ratio $I_{f}=\frac{M}{1+\alpha_{f}}=\frac{\mu_{N}}{\mu_{f}}q_{f}m_{p}$, with $M$ the  constituent (effective) quark mass\footnote{Assuming the isospin symmetry, we have $M_{u}=M_{d}\equiv M$.}, $\mu_{N}\equiv \frac{e}{2m_{p}}$ the nuclear magneton, $\mu_{f}$ the magnetic moment of $f$-th quark flavor\footnote{The magnetic moment of quarks will be determined by the magnetic moments of protons and neutrons.}, and $m_{p}$ the proton mass (see Appendix \ref{appA} for more details). As it is argued in \cite{AMM1998}, choosing the experimental values for $\mu_{N}, \mu_{f}, q_{f}$ and $m_{p}$, the ratio $I_{f}$ is fixed to be $I_{u}\sim 338$ MeV for the up quark and $I_{d}\sim 322$ MeV for the down quark \cite{AMM1998}. A phenomenological constant value for $\kappa_{f}$, compatible with the constituent quark model, can then be determined by choosing an appropriate constant value for the constituent (effective) mass $M$, that yields $\alpha_{f}$, and consequently $\kappa_{f}$ through given values of $I_{f}, f=u,d$ and the definition $\kappa_{f}=\frac{\alpha_{f}}{2M}$.
\par
In this paper, we will consider three different sets for the dimensionful coupling $\kappa_{f}, f=u,d$ in $T_{f}=\kappa_{f}q_{f}eB$; two of them will be $(T,\mu,eB)$ independent, and arise, in the framework of constituent quark model, by separately choosing $M=420$ MeV and $M=340$ MeV, and fixing $\alpha_{f}$ through the phenomenologically given ratio  $I_{f}$, as described above.
The third set of $\kappa_{f}, f=u,d$ will depend on $T,\mu$ and $eB$, and include the $T$-dependent one-loop perturbative correction to $\alpha_{f}$, generalized from the QED results presented in  \cite{peressutti1982, masood2012}. Plugging first (\ref{Z1}) with given values of $\kappa_{f}$ into the one-loop effective potential of our model, and then minimizing the resulting expression with respect to $m$, we will determine the $T,\mu$ and $eB$ dependence of the constituent quark mass for each fixed set of $\kappa_{f}, f=u,d$. The complete $T$-$\mu$, $\mu$-$eB$ and $T$-$eB$ phase portrait of the model for various fixed $eB,T,\mu$ and $\kappa_{f}$ will also be explored.  We will show that for large enough phenomenological value for $\kappa_{f}$, the phenomenon of IMC occurs, i.e. in certain regimes of the parameter space, the critical temperature (chemical potential) $T_{c}$ ($\mu_{c}$) decreases with increasing $eB$. Moreover, for $(T,\mu,eB)$ dependent $\kappa_{f}$, including the one-loop perturbative correction for $\alpha$, the phenomenon of reentrance of $\chi$SB occurs for certain fixed values of $T$ and $\mu$, and in the regime of weak magnetic fields.
\par
The organization of this paper is as follows: In Sec. \ref{sec2}, we will introduce the quark AMM in the magnetized two flavor NJL model and will determine the one-loop effective action of this model at finite $T, \mu$ and $eB$. We will also introduce three different sets of the factor $\kappa_{f}, f=u,d$, mentioned above. In Sec. \ref{subsec3A}, the $T,\mu$ and $eB$ dependence of the constituent quark mass $m$ will be presented for each fixed $\kappa_{f}$. In Sec. \ref{subsec3B}, the complete phase portrait of the model will be explored for various fixed $\kappa_{f}$. We will also study, in Sec. \ref{subsec3C}, the effect of different sets of $\kappa_{f}$ on the pressure anisotropies in the longitudinal and transverse directions with respect to the direction of the magnetic field. Eventually, the $eB$ dependence of the magnetization $M$ will be demonstrated, and the effect of $\kappa_{f}$ on the product of $MB$ as a function of $eB$ will be studied. Section \ref{sec4} is devoted to our concluding remarks. In Appendix \ref{appA}, we will use the constituent quark model, and argue how the constant values of $\kappa_{f}, f=u,d$ can be determined by the phenomenological data of the magnetic moment of protons and neutrons.
\section{The model}\label{sec2}
\par\noindent
\setcounter{equation}{0}
We start with the Lagrangian density of a two-flavor NJL model in the presence of a uniform magnetic field,
\begin{eqnarray}\label{F1}
\lefteqn{
{\cal{L}}=\bar{\psi}(x)\left(i\gamma^{\mu}D_{\mu}-m_0+\frac{1}{2}\hat{a}\sigma^{\mu\nu}F_{\mu\nu}\right)
\psi(x)}\nonumber\\
&&+G\bigg\{[\bar{\psi}(x)\psi(x)]^2
+[\bar{\psi}(x)i\gamma_5\boldsymbol{\tau}\psi(x)]^2\bigg\}.
\end{eqnarray}
Here, the Dirac field $\psi_{f}^{c}$ carries two flavors, $f\in (1,2)=(u,d)$, and three colors, $c\in (1,2,3)=(r,g,b)$. In the presence of a constant magnetic field, the covariant derivative $D_{\mu}$ is defined by $D_{\mu}=-\partial_{\mu}+ie\hat{Q}A_{\mu}^{\mbox{\tiny{ext.}}}$, where the quark charge matrix is given by $\hat{Q}\equiv\mbox{diag}(q_{u},q_{d})=\mbox{diag}\left(2/3,-1/3\right)$,\footnote{In this paper, the hat symbol on each quantity denotes its matrix character in the two-dimensional flavor space.} and the gauge field $A_{\mu}^{\mbox{\tiny{ext.}}}=(0,0,Bx_{1},0)$ is chosen so that it leads to a uniform magnetic field $\mathbf{B}=B\mathbf{e}_{3}$, aligned in the third direction.
The up and down current quark masses are assumed to be equal, and denoted by $m_{0}\equiv m_{u}=m_{d}$. This guarantees the isospin symmetry of the theory in the limit of vanishing magnetic field. The latter is introduced by the term proportional to $\sigma_{\mu\nu}F^{\mu\nu}$ in the fermionic kernel of the Lagrangian density ${\cal{L}}$. Here, $\sigma_{\mu\nu}=\frac{i}{2}[\gamma_{\mu},\gamma_{\nu}]$, and the Abelian field strength tensor is given by  $F_{\mu\nu}=\partial_{[\mu}A_{\nu]}^{\mbox{\tiny{ext.}}}$, with $A_{\mu}^{\mbox{\tiny{ext.}}}$ given above.  The proportionality factor $\hat{a}$ is related to the quark AMM (for more details, see Appendix A). In the present two-flavor NJL model, $\hat{a}$ is defined by $\hat{a}\equiv \hat{Q}\hat{\alpha}\mu_{B}$, where $\hat{Q}$ and $\hat{\alpha}\equiv \mbox{diag}(\alpha_{u},\alpha_{d})$ are $2\times 2$ matrices in the flavor space. At one-loop level, and for a system with isospin symmetry $\alpha_{f}=\frac{\alpha_{e}q_{f}^{2}}{2\pi}$ for both $f=u,d$ flavors. Here, $\alpha_{e}=\frac{1}{137}$ is the electromagnetic fine structure constant, and  $\mu_{B}\equiv \frac{e}{2m}$ is given in term of the electric charge $e$ and the quark constituent mass $m\equiv m_{0}+\sigma_{0}$, where $m_{0}$ is the current quark mass and $\sigma_{0}$ the chiral condensate.\footnote{This is in contrast to \cite{martinez2013, martinez2014-1}, where the Bohr magneton is defined in term of the current quark mass $m_{0}$ (see below).} To introduce the chiral condensate $\sigma_{0}$, let us rewrite the NJL Lagrangian (\ref{F1}) in a semi-bosonized form
\begin{eqnarray}\label{F2}
\lefteqn{
{\cal{L}}_{sb}=\bar{\psi}(x)\left(i\gamma^{\mu}D_{\mu}-m_{0}+\hat{a}\sigma^{12}B\right)\psi(x)}\nonumber\\
&&-\bar{\psi}(x)\left(\sigma+i\gamma_{5}\boldsymbol{\tau}
\cdot \boldsymbol{\pi}\right)\psi(x)-\frac{(\sigma^{2}+\boldsymbol{\pi}^{2})}{4G},
\end{eqnarray}
where the mesonic fields $\sigma$ and $\boldsymbol{\pi}$ are defined by
\begin{eqnarray}\label{F3}
\sigma(x)=-2G\bar{\psi}(x)\psi(x), \qquad \boldsymbol{\pi}=-2G\bar{\psi}(x)i\gamma_{5}\boldsymbol{\tau}\psi(x). \nonumber\\
\end{eqnarray}
Here, as in (\ref{F1}), $G$ is the NJL mesonic coupling and $\boldsymbol{\tau}=(\tau_{1},\tau_{2},\tau_{3})$  are the Pauli matrices. To arrive at (\ref{F2}), $F_{12}=-F_{21}=B$ is used. Integrating out the fermionic degrees of freedom, and using
\begin{eqnarray}\label{F4}
e^{i\Gamma_{\mbox{\tiny{eff.}}}[\sigma,\boldsymbol{\pi}]}=\int {\cal{D}}\bar{\psi}{\cal{D}}\psi\exp\left(i\int d^{4}x{\cal{L}}_{sb}\right),
\end{eqnarray}
the effective action of mesons $(\sigma,\boldsymbol{\pi})$,  $\Gamma_{\mbox{\tiny{eff.}}}=\Gamma_{\mbox{\tiny{eff.}}}^{(0)}+
\Gamma_{\mbox{\tiny{eff.}}}^{(1)}$ is derived. It is given in terms of a tree-level action
\begin{eqnarray}\label{F5}
\Gamma_{\mbox{\tiny{eff.}}}^{(0)}[\sigma,\boldsymbol{\pi}]=-\int d^{4}x \frac{(\sigma^{2}+\boldsymbol{\pi}^{2})}{4G},
\end{eqnarray}
and a one-loop effective action
\begin{eqnarray}\label{F6}
\Gamma_{\mbox{\tiny{eff.}}}^{(1)}=-i\mbox{Tr}_{\{cf\sigma x\}}\ln[iS_{Q}^{-1}(\sigma,\boldsymbol{\pi})],
\end{eqnarray}
where the inverse fermion propagator is formally given by
\begin{eqnarray}\label{F7}
\hspace{-0.4cm}iS_{Q}^{-1}(\sigma,\boldsymbol{\pi})=i\gamma^{\mu}D_{\mu}+\hat{a}\sigma^{12}B-\left(\bar{m}
+i\gamma_{5}
\boldsymbol{\tau}\cdot \boldsymbol{\pi}\right).
\end{eqnarray}
Here, $\bar{m}=m_{0}+\sigma(x)$. Expanding the effective action around a constant configuration $(\sigma_{0},\boldsymbol{\pi}_{0})=(\mbox{const.},\mathbf{0})$ for the mesonic fields $(\sigma,\boldsymbol{\pi})$, the constituent mass $\bar{m}$ turns out to be constant, and can be given by $m=m_{0}+\sigma_{0}$. By carrying out the trace operation over color ($c$), flavor ($f$), spinor ($\sigma$) degrees of freedom, as well as over the four-dimensional space-time coordinate ($x$), the one-loop effective action of this model reads
\begin{eqnarray}\label{F8}
\Gamma_{\mbox{\tiny{eff.}}}^{(1)}=-6i\sum\limits_{q_{f}=\{\frac{2}{3},-\frac{1}{3}\}}\ln\mbox{det}_{x}
\left(E_{q_{f}}^{(p,s)2}-p_{0}^{2}\right).
\end{eqnarray}
Here, the energy spectrum of up and down quarks in the presence of external magnetic fields and for nonvanishing AMM is given by
\begin{eqnarray}\label{F9}
\lefteqn{E_{q_{f}}^{(p,s)}
}\nonumber\\
&&\hspace{-0.5cm}=\sqrt{p_{3}^{2}+[\left(|q_{f}eB|[2p+1-s\xi_{f}]+m^{2}\right)^{1/2}-s\kappa_{f}q_{f}eB
]^{2}},\nonumber\\
\end{eqnarray}
where $p\geq 0$ labels the Landau levels, and $s=\pm 1$ stands for the spin of the quarks. Moreover,  $\xi_{f}\equiv \mbox{sgn}(q_{f}eB)$ and $\kappa_{f}\equiv \frac{\alpha_{f}}{2m}$, with $f=u,d$. The energy $E_{q_{f}}^{(n,s)}$ arises either by computing the fermion determinant of the present two-flavor NJL model including the quark AMM, or by solving the modified Dirac equation $(\gamma\cdot\Pi-m+\frac{1}{2}\hat{a}\sigma^{\mu\nu}F_{\mu\nu})\psi(x)=0$, with $\Pi_{\mu}\equiv i\partial_{\mu}-eQA_{\mu}$, using \textit{e.g.}, the Ritus eigenfunction method \cite{ritus}. Following this method, which is descibed also in \cite{taghinavaz2012} for two- and four-dimensional space-time, it turns out that for nonvanishing fermion AMM, only the energy eigenvalues of the Dirac operator are modified, while the Ritus eigenfunctions remain the same as for vanishing AMM. Let us notice that, the additional $\hat{a}\sigma^{\mu\nu}F_{\mu\nu}$ term in the Dirac operator  commutes with $(\gamma\cdot \Pi)^{2}$. This is why the modified Dirac operator $\gamma\cdot\Pi-m+\frac{1}{2}\hat{a}\sigma^{\mu\nu}F_{\mu\nu}$ has the same eigenfunctions as the ordinary Dirac $\gamma\cdot\Pi-m$ for vanishing AMM.
\par
Let us compare, at this stage, the energy dispersion (\ref{F9}), with the energy dispersion (\ref{Z1}) appearing also in \cite{ferrer2009,martinez2013, martinez2014-1}.\footnote{We have slightly changed the notations of Ferrer et al. in \cite{ferrer2009,martinez2013, martinez2014-1}, to be able to compare their results with ours. In particular,  to consider the multi-flavor nature of the NJL model in the present paper, we have inserted $q_{f}$ wherever it was necessary.}  According to Schwinger \cite{schwinger1948}, $T_{f}$ is linear in $B$, and is given by $T_{f}^{sch.}\equiv \alpha_{f}q_{f}\mu_{B}B$. Plugging $\mu_{B}=\frac{e}{2m}$ into this relation and using $\kappa_{f}=\frac{\alpha_{f}}{2m}$, we obtain $T_{f}^{sch.}=\kappa_{f}q_{f}eB$, which appears also in (\ref{F9}). Thus, starting from the above $T_{f}^{sch.}$, the two energy dispersion relations (\ref{F9}) and (\ref{Z1}) are equal. The Schwinger linear-in-$B$ ansatz for AMM is recently discussed in the literature \cite{martinez2013,martinez2014-1}. In particular, it is shown, that this ansatz is only valid in the weak-field limit $eB\ll m_{0}^{2}$. In the present paper, however, we will use it for the whole range of $eB$. To explain this apparent discrepancy, let us notice that the crucial difference between our approach and the one presented in \cite{martinez2013, martinez2014-1}, is that in our approach the Bohr magneton $\mu_{B}$ is defined in term of the constituent quark mass $m=m_{0}+\sigma_{0}$, with $m_{0}$ the current quark mass and $\sigma_{0}$ the chiral condensate, while in \cite{martinez2013, martinez2014-1}, the Schwinger ansatz is defined in term of $\mu_{B}^{0}\equiv\frac{e}{2m_{0}}$, and is independent of $\sigma_{0}$. Having this in mind, we expect, that the difficulties related to the Schwinger linear ansatz with $\mu_{B}^{0}$, described in \cite{martinez2013,martinez2014-1}, do not occur in our work. 
\par
As described in Sec. \ref{sec1}, in the present paper, the quark AMM will be introduced in the framework of a constituent quark model. This is in contrast to
\cite{ferrer2009, ferrer2013, martinez2013, martinez2014-1}, where a dynamical symmetry breaking is responsible for its generation. In \cite{ferrer2013}, for instance, the quark AMM is dynamically induced through a nonvanishing spin-one condensate. Here, starting from a massless theory, it is shown that since
a nonperturbative mechanism of quark pairing, mainly in the LLL, is responsible for the dynamical generation of AMM, the Schwinger linear-in-$B$ ansatz, $T_{f}^{sch.}=\alpha_{f}q_{f}\mu_{B}^{0}B$, is not even valid in the aforementioned weak-field approximation.\footnote{The LLL approximation is allowed only in the strong-field limit $eB\gg m_{0}^{2}$. Moreover, the nonperturbative result for AMM, arising in the LLL approximation \cite{ferrer2013} is non-analytic, and cannot be Taylor expanded in the orders of $B$.} In the present paper, however, in contrast to \cite{ferrer2013}, we do not start with a massless Dirac theory. Moreover, we will work, in contrast to \cite{ferrer2013}, in the supercritical regime of the NJL model, i.e., we will choose the NJL coupling $G$ in such a way that the model exhibits a dynamical mass, even for zero $eB$.
We will consider the contributions of all Landau level, and will not restrict ourselves to LLL, as in \cite{ferrer2013}, nor to one-loop approximation in the LLL, as in \cite{martinez2013, martinez2014-1}. Hence, starting from the Schwinger linear ansatz with $\mu_{B}$ instead of $\mu_{B}^{0}$ seems to be reasonable within the constituent quark model. Here, although the quark AMM is not dynamically generated, as in \cite{ferrer2013, martinez2013,martinez2014-1}, but it is related to the dynamically generated quark mass $m$ through the nonperturbative (effective) Bohr magneton \cite{ferrer2009} $\mu_{B}$ in  $T_{f}^{sch.}\equiv \alpha_{f}q_{f}\mu_{B}B$.
\par
The constituent quark mass $m$ is determined by minimizing the thermodynamic (one-loop effective) potential, arising from (\ref{F8}).  For a constant configuration $\sigma_{0}$, the one-loop effective potential of the theory is given by performing the remaining determinant over the coordinate space in (\ref{F8}). This leads to the one-loop effective potential $\Omega_{\mbox{\tiny{eff.}}}$, defined by
$\Omega_{\mbox{\tiny{eff.}}}^{(1)}\equiv -{\cal{V}}^{-1}\Gamma_{\mbox{\tiny{eff.}}}^{(1)}$, where ${\cal{V}}$ denotes the four-dimensional space-time volume. In momentum space, the aforementioned determinant is evaluated by the standard replacement
\begin{eqnarray}\label{F10}
\lefteqn{\hspace{-0.9cm}\int\frac{d^{4}p}{(2\pi)^{4}}f(p_{0}, \mathbf{p})\to \frac{|q_{f}eB|}{\beta}}
\nonumber\\
&&\hspace{-1.2cm}\times\sum\limits_{n=-\infty}^{+\infty}\sum\limits_{p=0}^{\infty}\sum\limits_{s=\pm 1}\int\frac{dp_{3}}{8\pi^{3}}f(i\omega_{n}-\mu,p,s,p_{3}),
\end{eqnarray}
where temperature $T$ and chemical potential $\mu$ are introduced by replacing $p_{0}$ with $p_{0}=i\omega_{n}-\mu$. Here, the Matsubara frequencies $\omega_{n}\equiv \frac{(2n+1)\pi}{\beta}$ with $\beta\equiv T^{-1}$ are labeled by $n$. After performing the sum over $n$, the effective potential of the two-flavor NJL model, including the tree-level and the one-loop part  is given by
\begin{widetext}
\begin{eqnarray}\label{F11}
\lefteqn{\hspace{-0.5cm}\Omega_{\mbox{\tiny{eff.}}}(m;T,\mu,eB)=\frac{(m-m_{0})^{2}}{4G}
}\nonumber\\
&&\hspace{-0.5cm}-3\sum\limits_{q_{f}=\{\frac{2}{3},-\frac{1}{3}\}}\frac{|q_{f}eB|}{\beta}\sum\limits_{p=0}^{\infty}\sum\limits_{s=\pm 1}\int\frac{dp_{3}}{4\pi^{2}}\bigg\{\beta E_{q_{f}}^{(p,s)}
+\ln\left(1+e^{-\beta(E_{q_{f}}^{(p,s)}+\mu)}\right)+
\ln\left(1+e^{-\beta(E_{q_{f}}^{(p,s)}-\mu)}\right)\bigg\}.
\end{eqnarray}
\end{widetext}
In the rest of this paper, we will use the above effective potential to study the phase diagram of the magnetized two-flavor NJL model at finite $(T,\mu,eB)$ and for a given value of $\kappa_{f}, f=u,d$. To do this, we will follow two different approaches:
\par
In the first approach, we assume that the value of $\kappa_{f}=\frac{\alpha_{f}}{2m}$, appearing in the quark energy dispersion relation $E_{q_{f}}^{(p,s)}$ from (\ref{F9}), is constant and independent of $(T,\mu,eB)$. Minimizing the effective potential $\Omega_{\mbox{\tiny{eff}}}$ from (\ref{F11}) with respect to $m$, we will numerically determine the constituent quark mass $m$ as a function of $(T,\mu,eB)$ and for a fixed value of $\kappa_{f}, f=u,d$. In this approach, $\alpha_{f}=2m\kappa_{f}$ will depend on $(T,\mu,eB)$ only through the constituent mass $m$.
To fix $\hat{\kappa}$, we will follow the method described in Appendix A. Here, two  different sets of constant $\hat{\kappa}_{i}=\mbox{diag}(\kappa_{u}^{(i)},\kappa_{d}^{(i)}), i=1,2$ are introduced, which are compatible with the constituent quark model [see (\ref{appA10}) and (\ref{appA12})]. To have a sizable quark AMM, we will use
\begin{eqnarray}\label{F12}
\kappa_{u}^{(1)}&=&0.29016~\mbox{GeV}^{-1},\nonumber\\ \kappa_{d}^{(1)}&=&0.35986~\mbox{GeV}^{-1},
\end{eqnarray}
[see also (\ref{appA9})]. The second set yields smaller values of $\alpha_{f}, f=u,d$,
\begin{eqnarray}\label{F13}
\kappa_{u}^{(2)}&=&0.00995~\mbox{GeV}^{-1},\nonumber\\ \kappa_{d}^{(2)}&=&0.07975~\mbox{GeV}^{-1},
\end{eqnarray}
[see also (\ref{appA11})]. In Sec. \ref{sec3}, we will in particular show, that once $\hat{\kappa}$ in (\ref{F9}) is chosen to be $\hat{\kappa}_{1}$ from (\ref{F12}), the phenomenon of IMC occurs in the phase diagram of our two-flavor magnetized NJL model.
\par
In the second approach, we will consider the leading one-loop correction to the quark (electromagnetic) AMM at nonzero $T$ and vanishing $\mu$ and $eB$. Let us denote it   by $\alpha_{f}=\alpha_{f}^{\mbox{\tiny{1-loop}}}(T,\mu=0,eB=0)$ (see below). To determine the constituent quark mass $m$, we will insert $\kappa_{f}=\frac{\alpha_{f}}{2m}$ with $\alpha_{f}=\alpha_{f}^{\mbox{\tiny{1-loop}}}$ into (\ref{F9}), and then (\ref{F9}) into the one-loop effective potential (\ref{F11}), whose minima will then lead to $(T,\mu,eB)$ dependent $m$.
\par
To determine the one-loop correction $\alpha_{f}^{\mbox{\tiny{1-loop}}}$, we use the result presented in \cite{peressutti1982,masood2012} for thermal QED, and generalize it to the case of QCD. In \cite{masood2012}, the anomalous magnetic moment of electrons at one- and two-loop orders are computed at finite $T$, zero $\mu$ and for vanishing $eB$. Since in our model the electromagnetic coupling between the quarks and the external photons is to be considered, the QED results presented in \cite{masood2012} are directly applicable for our QCD-like model, including quarks with different flavors $f=u,d$. It is enough to multiply the corresponding QED result with $q_{f}^{2}$, where $q_{f}$ is the charge of the $f$-th quark flavor. We therefore have
\begin{eqnarray}\label{F14}
\kappa_{f}(T,\mu,eB)=\frac{\alpha_{f}^{\mbox{\tiny{1-loop}}}(T,\mu=0,eB=0)}{2m(T,\mu,eB)},
\end{eqnarray}
with the one-loop contribution
\begin{eqnarray}\label{F15}
\alpha_{f}^{\mbox{\tiny{1-loop}}}(T,\mu=0,eB=0)=q_{f}^{2}
{\cal{F}}(m^{(0)}\beta),
\end{eqnarray}
for $f=u,d$. Here, $m^{(0)}\equiv m(T=\mu=eB=0)$ is the constituent quark mass at zero $(T,\mu,eB)$, and
\begin{eqnarray}\label{F16}
\lefteqn{{\cal{F}}(z,\alpha_{e})=\frac{\alpha_{e}}{2\pi}}\nonumber\\
&&-\frac{2\alpha_{e}}{3}\left(
\frac{\pi }{2z^{2}}\big[1-{\cal{C}}(z)\big]+{\cal{A}}(z)-{\cal{B}}(z)
\right),
\end{eqnarray}
with $\alpha_{e}$ the electromagnetic fine structure constant, and the functions ${\cal{A}}, {\cal{B}}$ and ${\cal{C}}$ given by
\begin{eqnarray}\label{F17}
{\cal{A}}(z)&=&\frac{2T}{\pi m}\ln(1+e^{-z}),\nonumber\\
{\cal{B}}(z)&=&\sum\limits_{n=1}^{\infty}(-1)^{n}\mbox{Ei}(-nz),\nonumber\\
{\cal{C}}(z)&=&\frac{6}{\pi}\sum\limits_{n=1}^{\infty}\frac{(-1)^{2}}{n^{2}}e^{-nz}.
\end{eqnarray}
Here, $\mbox{Ei}(z)$ is the well-known exponential integral function, defined by
\begin{eqnarray}\label{F18}
\mbox{Ei}(z)=-\int_{-z}^{\infty}\frac{e^{-t}}{t}~dt.
\end{eqnarray}
Let us notice, that since $\alpha_{e}\ll 1$, the one-loop contribution $\alpha_{f}^{\mbox{\tiny{1-loop}}}$ is indeed the dominant radiative correction. This is why, the higher order loop corrections will be neglected in the present paper.
\section{Numerical Results}\label{sec3}
\setcounter{equation}{0}
\par\noindent
The aim of this paper is to study the effect of the quark AMM on a hot and magnetized quark matter. To this purpose, the thermodynamic potential of a two-flavor NJL model is determined in the previous section. It mimics the thermodynamic properties of quark matter at high temperature, finite density and in the presence of an external magnetic field. The quantity, $\hat{\alpha}=\mbox{diag}(\alpha_{u},\alpha_{d})$, which is introduced in the original Lagrangian (\ref{F1}) as the coefficient corresponding to the spin-field interaction term $\sigma^{\mu\nu}F_{\mu\nu}$, appears in the quark energy dispersion relation $E_{q_{f}}^{(p,s)}$ from (\ref{F9}) in the one-loop effective potential (\ref{F11}), essentially in the combination with the constituent quark mass in $\kappa_{f}=\frac{\alpha_{f}}{2m}, f=u,d$. Using (\ref{F11}), it is now possible to determine the thermodynamic properties of the present quark model, and, in particular, to study various effects of the quark AMM on hot and magnetized quark matter.
\par
As we have discussed in Sec. \ref{sec1}, the most important effect of the presence of an uniform magnetic field on a system of charged fermions is the phenomenon of MC of chiral symmetry breaking. The first signature of the occurrence of this phenomenon is that the value of the chiral condensate $\sigma_{0}$, and consequently the constituent quark mass $m=m_{0}+\sigma_{0}$ increase with increasing the strength of the external magnetic field $eB$ (see \cite{fayazbakhsh2, fayazbakhsh3, fayazbakhsh4} for detailed discussions).  To study the impact of the quark AMM on this specific effect, we will first determine, in Sec. \ref{subsec3A1}, the $eB$ dependence of $m$ for zero chemical potential, at fixed temperature and for three different sets of $\hat{\kappa}$. They will be denoted by $\kappa_{1}, \kappa_{2}$ and $\kappa'$. Here, $\kappa_{1}$ and $\kappa_{2}$ correspond to the pair $\hat{\kappa}_{i}=(\kappa_{u}^{(i)}, \kappa_{d}^{(i)}), i=1,2$ from (\ref{F12}) and (\ref{F13}), and, $\kappa'$ corresponds to the $T$-dependent $\kappa_{f}$ given in (\ref{F14}). The latter includes one-loop perturbative correction to the AMM of quarks and is written as a function of the constituent quark mass $m$. We will show that whereas for $\kappa_{2}$ and $\kappa'$, $m$ increases with increasing $eB$, for $\kappa_{1}$, it decreases with increasing $eB$. Moreover, in the latter case, the $eB$ dependence of $m$ indicates a first order phase transition at certain critical $eB_{c}$ even at $T=\mu=0$ MeV. This preliminary, unexpected result can be regarded as an indication of the phenomenon of IMC.
As it is shown in \cite{fayazbakhsh2, fayazbakhsh3}, the formation of the chiral condensate is suppressed by increasing temperature. In Sec. \ref{subsec3A2}, we will study the effect of finite temperature on $m$ for zero chemical potential and various values of $eB$ and $\hat{\kappa}$. We will then compare the corresponding results to $\kappa_{1}, \kappa_{2}$ and $\kappa'$, and elaborate on the difference between the effect of these three different choices for $\hat{\kappa}$.
\par
In Sec. \ref{subsec3B}, we will then explore the complete phase portrait of our two-flavor hot and magnetized NJL model. The effect of external magnetic field on the nature of phase transition for $\hat{\kappa}=0$ GeV$^{-1}$ is previously studied in \cite{fayazbakhsh2, fayazbakhsh3, fayazbakhsh4}. It is, in particular, shown that for zero chemical potential, the critical temperature $T_c$ essentially increases with increasing $eB$. This can be regarded as the another indication of the phenomenon of MC. In Sec. \ref{subsec3B1}, we will first study the $T$--$eB$ phase diagram of our two-flavor hot and magnetized NJL model for zero and nonzero $\mu$ and for different $\hat{\kappa}$. We will, in particular, show that
the critical temperature of the phase transition decreases with increasing $eB$ once $\hat{\kappa}$ is chosen to be $\kappa_{1}$. This can be regarded as another signature of the aforementioned phenomenon of IMC.
In Secs. \ref{subsec3B2} and \ref{subsec3B3}, we will then study the $\mu$--$eB$ and $T$--$eB$ phase diagrams for various fixed temperatures and magnetic fields, respectively. We will, in particular, compare the results for $\kappa_{1}$ and $\kappa_{2}$ with $\hat{\kappa}=0$ GeV$^{-1}$, previously discussed in \cite{fayazbakhsh2,fayazbakhsh3}.
\par
In Sec. \ref{subsec3C}, we will finally study the pressure anisotropy between the longitudinal and transverse pressures with respect to the direction of the external magnetic field. Moreover, the $eB$ dependence of the magnetization of the quark matter will be determined for zero and nonzero chemical potential, temperature and for $\kappa_{1}, \kappa_{2}$ and $\kappa'$. We will show that whereas $\hat{\kappa}$ has essentially no effect on the pressure anisotropy, larger values of $\kappa$ leading to sizable values of the quark AMM, suppress the product $MB$, where $M$ is the quark matter magnetization. The relation between discontinuities appearing in $M$ and the first order phase transitions will be also discussed.
\par
To perform the numerical analysis in this section, we will essentially use the same method as described in \cite{fayazbakhsh2} and \cite{fayazbakhsh3}. In order to determine the constituent quark mass $m$, we will numerically solve the gap equation corresponding to the thermodynamic potential (\ref{F11})
\begin{eqnarray}\label{A1}
\frac{\partial\Omega_{\mbox{\tiny{eff}}}(\tilde{m};T,\mu,eB)}{\partial \tilde{m}}\bigg|_{\tilde{m}=m}=0.
\end{eqnarray}
Here, $m=m_{0}+\sigma_{0}$.
Our specific choice for free parameters of our model, the ultraviolet (UV) momentum cutoff $\Lambda$, the NJL (chiral) coupling constant $G$, and the current quark mass $m_{0}$, is as follows:
\begin{eqnarray}\label{A2}
&&\Lambda=0.6643~\mbox{GeV},\qquad G=4.668~\mbox{GeV}^{-2},\qquad\mbox{and} \nonumber\\
&&m_{0}=5~\mbox{MeV}.
\end{eqnarray}
The numerical integration over $p_{3}$, appearing in (\ref{F11}) will be performed using a smooth cutoff function
\begin{eqnarray}\label{A3}
f_{\Lambda}=\frac{1}{1+\exp\left(\frac{|\mathbf{p}|-\Lambda}{\Lambda}\right)},
\end{eqnarray}
for vanishing $eB$, and
\begin{eqnarray}\label{A4}
\hspace{-0.5cm}f_{\Lambda,B}^{(p,s)}=\frac{1}{1+\exp\left(\frac{\sqrt{p_{3}^{2}+|q_{f}eBp|[2p+1-s\xi_{f}]}-
\Lambda}{A}\right)},
\end{eqnarray}
for nonvanishing $eB$. Here, $q_{f}$ is the electric charge of the $f$-th quark, $p$ labels the Landau levels, $s=\pm 1$ stands for positive and negative spins of the quarks and $\xi_{f}=\mbox{sgn}(q_{f}eB)$. Moreover $A$ is a free parameter, which is related to the sharpness of the cutoff scheme.  As in \cite{fayazbakhsh2, fayazbakhsh3, fayazbakhsh4}, $A$ is chosen to be $A=0.05\Lambda$, where the UV cutoff $\Lambda$ is given in (\ref{A2}).  To determine the constituent quark mass $m$, the global minima of $\Omega_{\mbox{\tiny{eff}}}$ from (\ref{F11}) are to be determined. In the chiral limit $m_{0}\to 0$, the $\chi$SB is characterized by nonvanishing chiral condensate $\sigma_{0}$, and the chiral symmetry restored ($\chi$SR) phase by $\sigma_{0}=0$. As it is shown in \cite{fayazbakhsh2,fayazbakhsh3,fayazbakhsh4}, in the limit of vanishing $m_{0}$ and $\hat{\kappa}$, the presence of external magnetic field induces a first order phase transition from the $\chi$SB into the $\chi$SR phase. For $m_{0}\neq 0$ and $\hat{\kappa}=0$ GeV$^{-1}$, however, we expect a smooth crossover from the chiral $\chi SB$, characterized by $\sigma_{0}\neq 0$, into the pseudo-chiral symmetry restored (p$\chi$SR) phase, characterized by $\sigma_{0}=0$ MeV and  $m=m_{0}$. In Sec. \ref{subsec3C}, the order of the phase transition for $m_{0}\neq 0$ and $\hat{\kappa}\neq 0$ will be elaborated.
\subsection{The $T$ and $eB$ dependence of the constituent quark mass $m$}\label{subsec3A}
\subsubsection{The $eB$ dependence of the constituent quark mass $m$}\label{subsec3A1}
\begin{figure}[hbt]
\includegraphics[width=8.5cm,height=6cm]{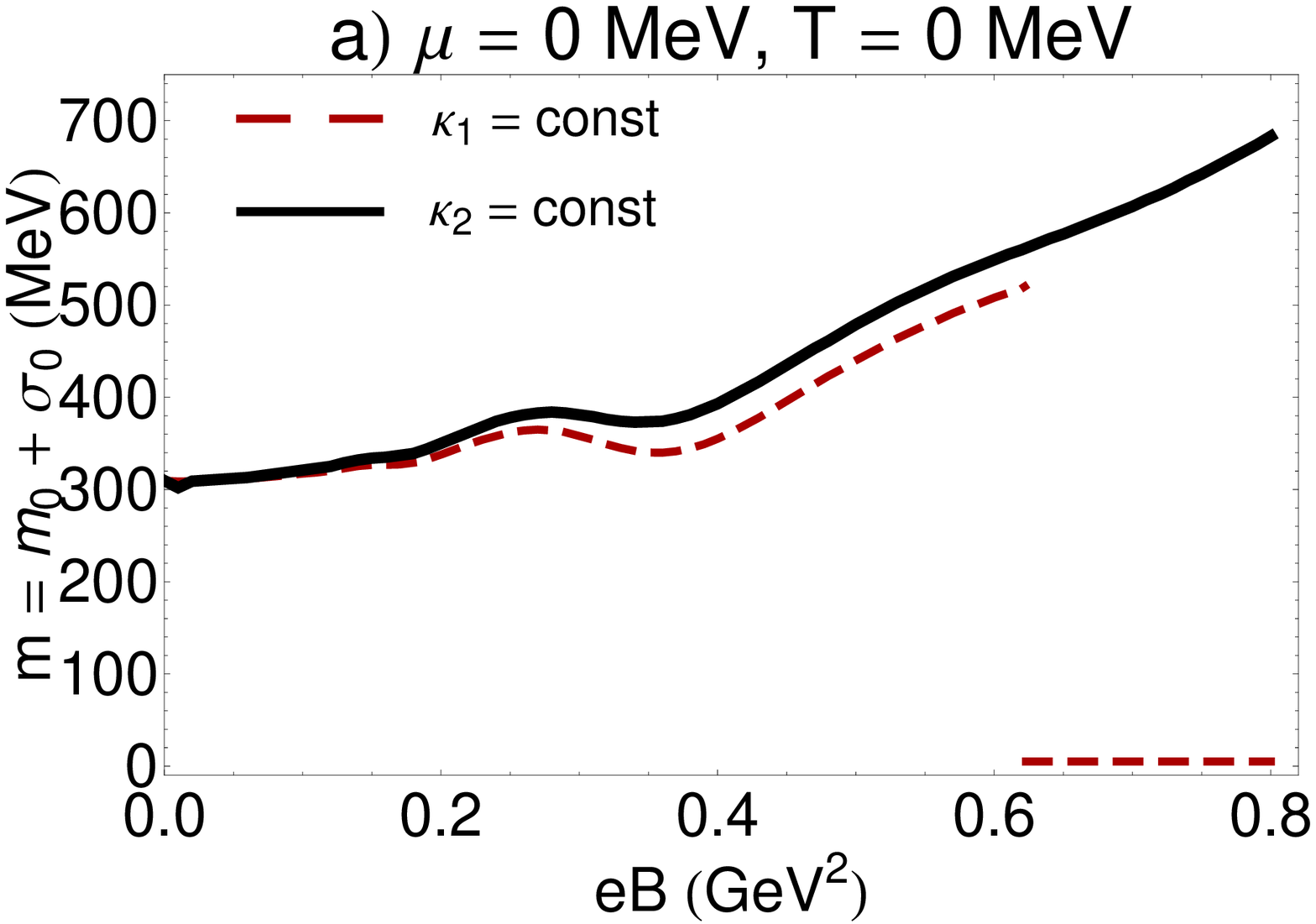}
\includegraphics[width=8.5cm,height=6cm]{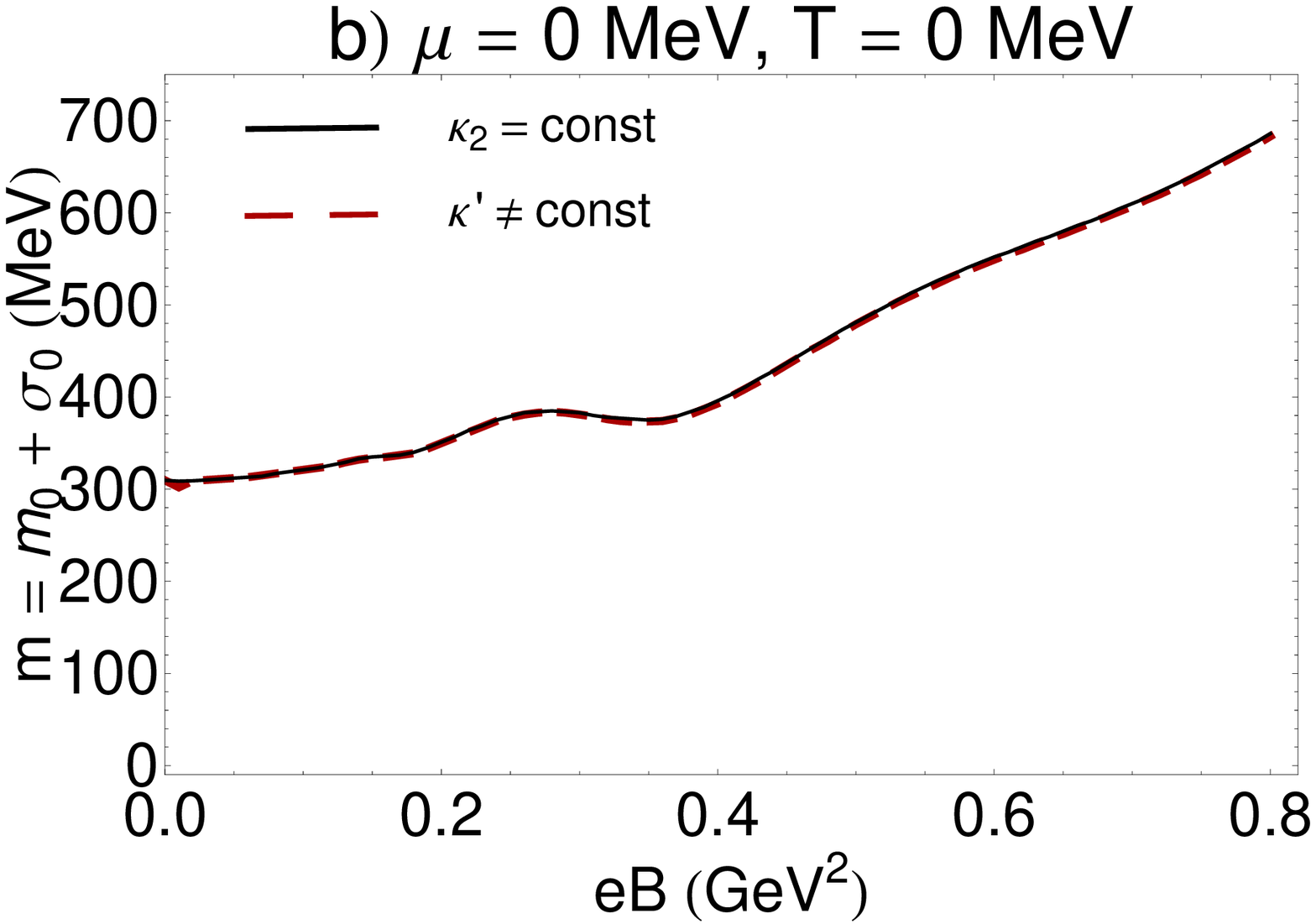}
\caption{(color online). (a) The $eB$ dependence of quark constituent mass $m=m_{0}+\sigma_{0}$ is demonstrated for $\kappa_{1}$ (red dashed line) and $\kappa_{2}$ (black solid line)  at $T=\mu=0$ MeV. The behavior of $m$ as a function of $eB$ for $\kappa_{1}$ suggests a first order phase transition at a critical $eB_{c}\sim 0.623$ GeV$^{2}$. This can be regarded as possible signature of the phenomenon of IMC. (b) The $eB$ dependence of $m$ is compared for $\kappa_{2}$ (black solid line) and $\kappa'$ (red dashed line) at $T=\mu=0$ MeV. The fact that $m$ increases with increasing $eB$ is related to the phenomenon of MC.}\label{fig1}
\end{figure}
\begin{figure}[hbt]
\includegraphics[width=8.5cm,height=6cm]{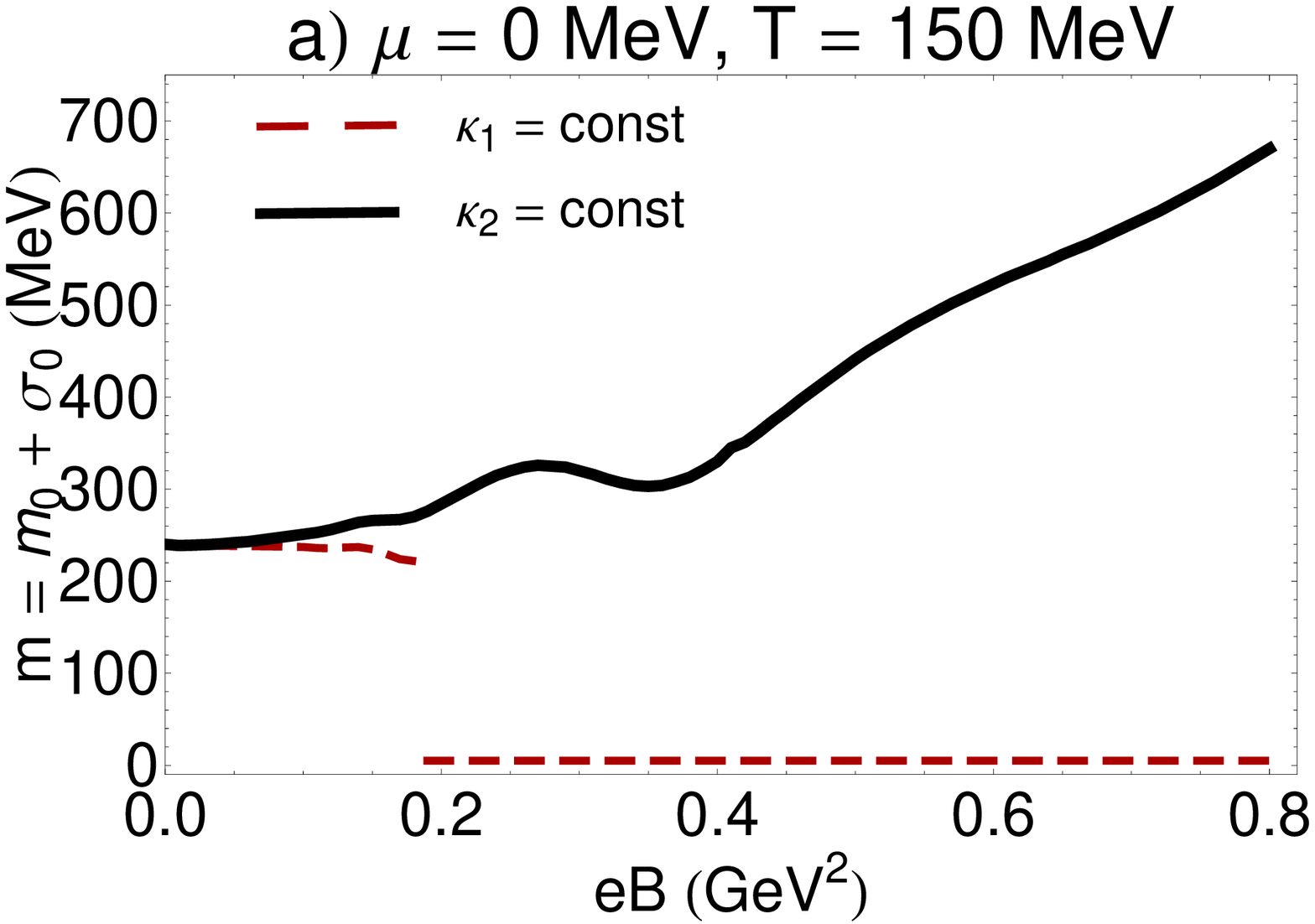}
\includegraphics[width=8.5cm,height=6cm]{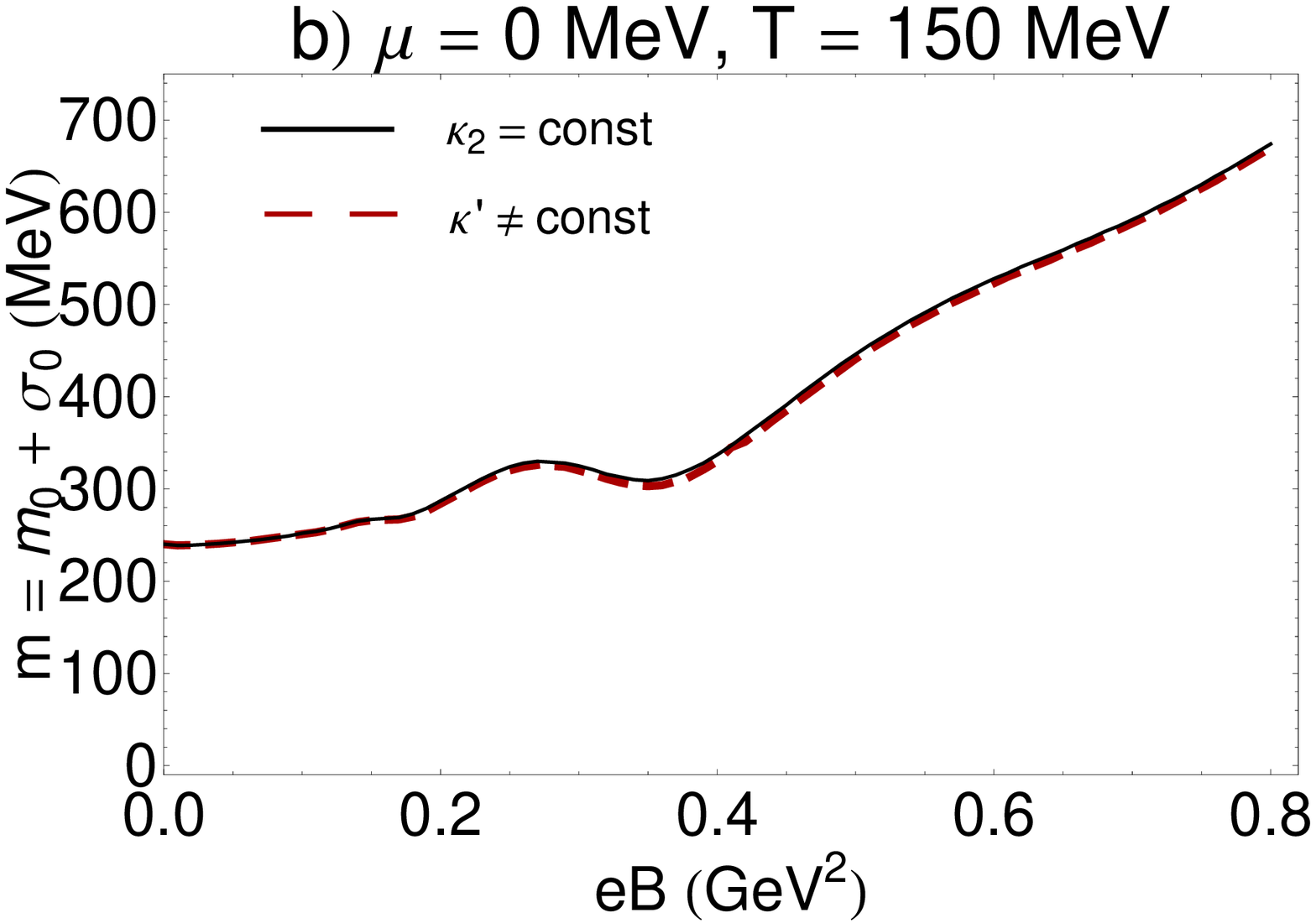}
\caption{(color online). (a) The $eB$ dependence of quark constituent mass $m=m_{0}+\sigma_{0}$ is demonstrated for $\kappa_{1}$ (red dashed line) and $\kappa_{2}$ (black solid line)  at $T=150$ MeV and $\mu=0$ MeV. As in the $T=0$ MeV case of Fig. \ref{fig1}, the behavior of $m$ as a function of $eB$ for $\kappa_{1}$ suggests a first order phase transition at a critical $eB_{c}\sim 0.190$ GeV$^{2}$. This is a possible signature of IMC. (b) The $eB$ dependence of $m$ is compared for $\kappa_{2}$ (black solid line) and $\kappa'$ (red dashed line) at $T=150$ MeV and $\mu=0$ MeV. }\label{fig2}
\end{figure}
\begin{figure}[hbt]
\includegraphics[width=8.5cm,height=6cm]{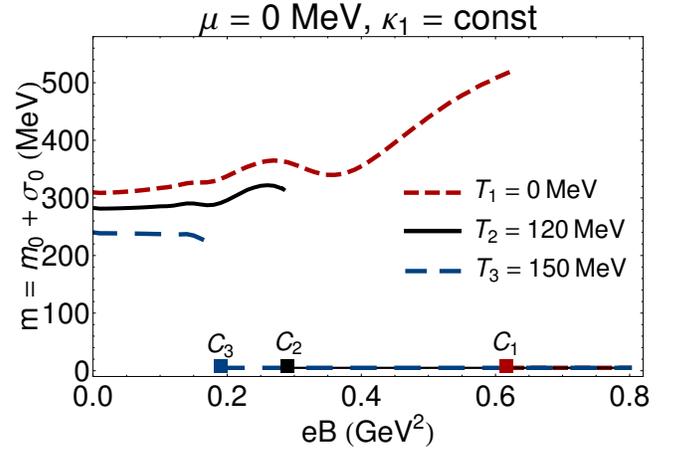}
\caption{(color online). The eB dependence of the constituent quark mass $m=m_{0}+\sigma_{0}$ is demonstrated for $\mu=0$ MeV and at different temperatures $T_{1}=0$ MeV, $T_{2}=150$ MeV and $T_{3}=150$ MeV. Here, $\hat{\kappa}$ is held fixed to be $\kappa_{1}$. The critical magnetic fields corresponding to $T_{i}, i=1,2,3$ MeV are denoted by $C_{1}$, $C_{2}$ and $C_{3}$, respectively. They are given by $eB_{c_1}=0.623$ GeV$^{2}$, $eB_{c_2}= 0.285$ GeV$^{2}$ and $eB_{c_{3}}=0.190$ GeV$^{2}$. As it turns out, $eB_{c}$ deceases with increasing $T$. This is an indication of IMC, that apparently occurs once the quark AMM is large enough.}\label{fig3}
\end{figure}
\begin{figure*}[hbt]
\includegraphics[width=5.9cm,height=4.4cm]{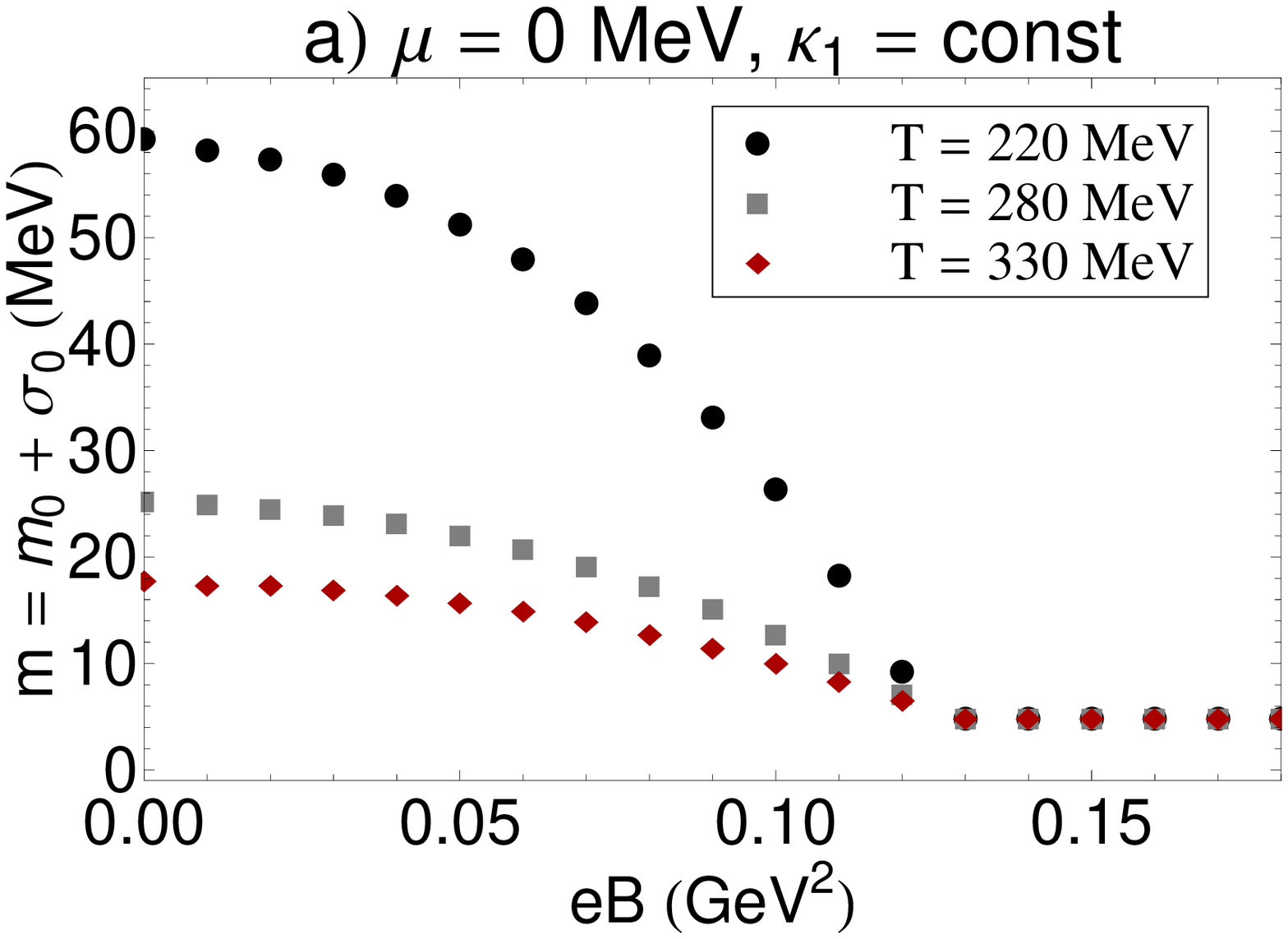}
\includegraphics[width=5.9cm,height=4.4cm]{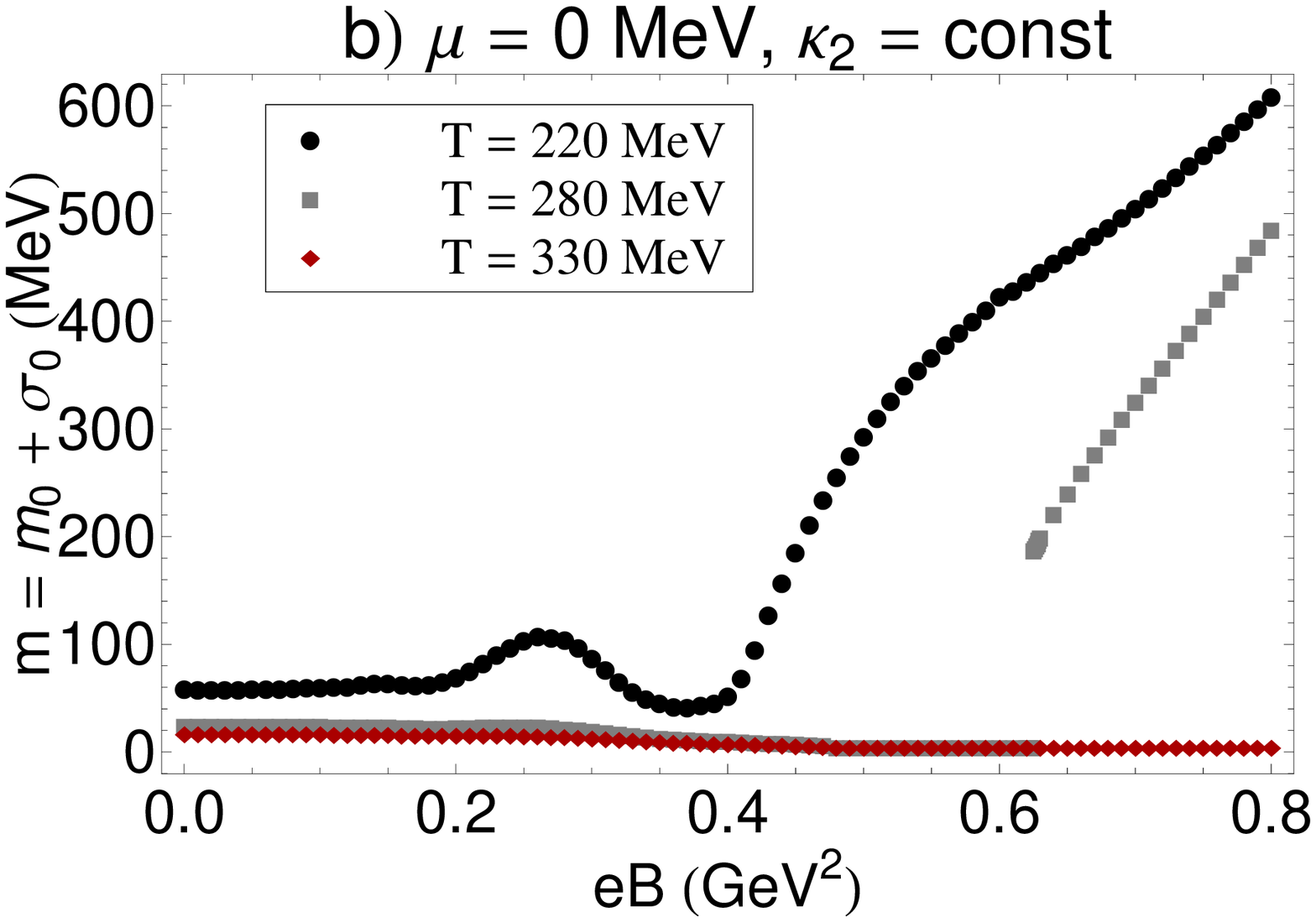}
\includegraphics[width=5.9cm,height=4.4cm]{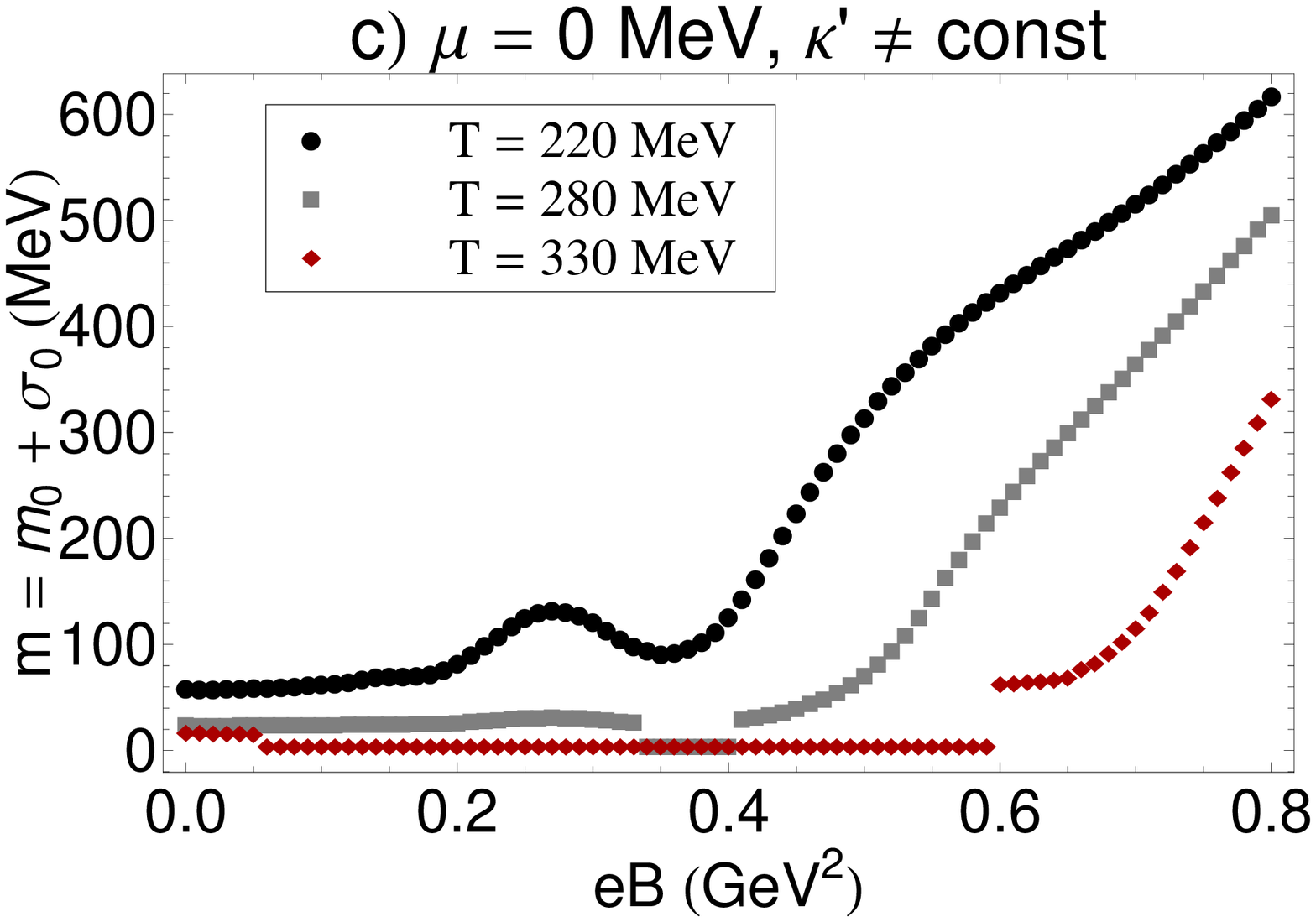}
\caption{(color online). The $eB$ dependence of the constituent quark mass $m=m_{0}+\sigma_{0}$ is demonstrated for $\mu=0$ MeV, at different temperatures $T=220$ MeV (black circles), $T=280$ MeV (gray squares), $T=330$ MeV (red diamonds), and for $\kappa_{1}$ (panel a), $\kappa_{2}$ (panel b) and $\kappa'$ (panel c).  The $eB$ dependence of $m$ for $\kappa_{2}$ and $\kappa'$ at $T>200$ MeV turns out to be different at $T<200$ MeV, demonstrated in Figs. \ref{fig1} and \ref{fig2}.  The data for $\kappa_2$, $\mu=0$ and $T=280$ MeV are characterized by a critical
$eB_{c}\sim 0.625$ GeV$^{2}$ (panel b), and those for $\kappa'$ are characterized by two critical $eB_{c_1}\sim 0.33$ GeV$^{2}$ and $eB_{c_2}\sim 0.41$ GeV$^{2}$ (panel c). Similarly, the data for $\kappa'$ for $\mu=0$ and $T=330$ MeV are characterized by two critical $eB_{c_1}\sim 0.05$ GeV$^{2}$ and $eB_{c_2}\sim 0.6$ GeV$^{2}$ (panel c).
The appearance of two critical $eB$s is, in particular, an indication of the phenomenon of reentrance from the p$\chi$SR into $\chi$SB phase.
}\label{fig4}
\end{figure*}
\begin{figure*}[hbt]
\includegraphics[width=5.9cm,height=4.4cm]{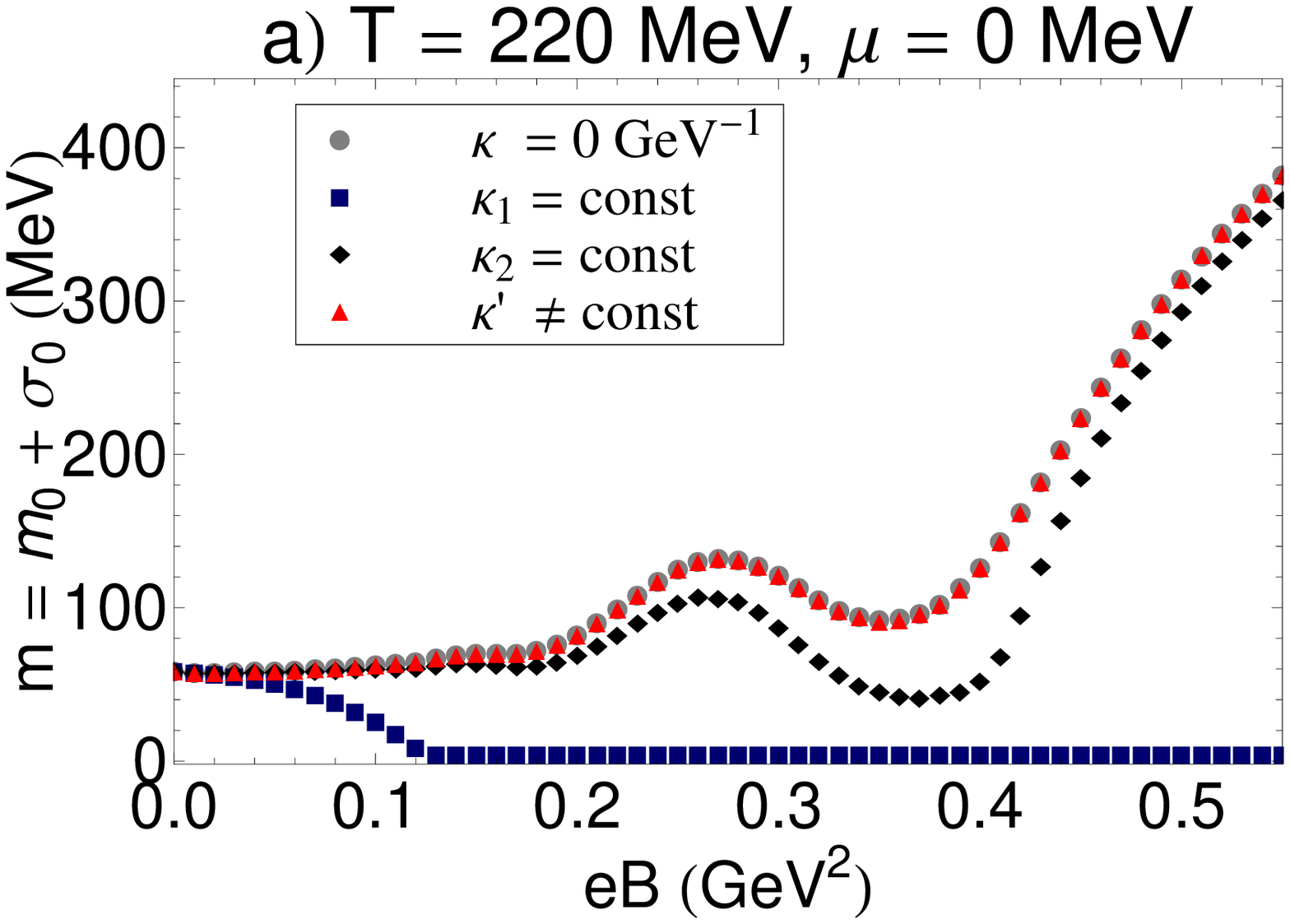}
\includegraphics[width=5.9cm,height=4.4cm]{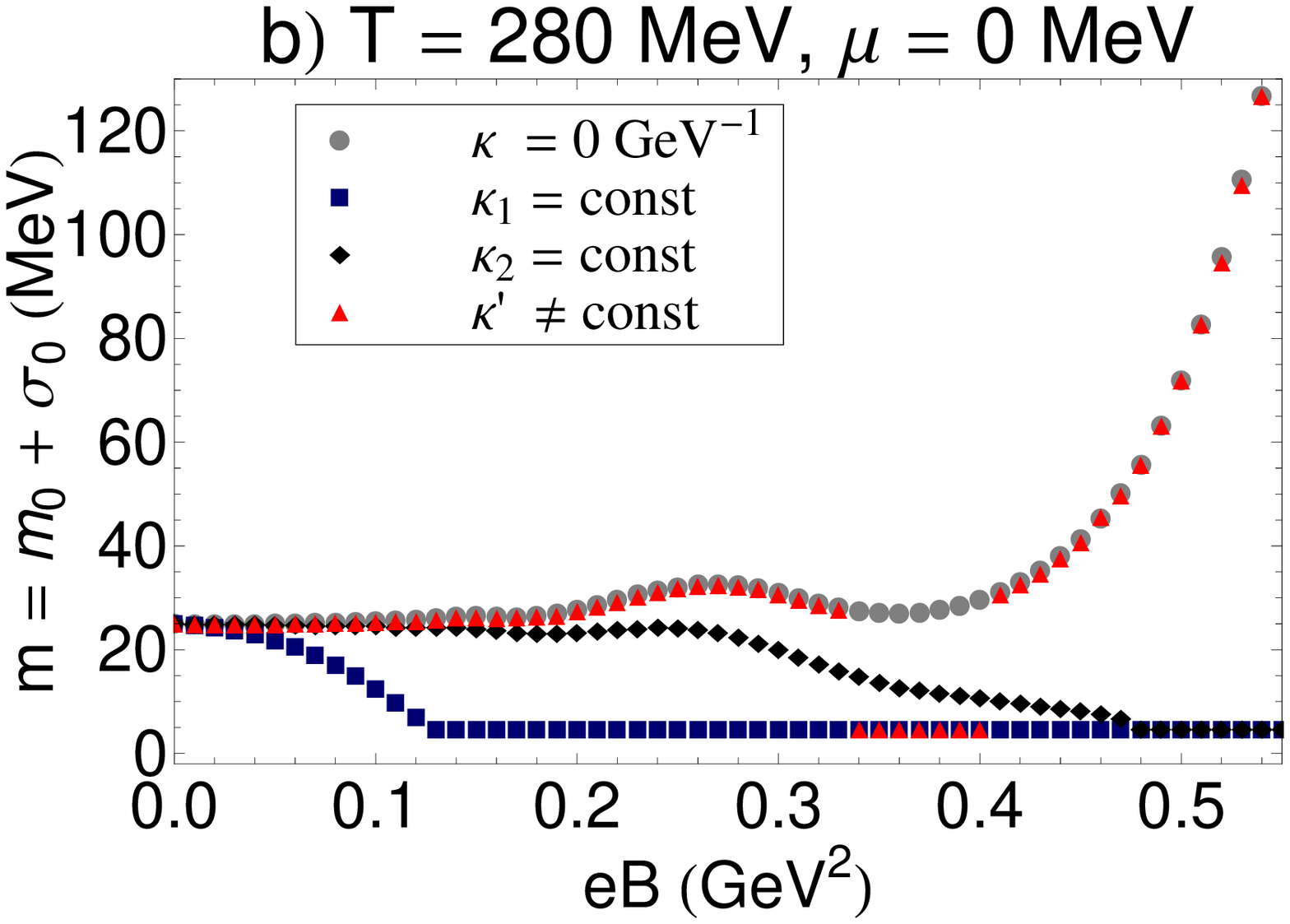}
\includegraphics[width=5.9cm,height=4.4cm]{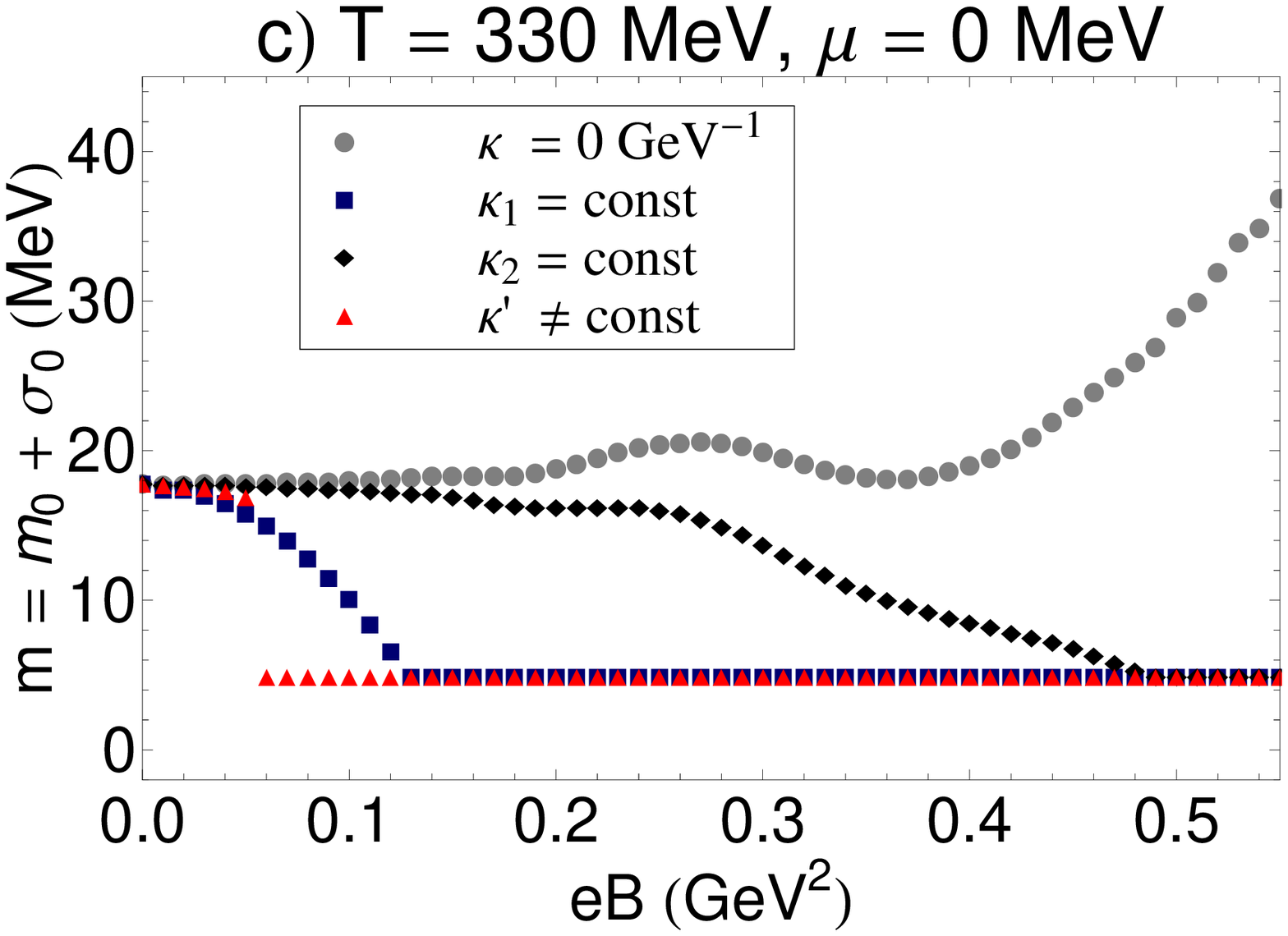}
\caption{(color online).  The $eB$ dependence of the constituent quark mass $m=m_{0}+\sigma_{0}$ is demonstrated for $\mu=0$, at $T=220$ MeV (panel a), $T=280$ MeV (panel b) and $T=330$ MeV (panel c). The data for different $\hat{\kappa}$, including $\hat{\kappa}=0$ GeV$^{-1}$ (gray circles), $\kappa_{1}$ (blue squares), $\kappa_{2}$ (black diamonds) and $\kappa'$ (red triangles), are compared. The difference between the data for $\hat{\kappa}=0$ GeV$^{-1}$ and $\kappa'$ maximizes with increasing temperature. The data for $\kappa_{2}$ for $T\geq 280$ MeV indicate a crossover transition from the $\chi$SB into the p$\chi$SR phase. The $eB$ dependence of $m$ for $\kappa'$ at $T=330$ MeV, $\mu=0$ MeV is similar to the case of $\kappa_{1}$. The data for $\kappa'$ are characterized by a critical magnetic field $eB_{c}\sim 0.05$ GeV$^{2}$. }\label{fig5}
\end{figure*}
\par\noindent
In Figs. \ref{fig1} and \ref{fig2}, the $eB$ dependence of the constituent quark mass $m=m_{0}+\sigma_{0}$ is demonstrated for $\kappa_{1}$, $\kappa_{2}$ and $\kappa'$, $\mu=0$ MeV and at $T=0$ MeV (Fig. \ref{fig1}) as well as $T=150$ MeV (Fig. \ref{fig2}). In Figs. \ref{fig1}(a) and \ref{fig2}(a), the results for $\kappa_{1}$ (red dashed lines) are compared with the corresponding data to $\kappa_{2}$ (black solid lines). The latter is then compared with the results for $\kappa'$  (red dashed lines) in Figs. \ref{fig1}(b) and \ref{fig2}(b). As it turns out that, whereas for $\kappa_{2}$ and $\kappa'$ the constituent quark mass $m$ increases with increasing $eB$, for $\kappa_{1}$, that yields a sizable quark AMM, $m$ abruptly decreases at a certain  critical magnetic field $eB_{c}$ to a value $m=m_{0}$.\footnote{As aforementioned, for nonzero current quark mass $m_{0}$, $m$ turns out to be equal to $m_{0}$ after the transition into the p$\chi$SR phase is occurred.} This indicates that, once $\hat{\kappa}$ is  chosen to be $\kappa_{1}$, for $\mu=0$ MeV a first order phase transition occurs at this specific $eB_{c}$, which is for $T=0$ MeV and $T=150$ MeV equal to $eB_{c}\sim 0.623$ GeV$^{2}$ and $eB_{c}=0.190$ GeV$^{2}$, respectively. As we have argued above, the fact that $m$ increases with increasing $eB$ can be regarded as one of the most important signatures of the phenomenon of MC, which is supposed to enhance the formation of the quark condensate in the supercritical regime of NJL coupling $G$. On the other hand, by definition, the phenomenon of IMC is related to the suppression of bound state formation even for large values of $eB$. Thus, the fact that $m$ decreases with increasing $eB$ for $\hat{\kappa}_{1}=(\kappa_{u}^{(1)},\kappa_{d}^{(1)})$ from (\ref{F12}) is related to the phenomenon of IMC for $\mu=0$ MeV and $T=0,150$ MeV.
\par
To study the effect of temperature on the aforementioned critical magnetic field $eB_{c}$ for $\kappa_{1}$, the $eB$ dependence of $m$ is considered in Fig. \ref{fig3} for three different temperatures, $T_{1}=0$ MeV (red dashed line), $T_{2}=120$ MeV (black solid line) and $T_{3}=150$ MeV (blue dashed line). The critical magnetic fields corresponding to these temperatures are denoted by $C_{1}, C_{2}$ and $C_{3}$. Numerically, they are given by $eB_{c_{1}}\sim 0.623$ GeV$^{2}$, $eB_{c_{2}}\sim 0.285$ GeV$^{2}$ and $eB_{c_{3}}\sim 0.190$ GeV$^{2}$. It turns out,  that $eB_{c}$ decreases with increasing $T$. This can again be considered as a signature of the phenomenon of IMC, which seems to occur once the quark AMM is large enough. Later, we will study the $T$--$eB$ phase diagram for $\mu=0$ MeV and  $\kappa_{1}, \kappa_{2}$ and $\kappa'$. We will show, that in contrast to the case of $\kappa_{2}$ and $\kappa'$, the critical temperature corresponding to $\kappa_{1}$ decreases with increasing $eB$ (see Fig. \ref{fig8} for more details).
\par
According to the results from \cite{fayazbakhsh2}, in the chiral limit $m_{0}\to 0$, the two-flavor NJL model at finite $T$ and zero $\mu$ and $eB$ exhibits a second order phase at $T_{c}\sim 200$ MeV, and for nonvanishing $eB$, $T_{c}$ increases with increasing $eB$ up to $T_{c}\sim 360$ MeV for $eB=0.7$ GeV$^{2}$ \cite{fayazbakhsh2}. Hence, the temperature interval $T\in [200, 360]$ MeV can be regarded as the regime of phase transition of this QCD-like model. On the other hand, as it is shown in \cite{fayazbakhsh3}, for $m_{0}\neq 0$, the above-mentioned second order phase transition turns into a smooth crossover. In this case the temperature interval
$T\in [200, 360]$ MeV  indicates the crossover region from the $\chi$SB into the p$\chi$SR phase. In Figs. \ref{fig4} and \ref{fig5}, the $eB$ dependence of the constituent quark mass $m$ is demonstrated at $T=220, 280, 330$ MeV, for $\mu=0$ MeV and different $\kappa$s in the crossover regime of our model. Comparing the results for $\kappa_{1}$ in Fig. \ref{fig3} at $T<200$ MeV with the results for $\kappa_{1}$ at $T>200$ MeV from Fig. \ref{fig4}(a), it turns out that the first order phase transition, appearing at $T<200$ MeV, turns into a smooth crossover at $T>200$ MeV. As concerns the results for $\kappa_{2}$ and $\kappa'$, we observed in Figs. \ref{fig1}(b) and \ref{fig2}(b), that at $T<200$ MeV, $m$ increases with increasing $eB$. At $T>200$ MeV, however, the situation changes. Whereas for $T=220$ MeV, $m$ increases with increasing $eB$ for both $\kappa_{2}$ and $\kappa'$ [see the black circles in Figs. \ref{fig4}(b) and \ref{fig4}(c)], at $T=280$ MeV and $T=330$ MeV, $m$ decreases first with $eB$ and then suddenly increases for certain critical magnetic fields [see gray squares for $T=280$ MeV and red diamonds for $T=330$ MeV in Figs. \ref{fig4}(b) and \ref{fig4}(c)]. Comparing the data for $\kappa_{2}$ and $\kappa'$ at $T=280$ MeV [gray squares in Figs. \ref{fig4}(b) and \ref{fig4}(c)], it turns out that for $\kappa_{2}$, one and for $\kappa'$ two critical magnetic fields exist. They are given by $eB_{c}\sim 0.625$ GeV$^{2}$ for $\kappa_{2}$ and $eB_{c_1}\sim 0.33$ GeV$^{2}$ and $eB_{c_2}\sim 0.41$ GeV$^{2}$ for $\kappa'$. This means that in the latter case, the system is first in the $\chi$SB phase, enters at $eB_{c_1}$ the p$\chi$SR phase, and reenters the $\chi$SB phase at $eB_{c_2}$ (see Fig. \ref{fig8} for more detailed analysis of the phase transitions for different $\hat{\kappa}$ and the discussions related to the phenomenon of reentrance from the $\chi$SB into the p$\chi$SR phase). Let us now compare the data corresponding to $\kappa_{2}$ and $\kappa'$ at $T=330$ MeV and $\mu=0$ MeV in Figs. \ref{fig4}(b) and \ref{fig4}(c) (red diamonds). It turns out, that for $\kappa_{2}$, $m$ decreases with increasing $eB$ in the whole range $eB\in [0,0.8]$ GeV$^{2}$, while, for $\kappa'$, there are two critical magnetic fields $eB_{c_1}\sim 0.05$ GeV$^{2}$ and $eB_{c_{2}}\sim 0.6$ GeV$^{2}$. In other words, for $\kappa'$, the quark matter is first in the $\chi$SB phase, enters the p$\chi$SR phase, and then reenters the $\chi$SB phase. Our findings in Fig. \ref{fig8} confirm this conclusion.
\par
In Fig. \ref{fig5}, we compare the $eB$ dependence of $m$ for different $\hat{\kappa}$: $\hat{\kappa}=0$ GeV$^{-1}$ (gray circles), $\kappa_1$ (blue squares), $\kappa_2$ (black diamonds) $\kappa'$ (red triangles), for $\mu=0$ MeV and at $T=220$ MeV [Fig. \ref{fig5}(a)], $T=280$ MeV [Fig. \ref{fig5}(b)] and $T=330$ MeV
 [Fig. \ref{fig5}(c)]. We limited ourselves to the regime $eB\in [0,0.5]$ GeV$^{2}$ in order to magnify the difference in the $eB$ dependence of $m$, especially for $\hat{\kappa}=0$ GeV$^{-1}$, $\kappa_{2}$ and $\kappa'$. According to these results, no difference between the data corresponding to $\hat{\kappa}=0$ GeV$^{-1}$ and $\kappa'$ occurs as long as $T\lesssim 250$ MeV. On the other hand, whereas, according to Figs. \ref{fig1}(b) and \ref{fig2}(b), there is no difference between the data corresponding to $\kappa_{2}$ and $\kappa'$ at $T\lesssim 150$ MeV, at $T>200$ MeV the $eB$ dependence of $m$ are different for $\kappa_{2}$ and $\kappa'$.  In particular, it turns out that at $T=330$ MeV, $\mu=0$ MeV and for $\kappa'$, similar to the case of $\kappa_{1}$, $m$ first decreases with increasing $eB$ up to a certain critical $eB_{c}\sim 0.05$ GeV$^{2}$, then suddenly falls down to a value $m=m_{0}=5$ MeV. This is again an indication of the phenomenon of IMC for $\mu=0$ MeV, at high temperature $T>280$ MeV and $eB<0.15$ GeV$^{2}$. Let us finally notice that the data for $\kappa_{2}$ in Figs. \ref{fig5}(b) and \ref{fig5}(c) indicates a smooth crossover from the $\chi$SB into the p$\chi$SR phase in the regime $220<T<360$ MeV and $eB<0.55$ GeV$^{2}$   and for $\mu=0$ MeV (see also Fig. \ref{fig8} for more details).
\subsubsection{The $T$ dependence of the constituent quark mass $m$}\label{subsec3A2}
\begin{figure*}[hbt]
\includegraphics[width=5.9cm,height=4.4cm]{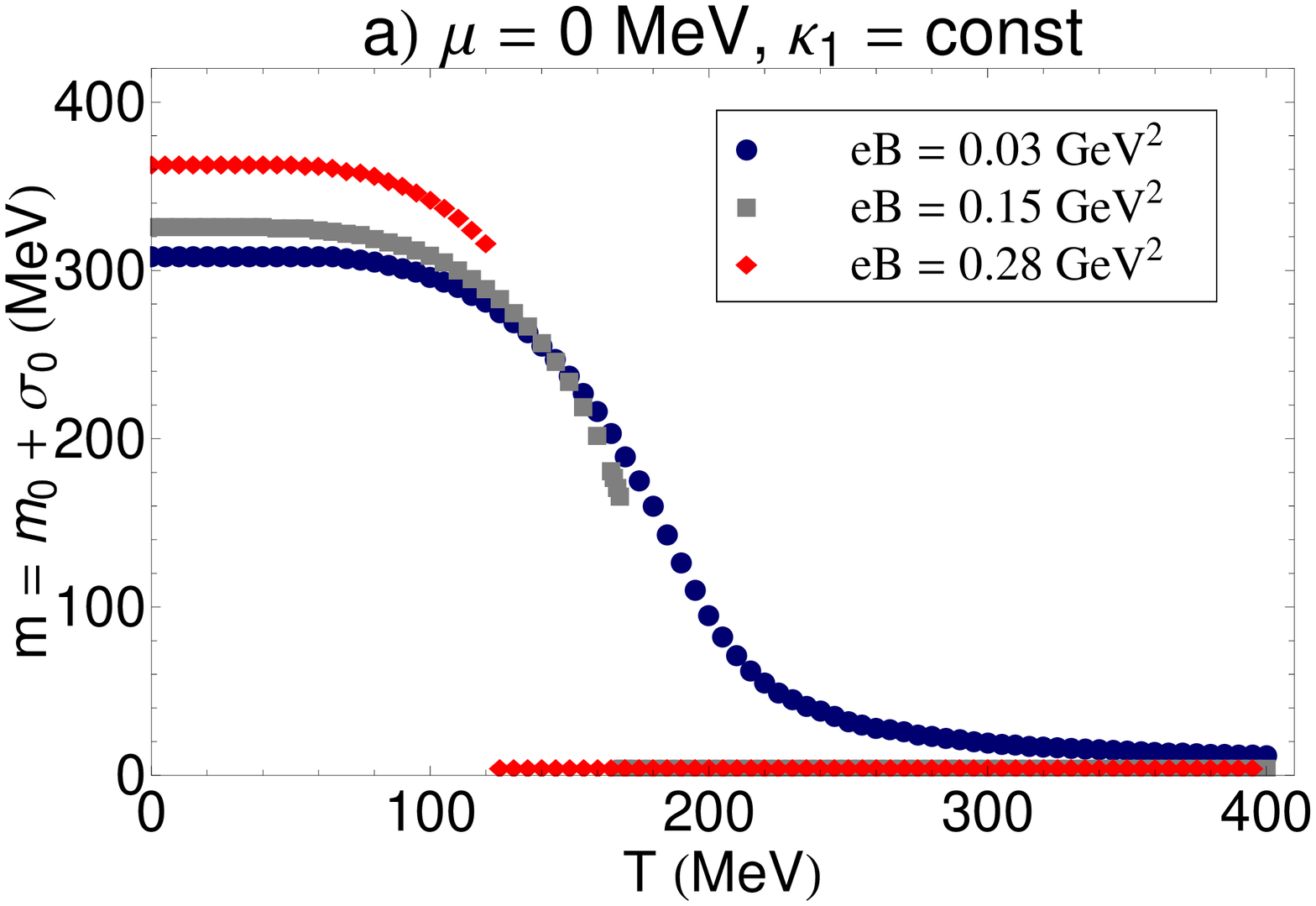}
\includegraphics[width=5.9cm,height=4.4cm]{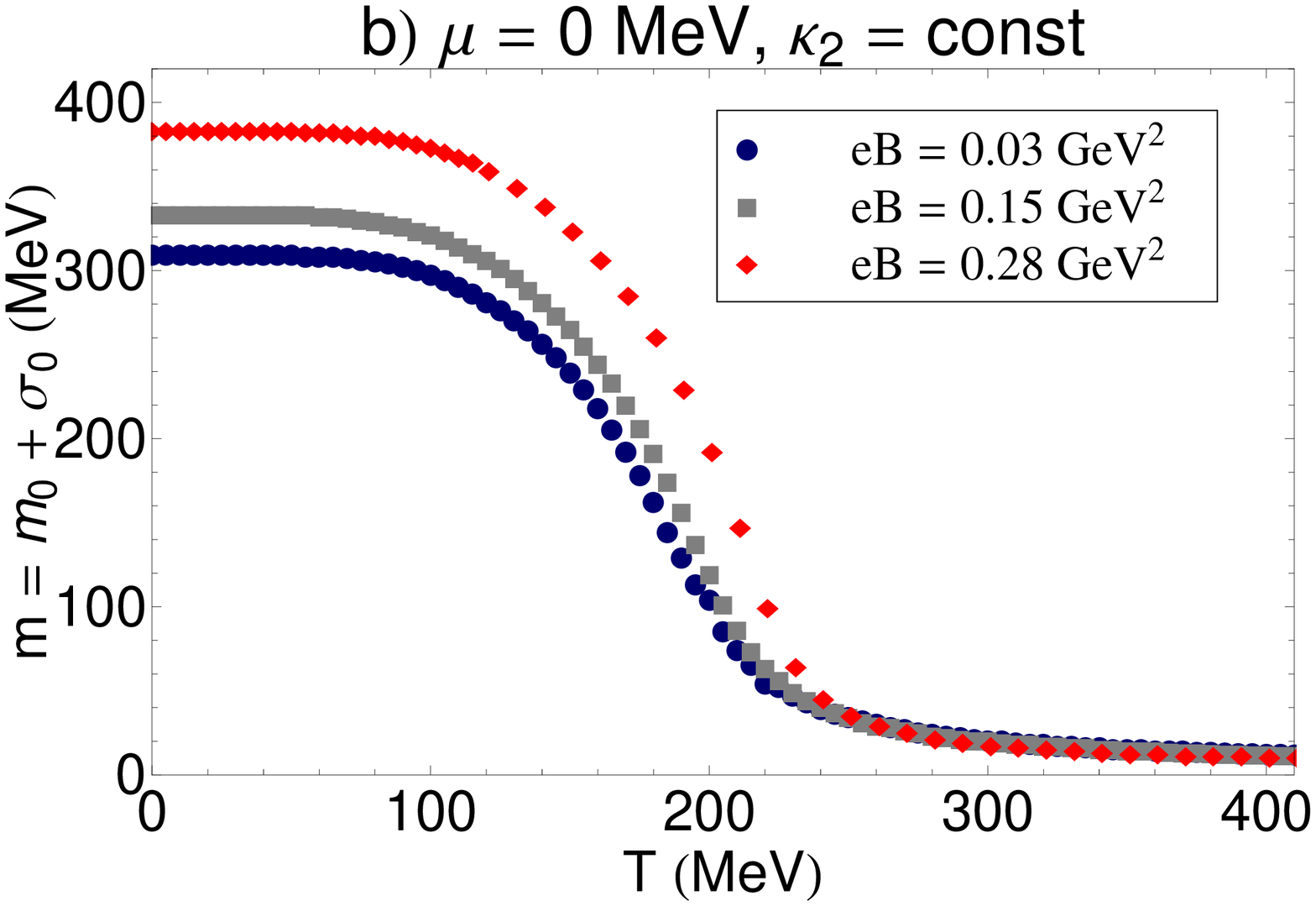}
\includegraphics[width=5.9cm,height=4.4cm]{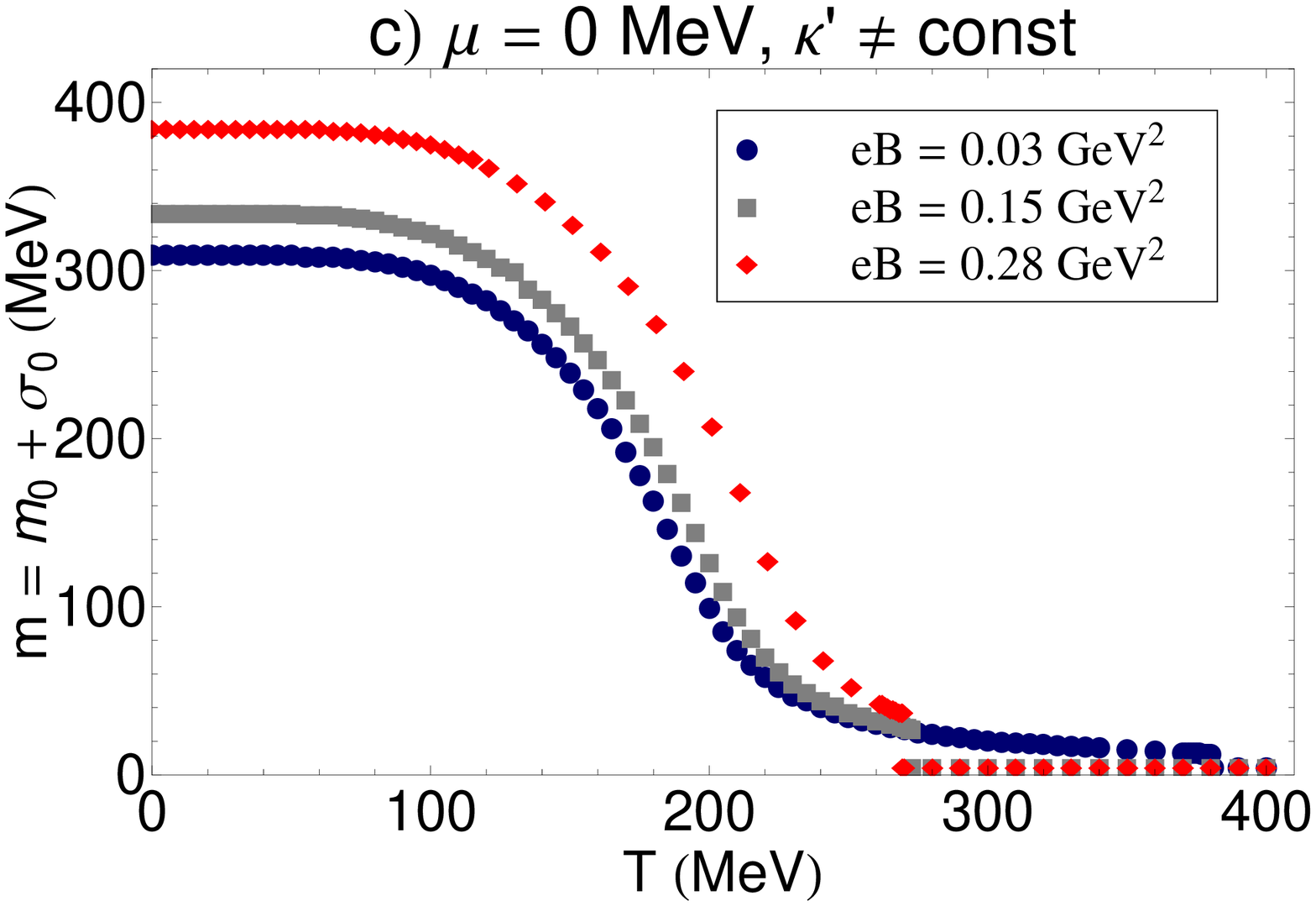}
\caption{(color online). The $T$ dependence of the constituent quark mass $m=m_{0}+\sigma_{0}$ is demonstrated for $\mu=0$ and $T$-independent $\kappa_{1}$ (panel a), $\kappa_{2}$ (panel b) and $\kappa'$ (panel c) for
 $eB=0.03$ GeV$^{2}$ (blue circles) $eB=0.15$ GeV$^{2}$ (gray squares) and $eB=0.28$ GeV$^2$ (red diamonds). Discontinuous decreasing of $m$ with increasing $T$ indicates a first order phase transition from the $\chi$SB into the p$\chi$SR phase. This occurs, in particular, for $\kappa_{1}$, that leads to a sizable quark AMM.  }\label{fig6}
\end{figure*}
\begin{figure*}[hbt]
\includegraphics[width=5.9cm,height=4.4cm]{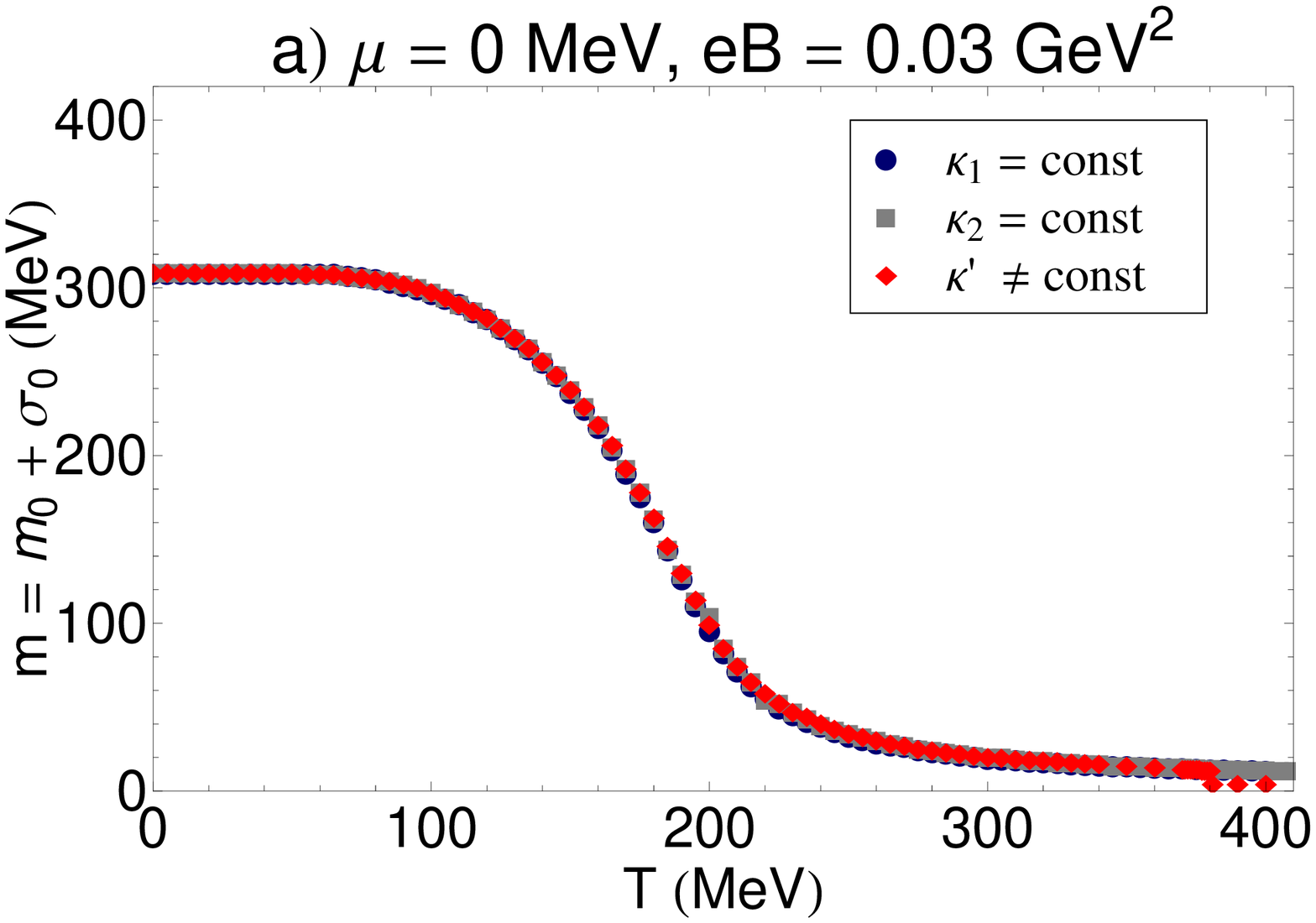}
\includegraphics[width=5.9cm,height=4.4cm]{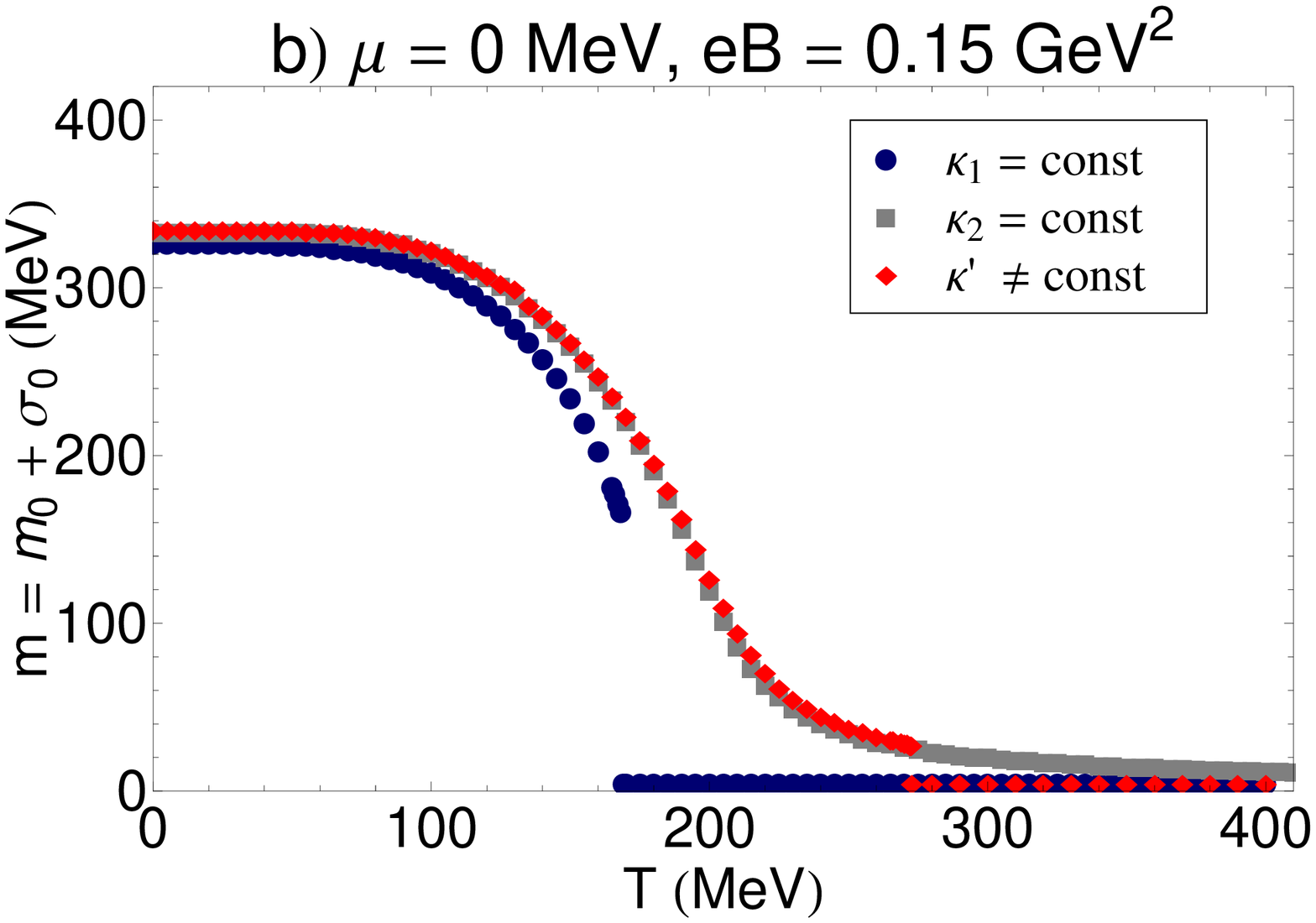}
\includegraphics[width=5.9cm,height=4.4cm]{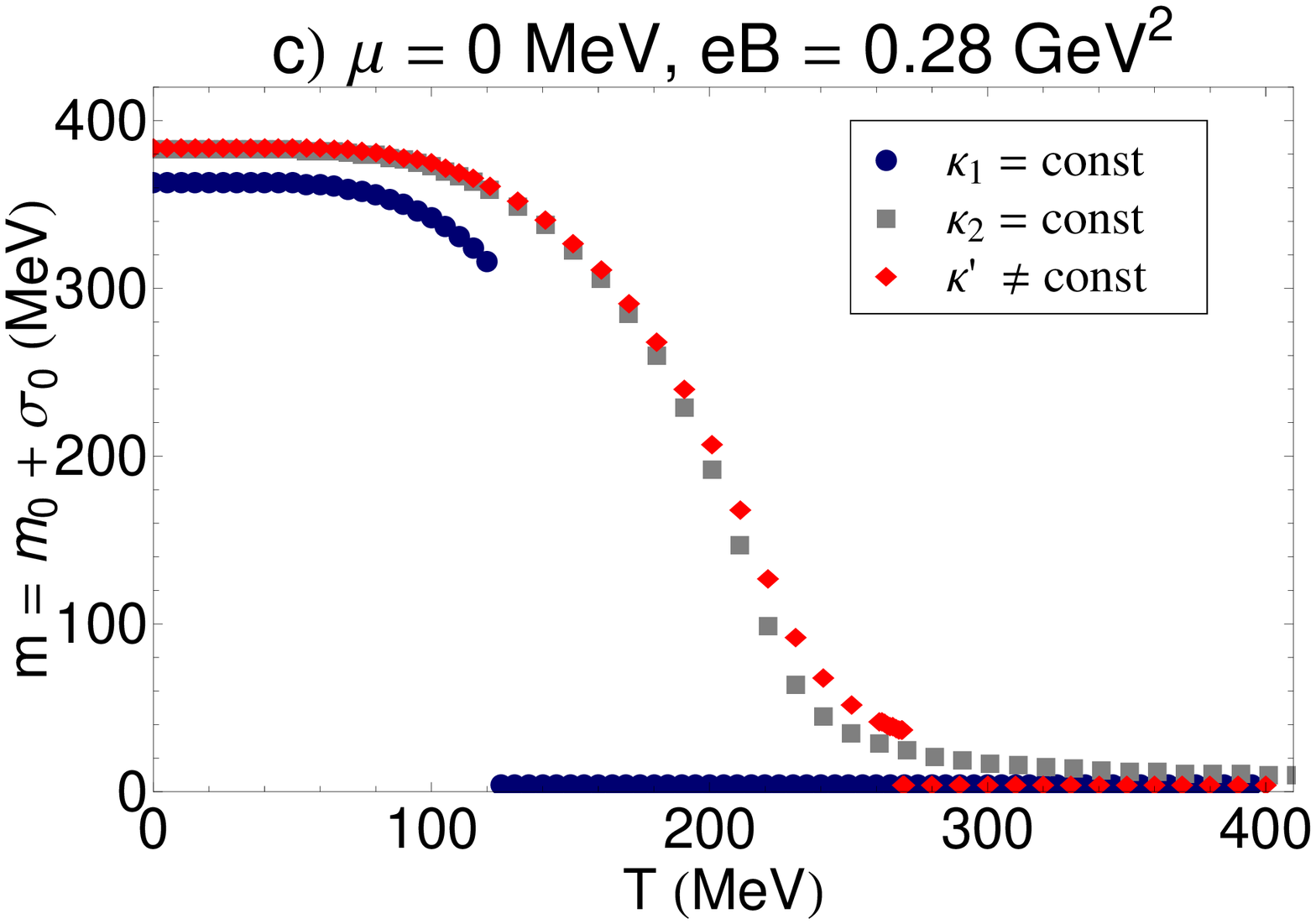}
\caption{(color online). The $T$ dependence of the constituent quark mass $m=m_{0}+\sigma_{0}$ is demonstrated for $\mu=0$ and $eB=0.03$ GeV$^{2}$ (panel a), $eB=0.15$ GeV$^{2}$ (panel b) and $eB=0.28$ GeV$^2$ (panel c), for  $\kappa_{1}$ (blue circles), $\kappa_{2}$ (gray squares) and $\kappa'$ (red diamonds). }\label{fig7}
\end{figure*}
\par\noindent
According to the results from Figs. \ref{fig1}-\ref{fig5} for fixed $\mu$, $eB$ and $\hat{\kappa}$, finite temperature suppresses the formation of the chiral condensate $\sigma_{0}$ (see also \cite{fayazbakhsh2,fayazbakhsh3}). Consequently, it is expected that $m$ decreases with increasing $T$. In Fig. \ref{fig6}, the $T$ dependence of the constituent quark mass is demonstrated for $\mu=0$ MeV, $eB=0.03$ GeV$^{2}$ (blue circles), $eB=0.15$ GeV$^{2}$ (gray squares), $eB=0.28$ GeV$^{2}$ (red diamonds) and different $\kappa_{1}$ (panel a), $\kappa_{2}$ (panel b) and $\kappa'$ (panel c). As it turns out from the results in Fig. \ref{fig6}(a), whereas for small values of $eB=0.03$ GeV$^{2}$, $m$ decreases continuously with increasing $T$, for larger values of $eB=0.15, 0.28$ GeV$^{2}$, $m$ decreases with increasing temperature up to a certain critical temperature $T_{c}$, which is $T_{c}\sim 168$ MeV for $eB=0.15$ GeV$^{2}$ and $T_{c}\sim 120$ MeV for $eB=0.28$ GeV$^{2}$. This behavior indicates a first order phase transition from the $\chi$SB into the p$\chi$SR phase for $\mu=0$ MeV and $\kappa_{1}$. Interestingly, the critical temperature corresponding to this phase transition decreases with increasing $eB$, which is, as aforementioned, another indication of the phenomenon of IMC once $\hat{\kappa}$ is large enough. For $\kappa_{2}$ and $\kappa'$, the situation is rather different. According to the results from Fig. \ref{fig6}(b), for $\kappa_{2}$, in contrast to $\kappa_{1}$, $m$ decreases smoothly with increasing $T$ for all values of $eB=0.03, 0.15, 0.28$ GeV. For $T$-dependent $\kappa'$, however, whereas for $eB=0.03$ GeV$^{2}$  the constituent quark mass $m$ decreases smoothly with increasing $T$ up to a certain critical temperature $T_{c}\sim 380$ MeV, for $eB=0.15$ GeV$^{2}$ and $eB=0.28$ GeV$^{2}$, discontinuities occurs at critical temperatures $T_{c}\sim 272.6$ MeV and $T_{c}\sim 264$ MeV, respectively. These kind of discontinuities in the $T$ dependence of $m$ are indications of first order phase transitions from the $\chi$SB into the p$\chi$SR phase.
\par
To compare the data for different $\hat{\kappa}$ for fixed $\mu$ and $eB$, the $T$ dependence of $m$ is plotted in Fig. \ref{fig7} for $\mu=0$ MeV and $eB=0.03$ GeV$^{2}$ (panel a),
 $eB=0.15$ GeV$^{2}$ (panel b) and $eB=0.28$ GeV$^{2}$ (panel c) for $T$ independent $\kappa_{1}$
  (blue circles), $\kappa_{2}$ (gray squares) and $\kappa'$ (red triangles).
As it turns out, for small value of $eB=0.03$ GeV$^{2}$, there is almost no difference between
 the $T$ dependence of $m$ for $\kappa_{1}, \kappa_{2}$ and $\kappa'$, and the transition
 from the $\chi$SB into the p$\chi$SR phase turns out to be a smooth crossover.
 In contrast, according to the results from Figs. \ref{fig7}(b) and \ref{fig7}(c), for
  $\mu=0$ MeV and $eB=0.15$ GeV$^{2}$ as well as $eB=0.28$ GeV$^{2}$, whereas in the case
   of  $\kappa_{2}$, $m$ decreases smoothly with increasing $T$, in two other cases of
   $\kappa_{1}$ and $\kappa'$, there exists a certain critical temperature below which $m$
   decreases smoothly and above which $m$ turns out to be $m=m_{0}$ (for the values of $T_{c}$, see above).
    This indicates a first order phase transition for $eB=0.15$ GeV$^{2}$ and $eB=0.28$ GeV$^{2}$. In what follows, we study the full phase portrait of the two-flavor NJL model for finite $(T,\mu,eB)$
    and nonzero $\hat{\kappa}$.
\subsection{The phase portrait of hot and magnetized two-flavor NJL model for nonvanishing $\hat{\kappa}$}\label{subsec3B}
\subsubsection{The $T$--$eB$ phase diagram for various $\mu$ and $\hat{\kappa}$}\label{subsec3B1}
\begin{figure}[hbt]
\includegraphics[width=8.5cm,height=6.0cm]{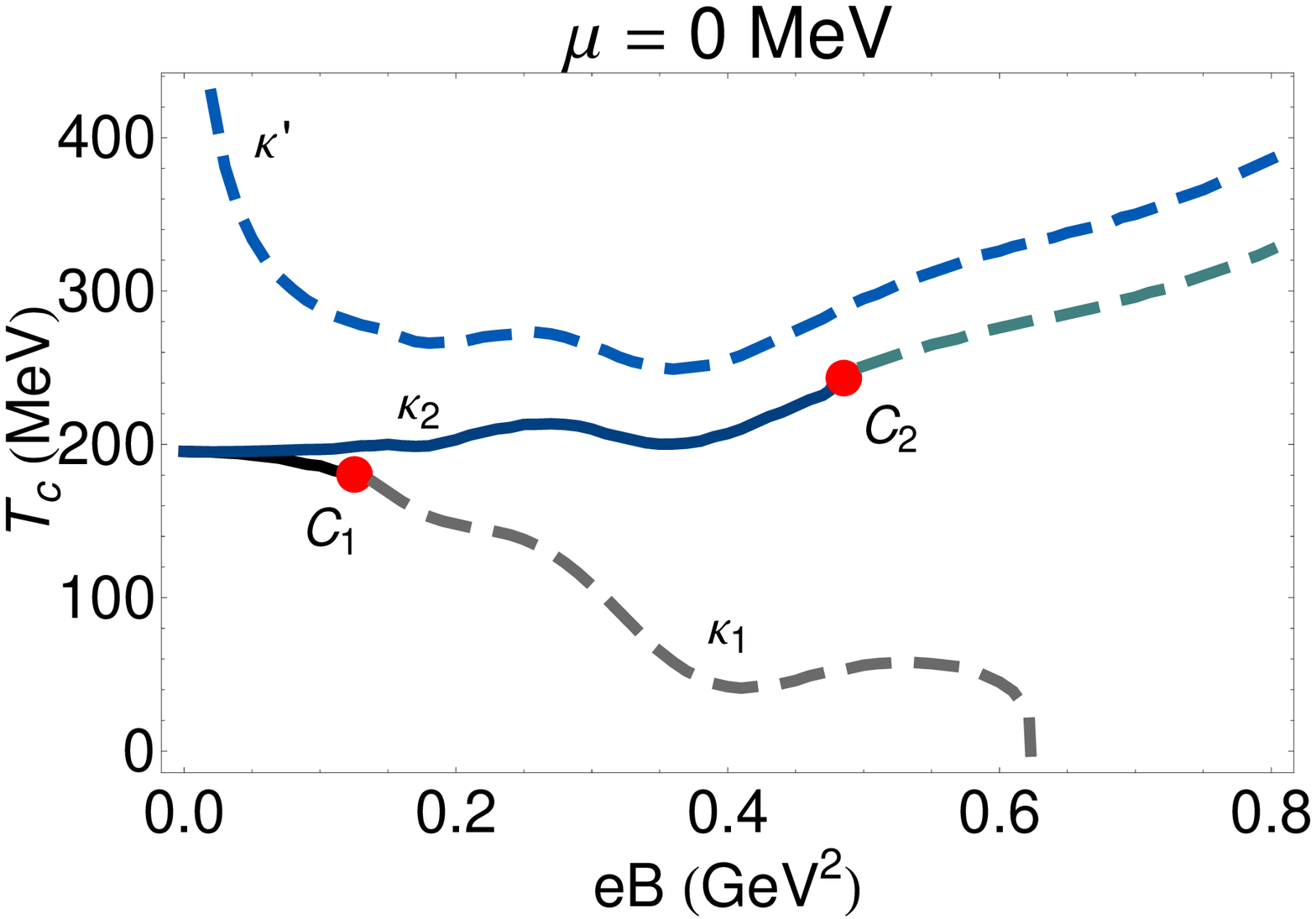}
\caption{(color online). The $T$--$eB$ phase diagram of a hot and magnetized two-flavor NJL model is presented for $\mu=0$ MeV, $\kappa_{1}, \kappa_{2}$ and $\kappa'$. The black (for $\kappa_{1}$) and dark blue (for $\kappa_{2}$) solid lines denote the smooth crossovers, and the gray, green and light blue dashed lines the first order phase transitions for $\kappa_{1},\kappa_{2}$ and $\kappa'$, respectively. The starting points of the first order transition lines are denoted by $C_{1}$ (for $\kappa_{1}$) and $C_{2}$ (for $\kappa_{2}$). The phenomenon of IMC occurs for $\kappa_{1}$ in $eB\in [0, 0.65]$ GeV$^{2}$ and $\kappa'$ in the interval $eB\in [0,0.25]$ GeV$^{2}$. For $\kappa'$, the dHvA oscillations in the regime $eB\in [0, 0.5]$ GeV$^{2}$ lead to the phenomenon of reentrance from the $\chi$SB into the p$\chi$SR phase.}\label{fig8}
\end{figure}
\begin{figure*}[hbt]
\includegraphics[width=5.9cm,height=4.4cm]{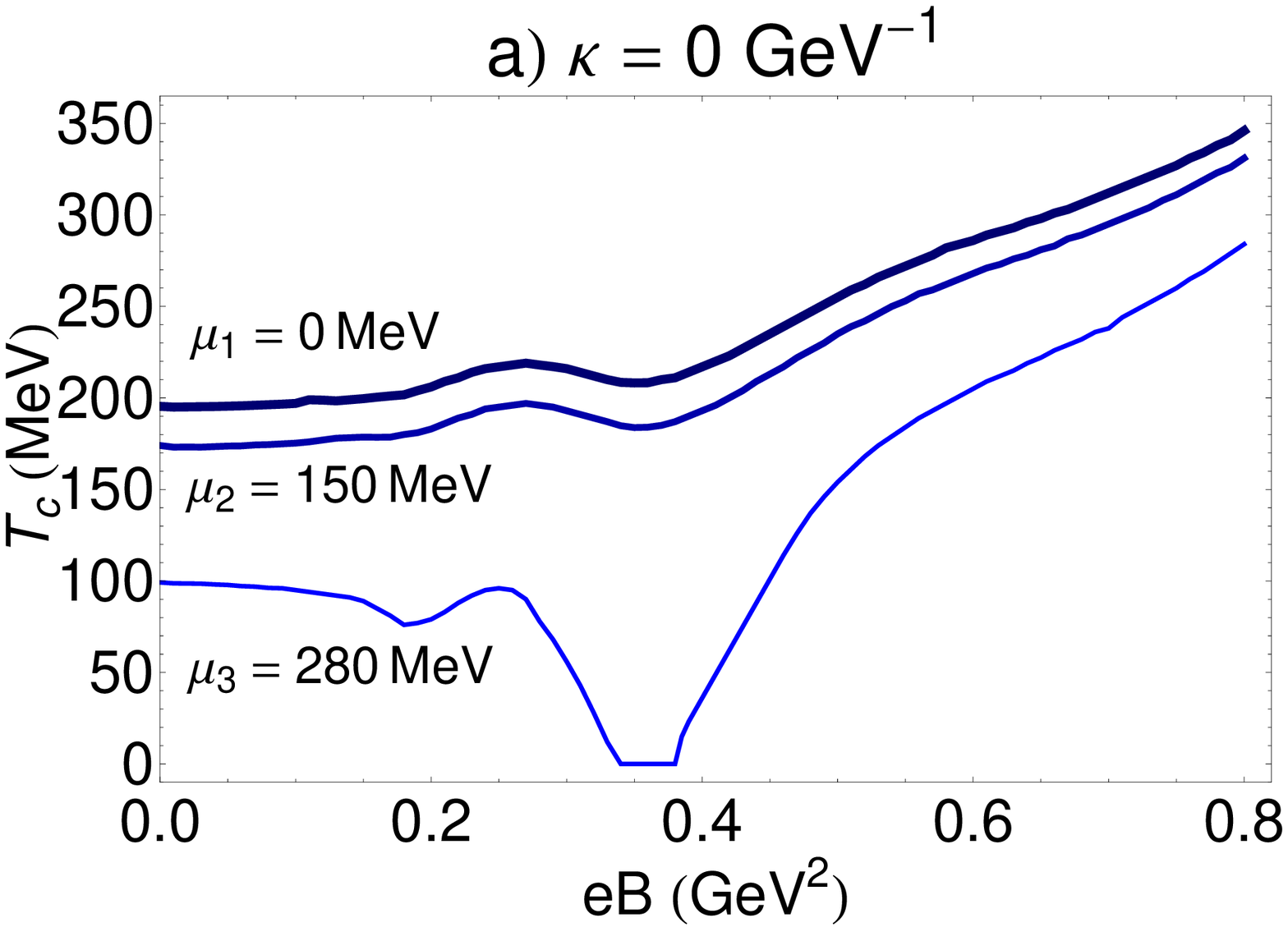}
\includegraphics[width=5.9cm,height=4.4cm]{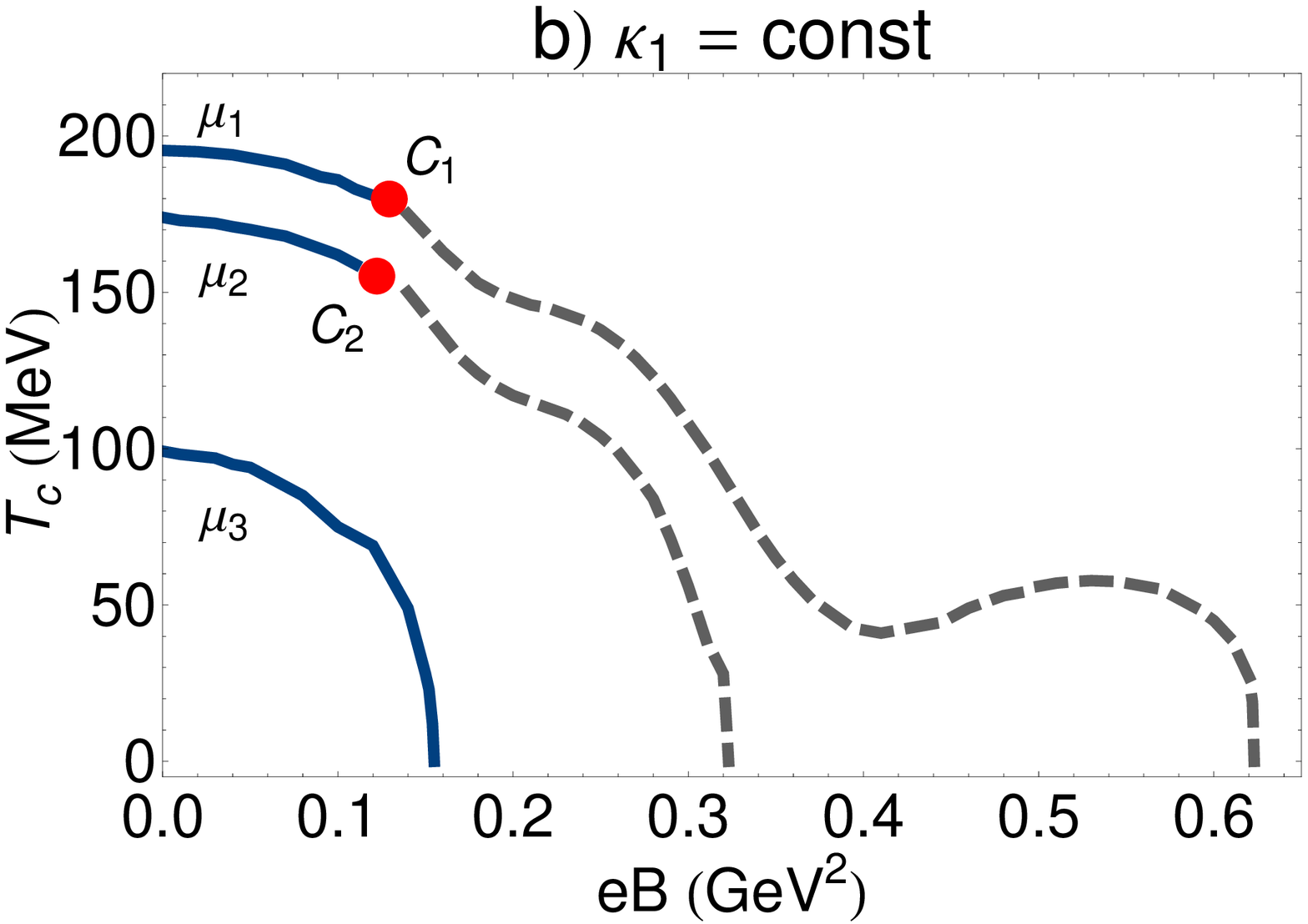}
\includegraphics[width=5.9cm,height=4.4cm]{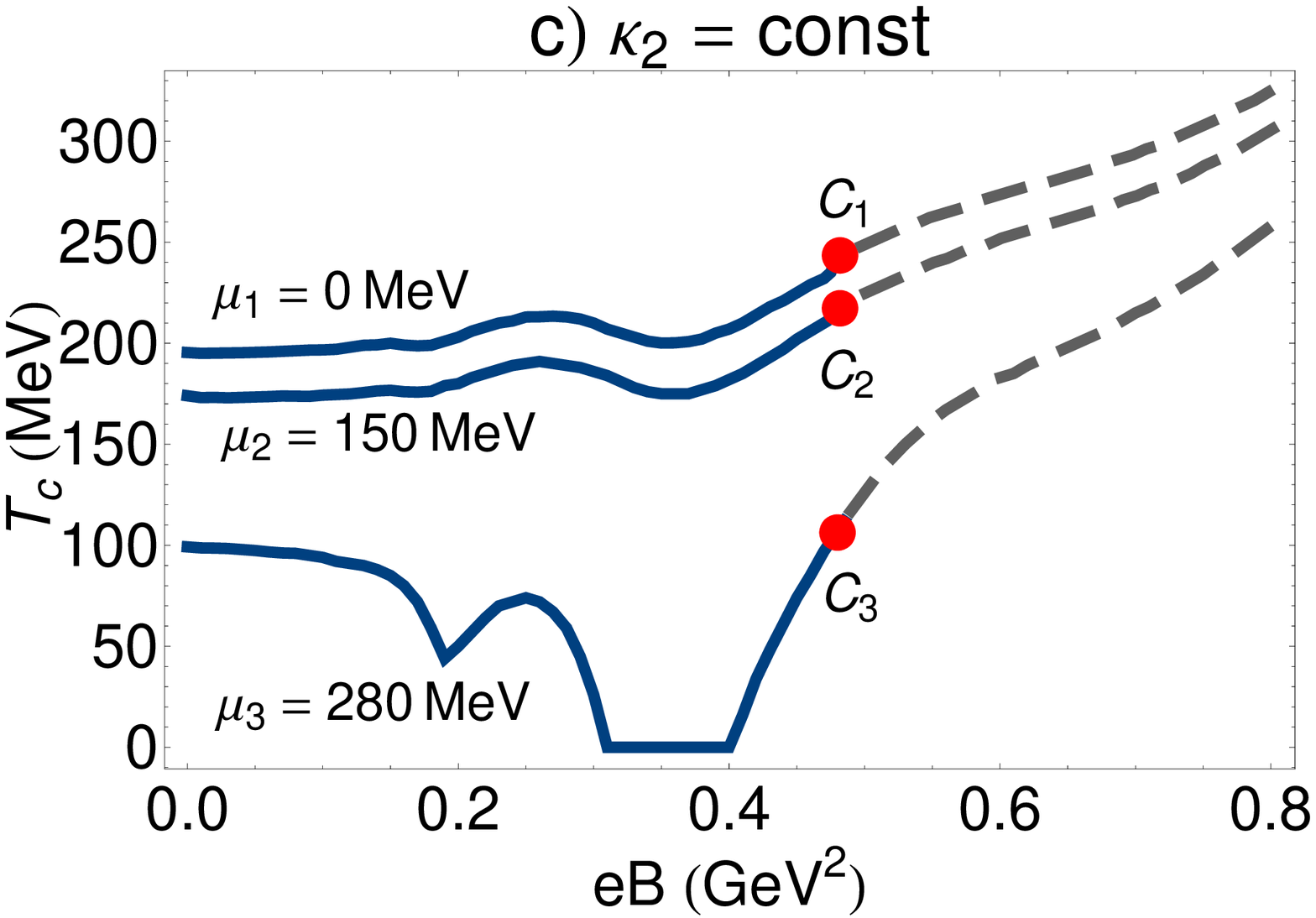}
\caption{(color online). The $T$--$eB$ phase diagram of a hot and magnetized two-flavor NJL model is presented for $\mu_{1}=0$ MeV, $\mu_{2}=150$ MeV, $\mu_{3}=280$ MeV and  $\hat{\kappa}=0$ GeV$^{-1}$ (panel a), $\kappa_{1}$ (panel b) and $\kappa_{2}$ (panel c). Blue solid lines denote the smooth crossovers from the $\chi$SB into the p$\chi$SR phase, and gray dashed lines, the first order phase transitions. The starting points of the first order transition lines are denoted by $C_{i}, i=1,2,3$ corresponding to $\mu_{i}, i=1,2,3$. For $\kappa_{1}$, the phenomenon of IMC occurs for $eB\in [0,0.65]$ GeV$^{2}$. For $\hat{\kappa}=0$ GeV$^{-1}$ and $\kappa_{2}$, the dHvA oscillations in the regime of weak magnetic fields lead to the phenomenon of reentrance from $\chi$SB into p$\chi$SR phase. }\label{fig9}
\end{figure*}
\begin{figure*}[hbt]
\includegraphics[width=5.9cm,height=4.4cm]{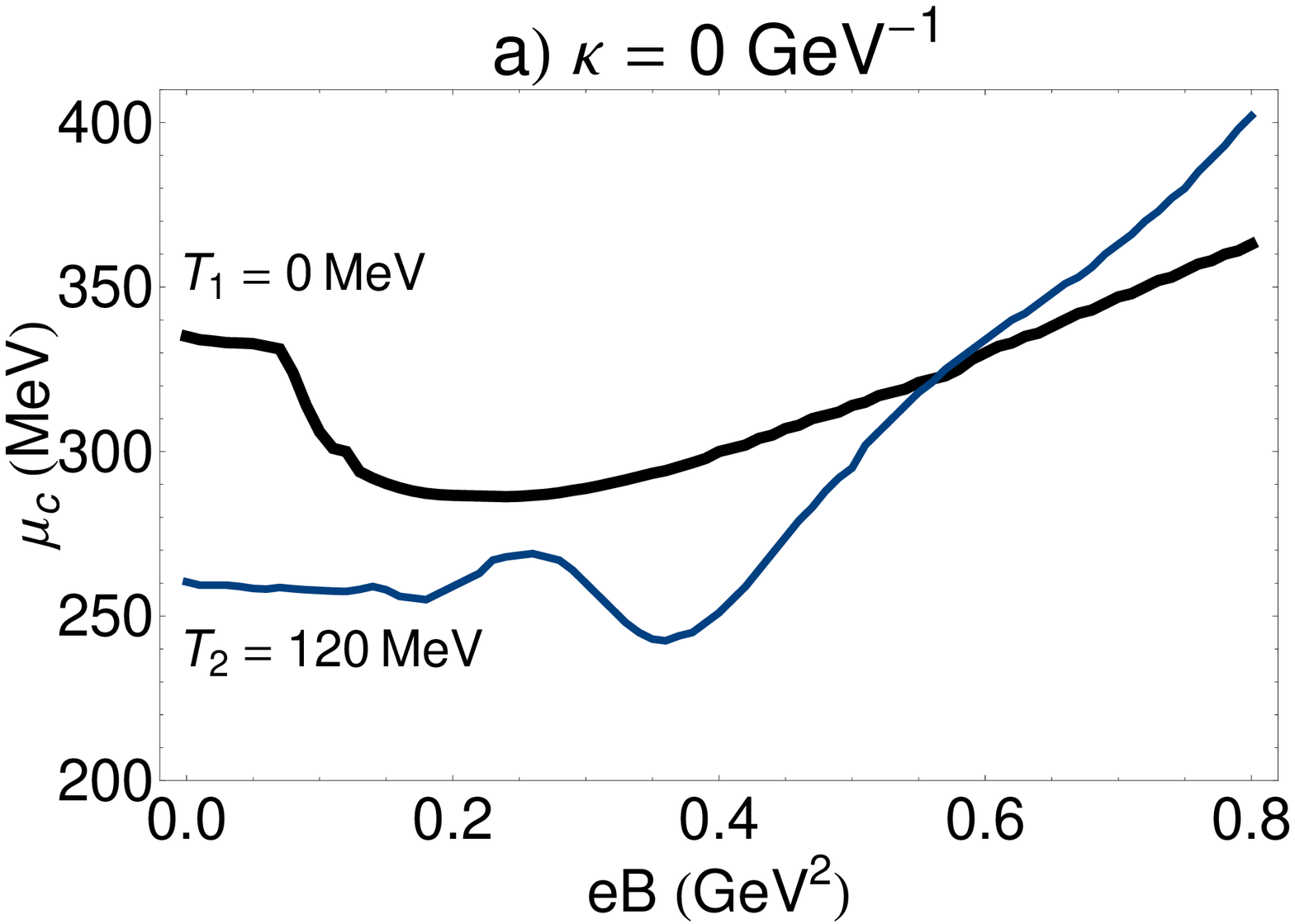}
\includegraphics[width=5.9cm,height=4.4cm]{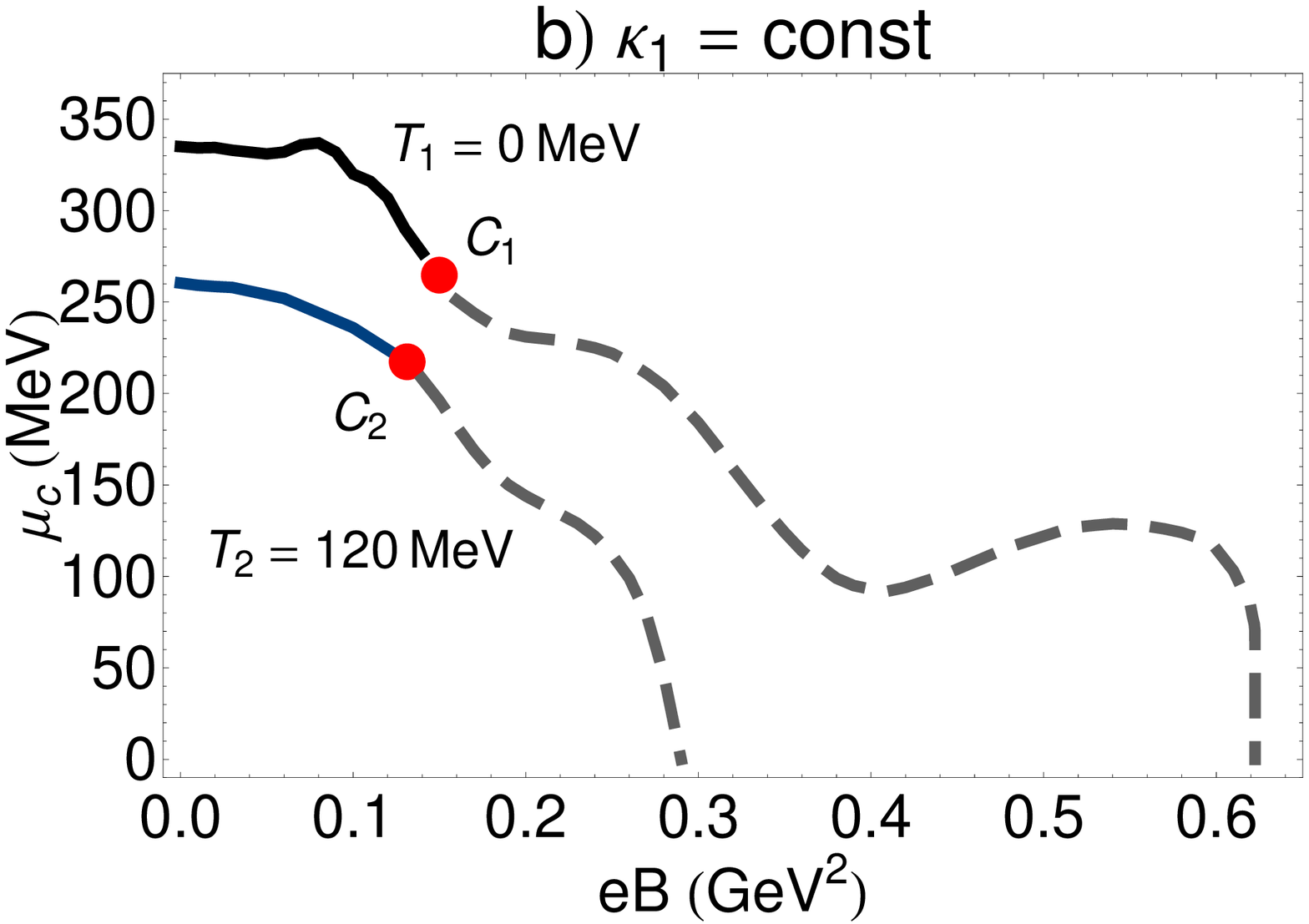}
\includegraphics[width=5.9cm,height=4.4cm]{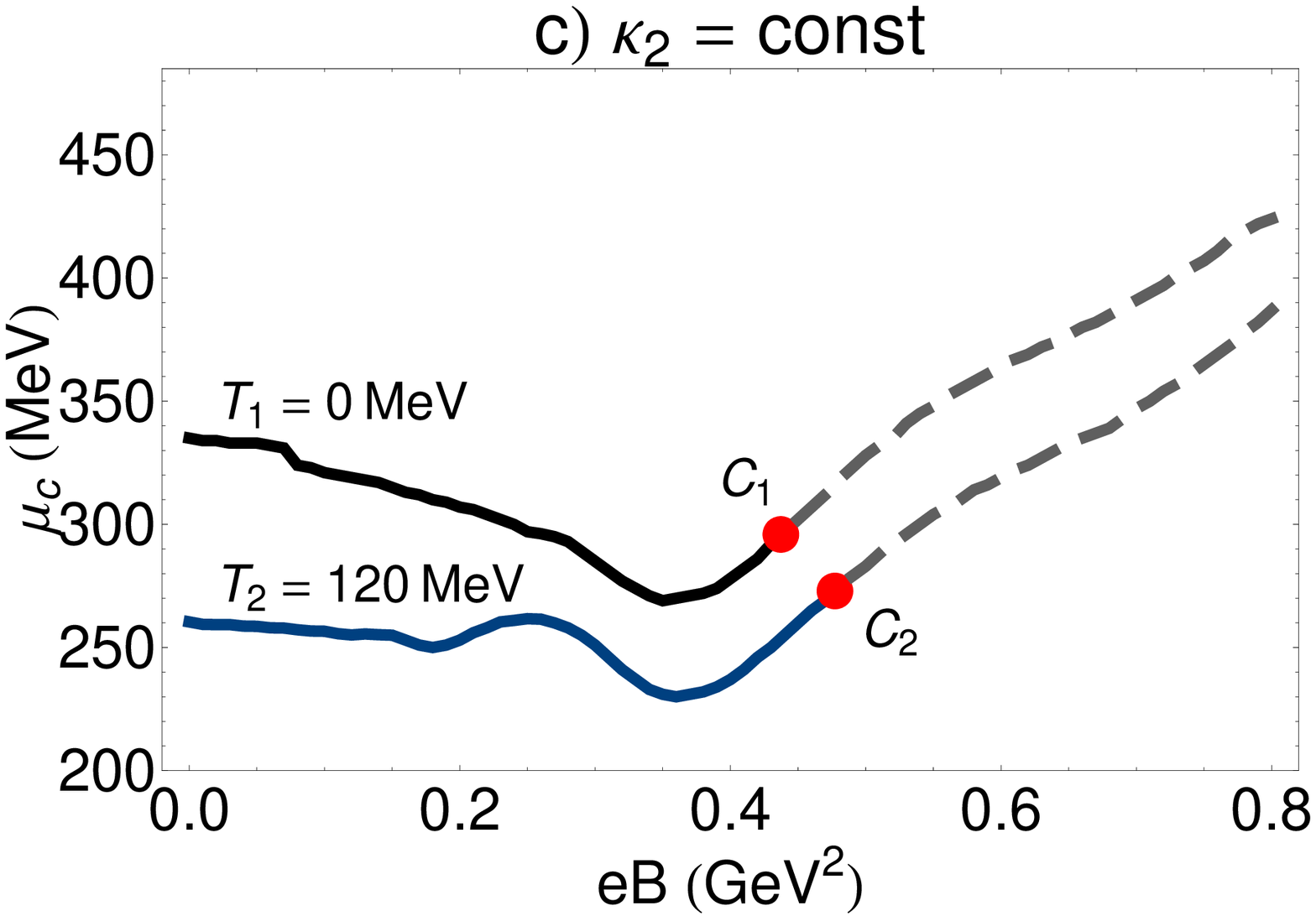}
\caption{(color online). The $\mu$--$eB$ phase diagram of a hot and magnetized two-flavor NJL model is presented for $T_{1}=0$ MeV, $T_{2}=120$ MeV and  $\hat{\kappa}=0$ GeV$^{-1}$ (panel a), $\kappa_{1}$ (panel b) and $\kappa_{2}$ (panel c). The black and blue solid lines denote the smooth crossovers from the $\chi$SB into the p$\chi$SR phase, and gray dashed lines, the first order phase transitions. The starting points of the first order transition lines are denoted by $C_{i}, i=1,2$ corresponding to $T_{i}, i=1,2$. For $\kappa_{1}$, the phenomenon of IMC for $eB\in [0,0.65]$ GeV$^{2}$. For $\hat{\kappa}=0$ GeV$^{-1}$ and $\kappa_{2}$, the dHvA oscillations in the regime of weak magnetic fields lead to the phenomenon of reentrance from $\chi$SB into p$\chi$SR phase.}\label{fig10}
\end{figure*}
\par\noindent
In Fig. \ref{fig8}, the $T$--$eB$ phase diagram of a hot and magnetized two-flavor NJL model is presented for $\mu=0$ MeV, constant $\kappa_{1}, \kappa_{2}$ and $(T,\mu,eB)$--dependent $\kappa'$. The black and dark blue solid lines denote smooth crossover for $\kappa_{1}$ and $\kappa_{2}$, respectively, and the gray, green and light blue dashed lines the first order phase transitions for $\kappa_{1}, \kappa_{2}$ and $\kappa'$, respectively. The starting points of the first order phase transitions are denoted by $C_{1}$ (for $\kappa_{1}$) and $C_{2}$ (for $\kappa_{2}$).
As aforementioned, for nonzero $m_{0}$, the $\chi$SB phase is characterized by $\sigma_{0}\neq 0$, and the p$\chi$SR phase by $\sigma_{0}=0$ MeV and $m=m_{0}$. Hence, by definition, the regimes below the critical lines denote the  $\chi$SB phases and the regimes above them, the p$\chi$SR phases.
To compare the effect of different choices for $\hat{\kappa}$, let us first consider the transition curve for $\kappa_{1}$. Starting from $eB=0$ GeV$^{2}$, the model exhibits a smooth crossover in the interval $eB\in [0,0.12]$ GeV$^{2}$ (see the black solid line). As it turns out, the crossover temperatures\footnote{In the present paper, the crossover temperature $T_{cr}$ is defined by $m(T_{cr}, \mu, eB)\leq e^{-1}m(T=0,\mu, eB)$.} decreases with increasing $eB$ from $T_{c}\sim 195.4$ MeV for $eB=0$ MeV to $T_{c}\sim 181$ MeV for $eB=0.12$ GeV$^{2}$. The end point of the crossover transition line is denoted  in Fig. \ref{fig8} by $C_{1}$. For larger values of $eB\in [0.12,0.623]$ GeV$^{2}$, a first order phase transition occurs (gray dashed line). The critical temperatures for this first order phase transition decreases with increasing $eB$. This result confirms our findings from Figs. \ref{fig1}-\ref{fig7}. Let us notice, e.g. that for $T=0$ MeV, the point $eB\sim 0.623$ GeV$^{2}$ on the $eB$ axis, is the same $eB_{c_{1}}$ appearing in Fig. \ref{fig3} for $\kappa_{1}$ at $T=\mu=0$ MeV.
The fact that, for $\kappa_{1}$, $T_{c}$ decreases with increasing $eB$ is related to the phenomenon of IMC. This is in contrast to what happens for $\kappa_{2}$. Here, as it turns out, the critical temperature essentially increases with increasing $eB$ in the interval $eB\in [0,0.8]$ GeV$^{2}$. First, in the regime $eB\in [0,0.478]$ GeV$^{2}$, the system exhibits a smooth crossover from the $\chi$SB into the p$\chi$SR phase (blue solid line). At a certain critical temperature $T=240$ MeV and magnetic field $eB=0.478$ GeV$^{2}$, denoted by $C_{2}$, the crossover transition line turns into a first order transition line in the regime $eB\in [0.478, 0.8]$ GeV$^{2}$ (green dashed line).  Let us notice at this stage, that, according to our arguments in \cite{fayazbakhsh1, fayazbakhsh2, fayazbakhsh3, fayazbakhsh4}, the regime $eB>0.5$ GeV$^{2}$ is the regime of LLL dominance. In this regime, in the most cases which we have considered in this paper,
the constituent quark mass $m$ as well as the critical temperature $T_{c}$
monotonically increase with increasing $eB$. As concerns the $T$--$eB$ phase diagram for $\kappa'$ (light blue dashed line). As it turns out, $T_{c}$ decreases with increasing $eB$ in the regime $eB\in [0.02, 0.24]$ GeV$^{2}$, exhibits some oscillations in the regime $eB\in [0.25, 0.37]$ GeV$^{2}$ and increases monotonically in the regime $eB\in [0.37, 0.8]$ GeV$^{2}$. Hence, an IMC is followed first by some oscillations and then the MC. Moreover, for $\kappa'$, the transition from the $\chi$SB into the p$\chi$SB phase is of first order in the whole regime $eB\in [0.02, 0.8]$ GeV$^{2}$ (light blue dashed line).
\par
For $\kappa'$, because of this special shape of the first order transition line, another interesting effect occurs. To describe this effect, let us assume the temperature to be $T\sim 300$ MeV. In this case, starting from $eB=0$ GeV$^{2}$, the system is first in the $\chi$SB phase, then for $eB\sim 0.1$ GeV$^{2}$ a first order phase transition into the $p\chi$SR phase occurs. With increasing the strength of the magnetic field up to $eB\sim 0.55$ GeV$^{2}$, the system remains in this phase, and then reenters the $\chi$SB phase for $eB>0.55$ GeV$^{2}$ phase. The same phenomenon occurs for all $T\gtrsim 300$ MeV. Let us notice, that the phenomenon of reentrance has been previously observed for nonvanishing $\mu$ and vanishing $\hat{\kappa}$ for the same hot and magnetized two-flavor NJL model, as considered in the present paper \cite{fayazbakhsh2} (see also Figs. \ref{fig9}(a) and \ref{fig9}(b) for $\mu=280$ MeV and $T\sim 90$ MeV). The results from Fig. \ref{fig8} show, that for $\kappa'$, the same phenomenon occurs also for $\mu=0$ MeV. We believe that this phenomenon, as well as the observed oscillations in the $eB$ dependence of $m$ appearing in Figs. \ref{fig1}, \ref{fig2} and \ref{fig5}, are essentially related to the
dHvA \cite{alphen1930} in the weak magnetic field regime $eB\in [0.2,0.5]$ GeV$^{2}$. These effect had been also studied in \cite{ebert1998, shovkovy2007, fayazbakhsh1, fayazbakhsh2} and most recently in \cite{simonov2014}. It occurs whenever Landau levels pass the quark Fermi level. As it turns out, the dHvA oscillations are weakened, once the system enters the LLL dominant regime $eB>0.5$ GeV$^{2}$.
\par
To compare the results for vanishing $\mu$ from Fig. \ref{fig8} with the results for nonvanishing $\mu$, the $T$--$eB$ phase diagram of hot and magnetized NJL model is demonstrated in Fig. \ref{fig9} for different $\hat{\kappa}=0$ GeV$^{-1}$ (panel a), $\kappa_{1}$ (panel b) and $\kappa_{2}$ (panel c) and for $\mu_{1}=0$ MeV, $\mu_{2}=150$ MeV and $\mu_{3}=280$ MeV. Solid lines denote the crossover transition lines, and dashed lines the first order phase transitions from $\chi$SB into p$\chi$SR phase. Let us first consider Fig. \ref{fig9}(a). We observe, that for $\hat{\kappa}=0$ GeV$^{-1}$ the crossover transition temperature decreases with increasing $\mu$. The aforementioned dHvA oscillations are stronger for larger $\mu$ and disappear in the LLL dominant regime $eB\gtrsim 0.5$ GeV$^{2}$. In this regime, $T_{c}$ increases with increasing $eB$. This can be regarded as a signature of MC, especially in the LLL dominant regime. For $\kappa_{1}$, however, $T_{c}$ decreases with increasing $eB$. Moreover, whereas for $\mu_{1}=0$ MeV and $\mu_{2}=150$ MeV the crossover transitions in the weak magnetic field regime turn into first order transition lines, for $\mu_{3}=280$ MeV only a crossover transition occurs in the regime $eB\in [0,0.155]$ GeV$^{2}$. The critical points corresponding to $\mu_{i}, i=1,2$ are denoted by $C_{i}, i=1,2$.  They are given by $C_{1}=(0.12~\mbox{GeV}^{2}, 181~\mbox{MeV})$ $C_{2}=(0.12~\mbox{GeV}^{2}, 157~\mbox{MeV})$. As in the case of $\hat{\kappa}=0$ GeV$^{-1}$, $T_{c}$ decreases with increasing $\mu$ for each fixed $eB$ (see also the $T$--$\mu$ phase diagram from Fig. \ref{fig11}).
Let us now compare the results for $\kappa_{2}$ from Fig. \ref{fig9}(c) with those for $\hat{\kappa}=0$ GeV$^{-1}$ from Fig. \ref{fig9}(a). We observe that although the $eB$
dependence of $T_{c}$ for $\kappa_{2}$, is in general similar to the case with $\hat{\kappa}=0$ GeV$^{-1}$, but for $\kappa_{2}$, the crossover transitions, appearing for
$\hat{\kappa}=0$ GeV$^{-1}$ in the whole range of $eB\in[0,0.8]$ GeV$^{2}$ and for all $\mu_{i}, i=1,2,3$, turn into a first order phase transitions in the LLL dominant regime
$eB\gtrsim 0.5$ GeV$^{2}$. The critical points corresponding to $\mu_{i}, i=1,2,3$  are given by $C_{1}=(0.478~\mbox{GeV}^{2}, 240~\mbox{MeV})$
$C_{2}=(0.48~\mbox{GeV}^{2}, 215~\mbox{MeV})$ and $C_{3}=(0.485~\mbox{GeV}^{2}, 112~\mbox{MeV})$. Similar to the case of $\hat{\kappa}=0$ GeV$^{-1}$, the aforementioned phenomenon of
reentrance from $\chi$SB into p$\chi$SR phase occurs also in the case $\kappa_{2}$, in particular, for $\mu=280$ MeV. Moreover, no dHvA oscillations occur in the LLL dominant regime.
\subsubsection{The $\mu$--$eB$ phase diagram for various $T$ and $\hat{\kappa}$}\label{subsec3B2}
\par\noindent
In Fig. \ref{fig10}, the $\mu$--$eB$ phase diagram of a hot and magnetized two-flavor NJL model is presented for $T_{1}=0$ MeV and $T_{2}=120$ MeV, as well as for $\hat{\kappa}=0$
GeV$^{-1}$ (panel a),  $\kappa_{1}$ (panel b) and  $\kappa_{2}$ (panel c). The black and blue solid lines denote the smooth crossovers for $T_{1}$ and $T_{2}$, respectively,
and the gray dashed lines the first order phase transitions from the $\chi$SB phase (the region below the critical lines) into the p$\chi$SR phase
(the region above the critical lines). The starting points of the first order phase transitions are denoted by $C_{i}, i=1,2$ for $T_{i}, i=1,2$. As it turns out,
for $\hat{\kappa}=0$ GeV$^{-1}$, crossover transitions occur in the whole regime of $eB\in [0,08]$ GeV$^{2}$. In contrast, for $\kappa_{1}$ and $\kappa_{2}$, crossover transitions
occur up to certain critical points $C_{i}, i=1,2$ and then with increasing $eB$, they turn into first order phase transitions. These critical points for $\kappa_{1}$ and $T_{1}=0$
MeV as well as $T_{2}=120$ MeV are, $C_{1}=(0.15~\mbox{GeV}^{2}, 261~\mbox{MeV})$ and $C_{2}=(0.13 ~\mbox{GeV}^{2},219~\mbox{MeV})$ [see Fig. \ref{fig10}(b)]. For $\kappa_{2}$
and $T_{1}=0$ MeV as well as $T_{2}=120$ MeV they are given by $C_{1}=(0.44~\mbox{GeV}^{2}, 296~\mbox{MeV})$ and $C_{2}=(0.48~\mbox{GeV}^{2}, 274~\mbox{MeV})$, respectively
[see Fig. \ref{fig10}(c)].  Here, similar to the $T$--$eB$ phase diagrams, for $\kappa_{1}$ the critical chemical potential $\mu_{c}$ decreases with increasing $eB$
[see Fig. \ref{fig10}(b)]. We conclude therefore that once the quark AMM is sizable enough an inverse magnetic catalysis occurs. Keeping $eB$ fixed, $\mu_{c}$ decreases
with increasing $T$ for all values of $\hat{\kappa}\neq 0$ (see also the $T$--$\mu$ phase diagram in Fig. \ref{fig11}). According to the results from Figs. \ref{fig10}(a) and
\ref{fig10}(c), for $\hat{\kappa}=0$ GeV$^{-1}$ and $\kappa_{2}$, the aforementioned dHvA oscillations lead to the phenomenon of reentrance in the regime $\mu\in [285, 335]$
MeV and $\mu\in [260, 330]$ MeV for $T_{1}=0$ MeV and $\hat{\kappa}=0$ GeV$^{-1}$ and $\kappa_{2}$, respectively. For $T_{2}=120$ MeV, these regimes are given by $\mu\in [245,270]$
MeV for $\hat{\kappa}=0$ GeV$^{-1}$ and $\mu\in [240, 265]$ MeV for $\kappa_{2}$.
\subsubsection{The $T$--$\mu$ phase diagram for various $eB$ and $\hat{\kappa}$}\label{subsec3B3}
\begin{figure*}[hbt]
\includegraphics[width=5.9cm,height=4.4cm]{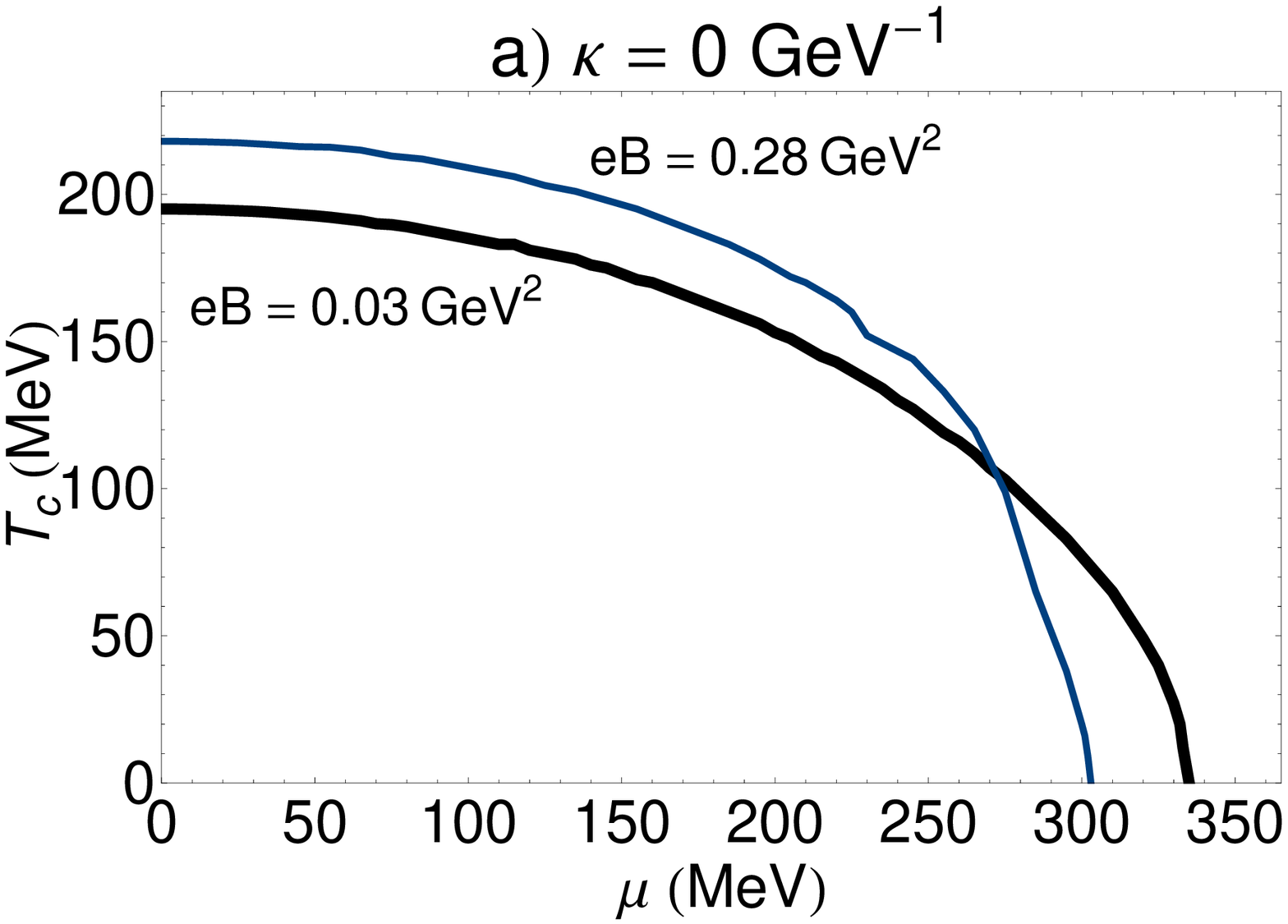}
\includegraphics[width=5.9cm,height=4.4cm]{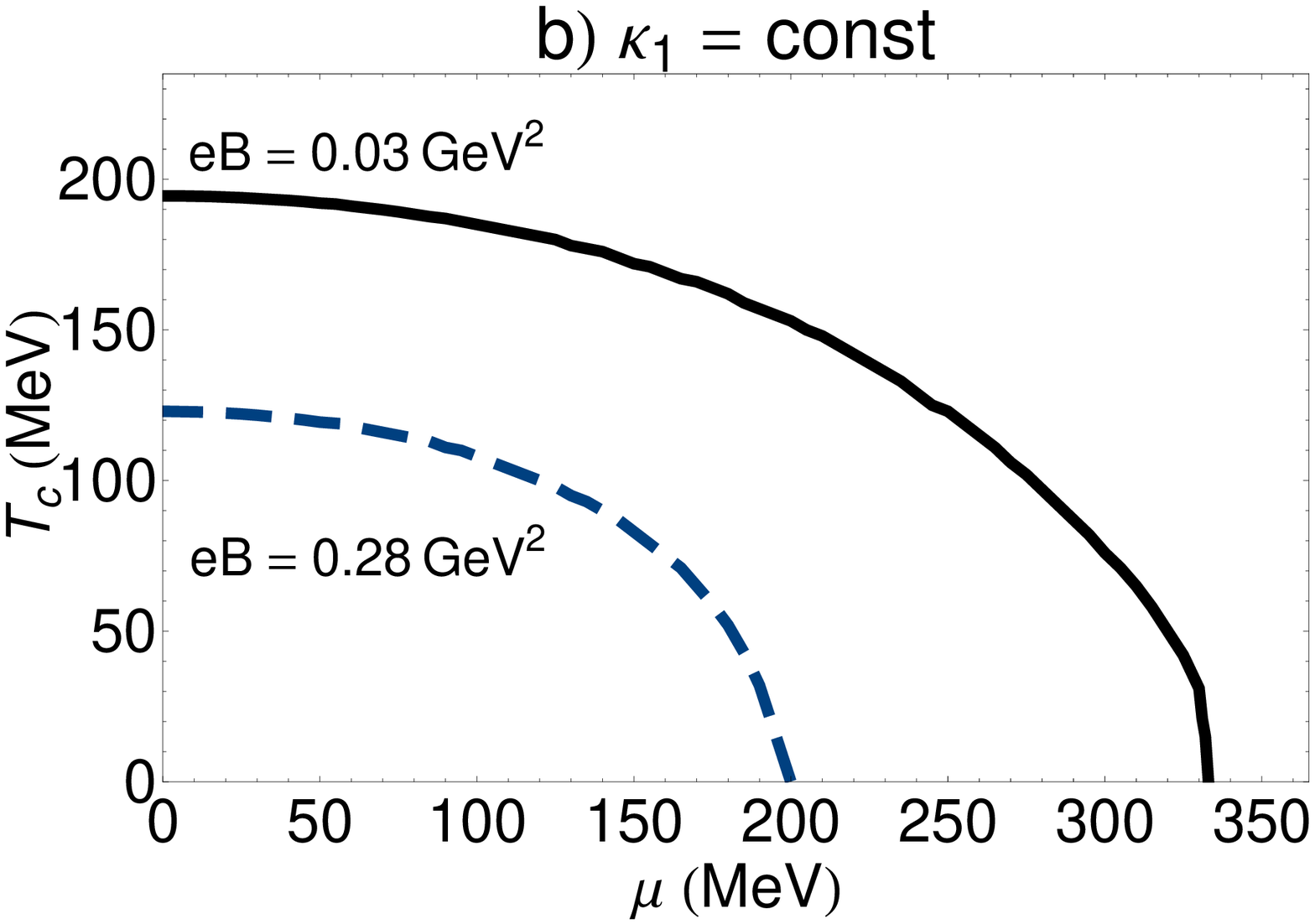}
\includegraphics[width=5.9cm,height=4.4cm]{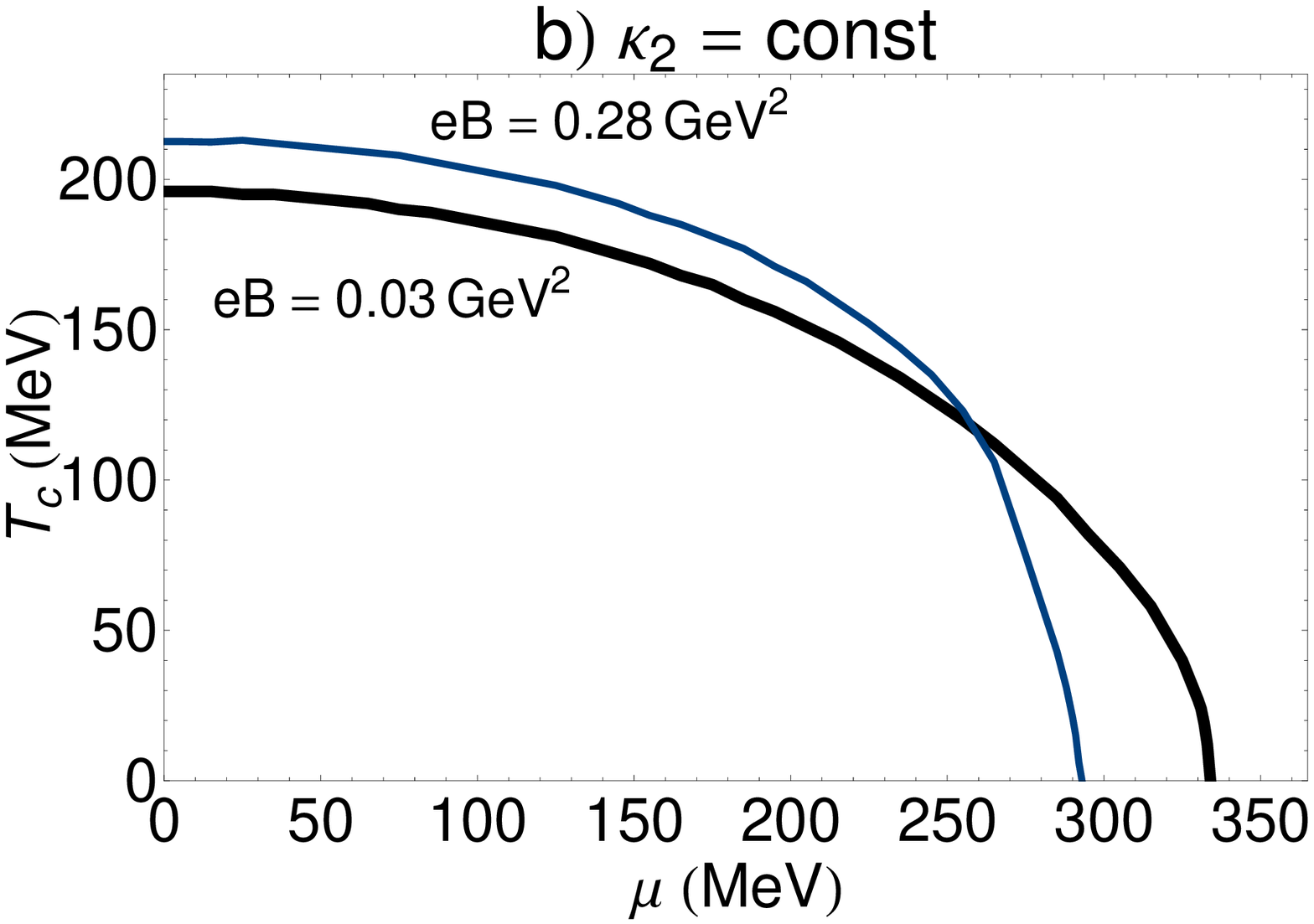}
\caption{(color online). The $T$--$\mu$ phase diagram of a hot and magnetized two-flavor NJL model is presented for $eB=0.03, 0.28$ GeV$^{2}$ and $\hat{\kappa}=0$ GeV$^{-1}$ (panel a), $\kappa_{1}$ (panel b) and $\kappa_{2}$ (panel c). The black and blue solid lines denote the crossover transitions for $eB=0.03, 0.28$ GeV$^{2}$, and the blue dashed line denotes the first order phase transition for $eB=0.28$ GeV$^{2}$.}\label{fig11}
\end{figure*}
\begin{figure*}[hbt]
\includegraphics[width=5.9cm,height=4.4cm]{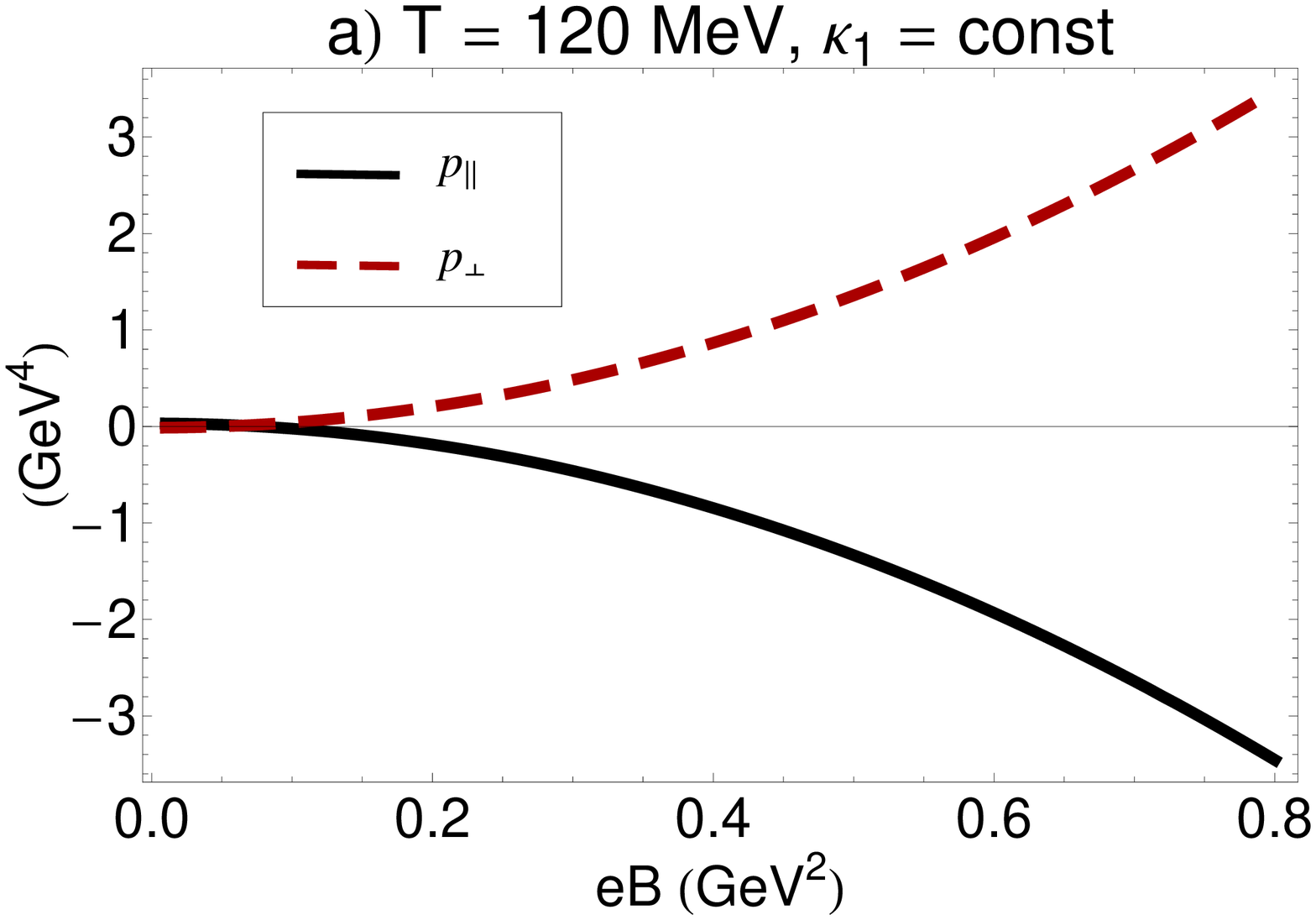}
\includegraphics[width=5.9cm,height=4.4cm]{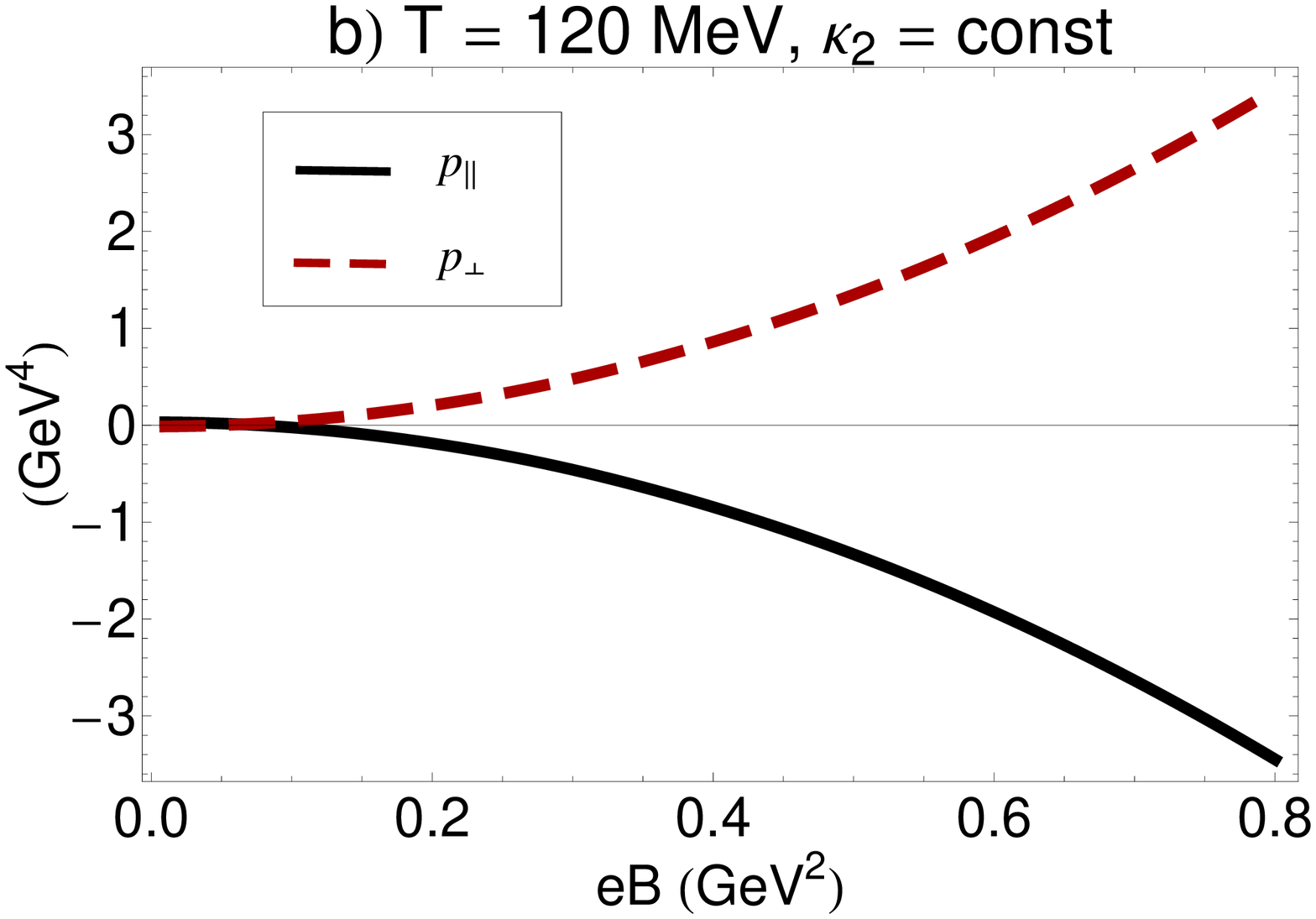}
\includegraphics[width=5.9cm,height=4.4cm]{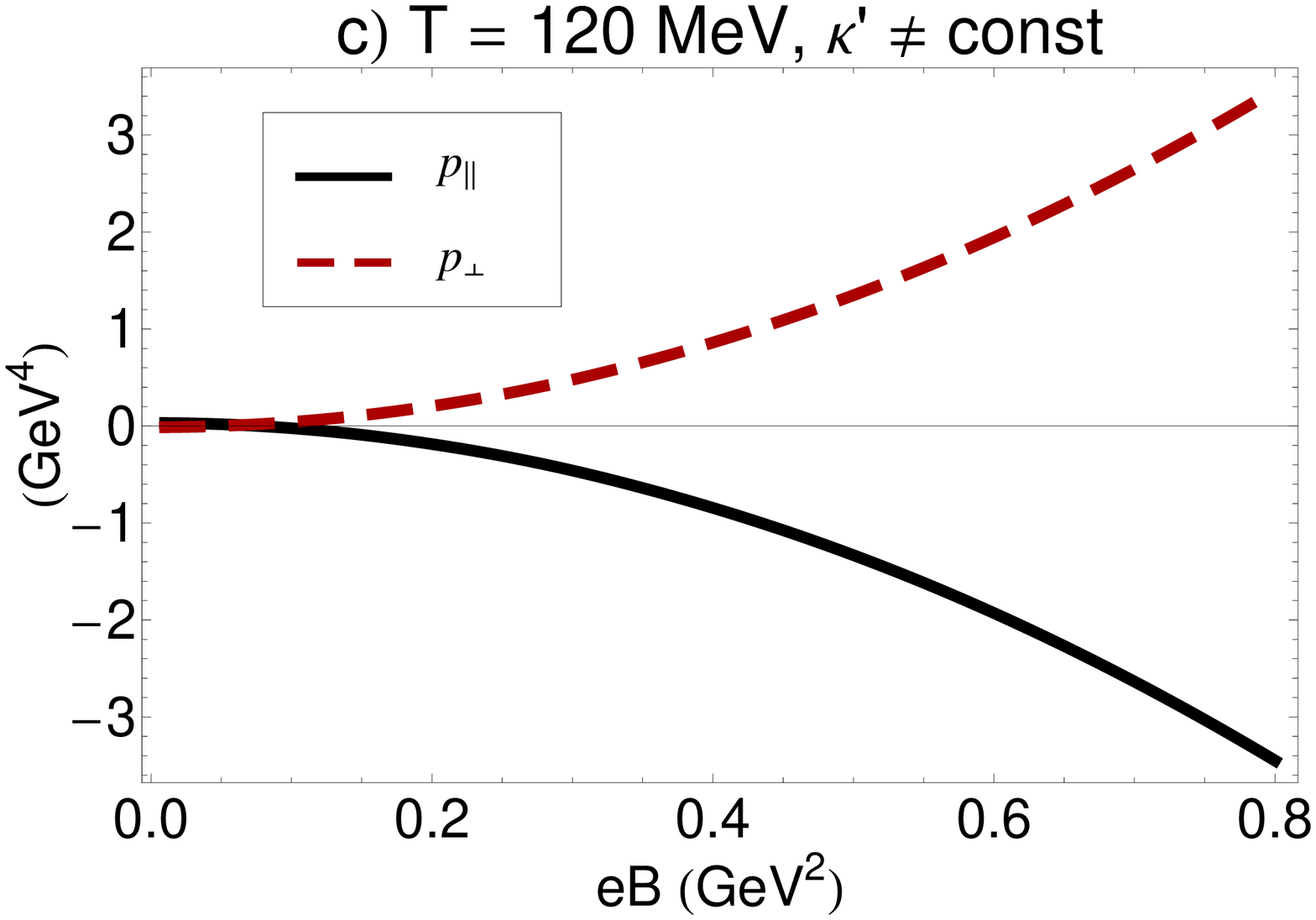}
\caption{(color online).  The longitudinal (black solid lines) and transverse (red dashed lines) pressures of a two-flavor NJL model, $p_{\|}$ and $p_{\perp}$, are demonstrated as a function of $eB$ for $T=120$ MeV, $\mu=0$ MeV and $\kappa_{1}$ (panel a) and $\kappa_{2}$ (panel b), as well as $\kappa'$ (panel c). As it turns out, the longitudinal (transverse) pressure decreases (increases) with increasing $eB$. Different choices for $\hat{\kappa}$ do not significantly affect this specific feature.}\label{fig12}
\end{figure*}
\begin{figure}[hbt]
\includegraphics[width=8.9cm,height=6.0cm]{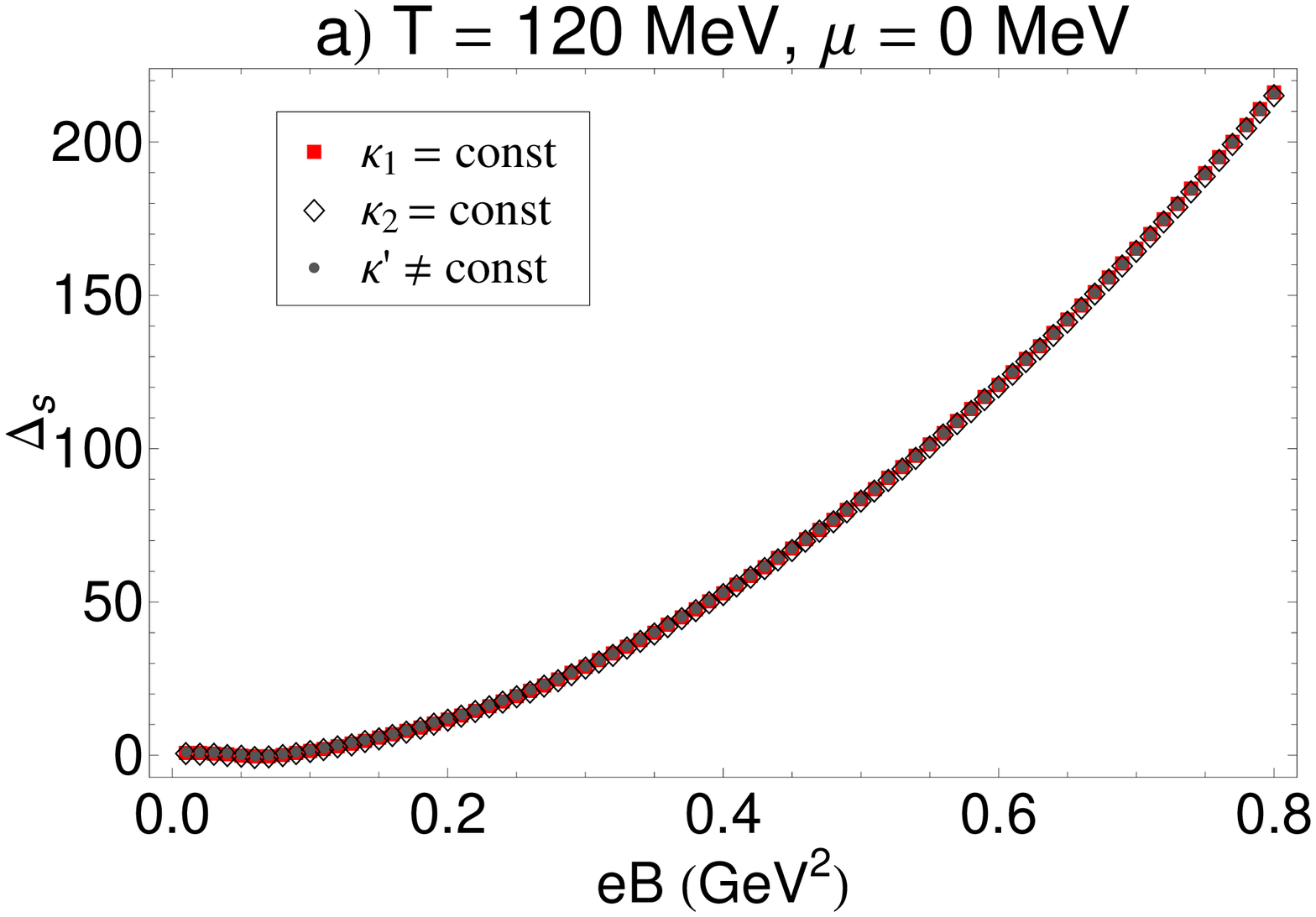}
\includegraphics[width=8.9cm,height=6.0cm]{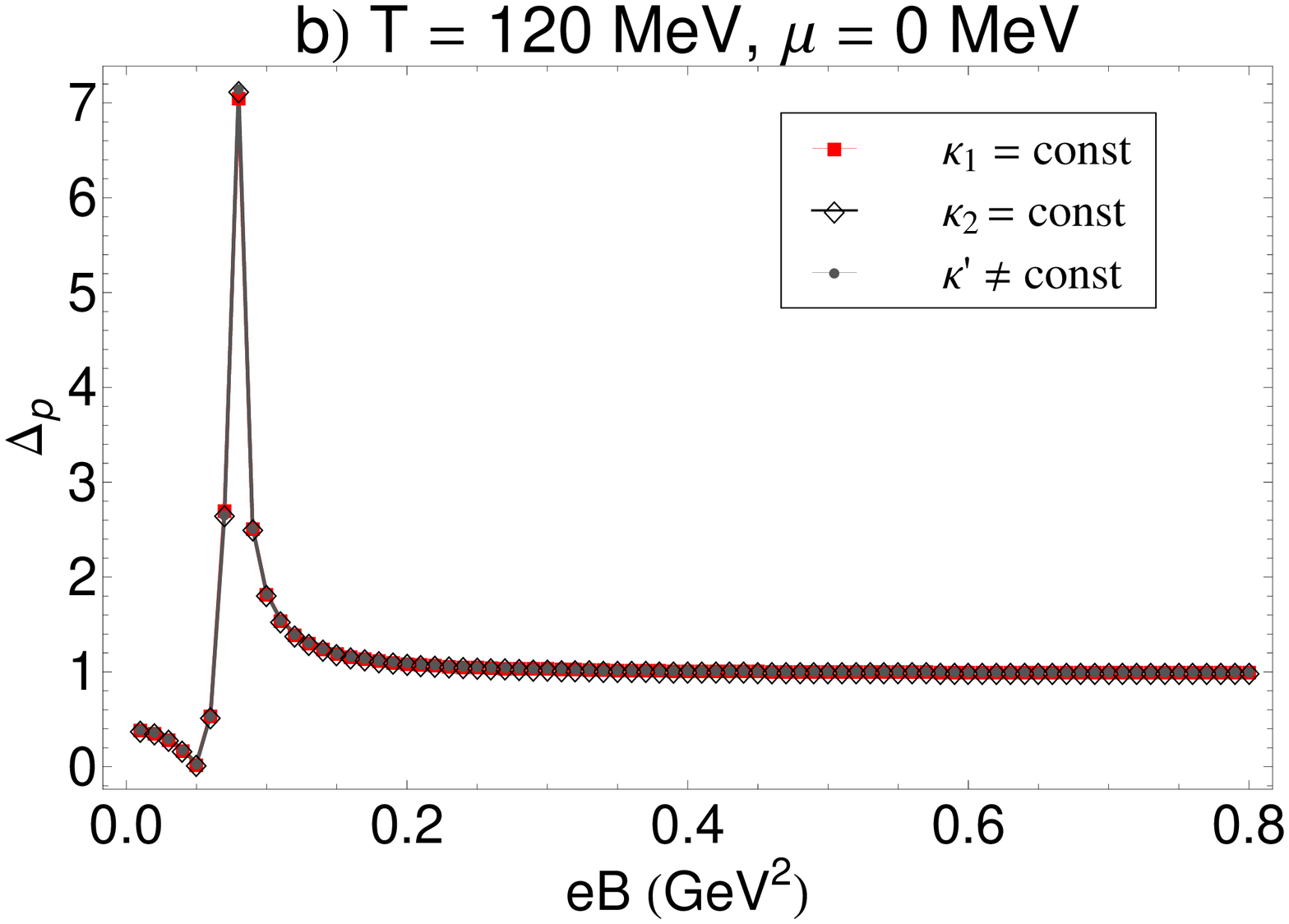}
\caption{(color online).  The quantities $\Delta_{s}$ and $\Delta_{p}$, defined in (\ref{A7}) and (\ref{A8}), are plotted as functions of $eB$, at  $T=120$ MeV and for $\mu=0$ MeV as well as $\kappa_{1}$ (red squares), $\kappa_{2}$ (empty diamonds) and $\kappa'$ (gray circles).  Different choices of $\hat{\kappa}$s have no significant effect on the $eB$ dependence of $\Delta_{s}$ and $\Delta_{p}$.}\label{fig13}
\end{figure}
\begin{figure*}[hbt]
\includegraphics[width=5.9cm,height=4.4cm]{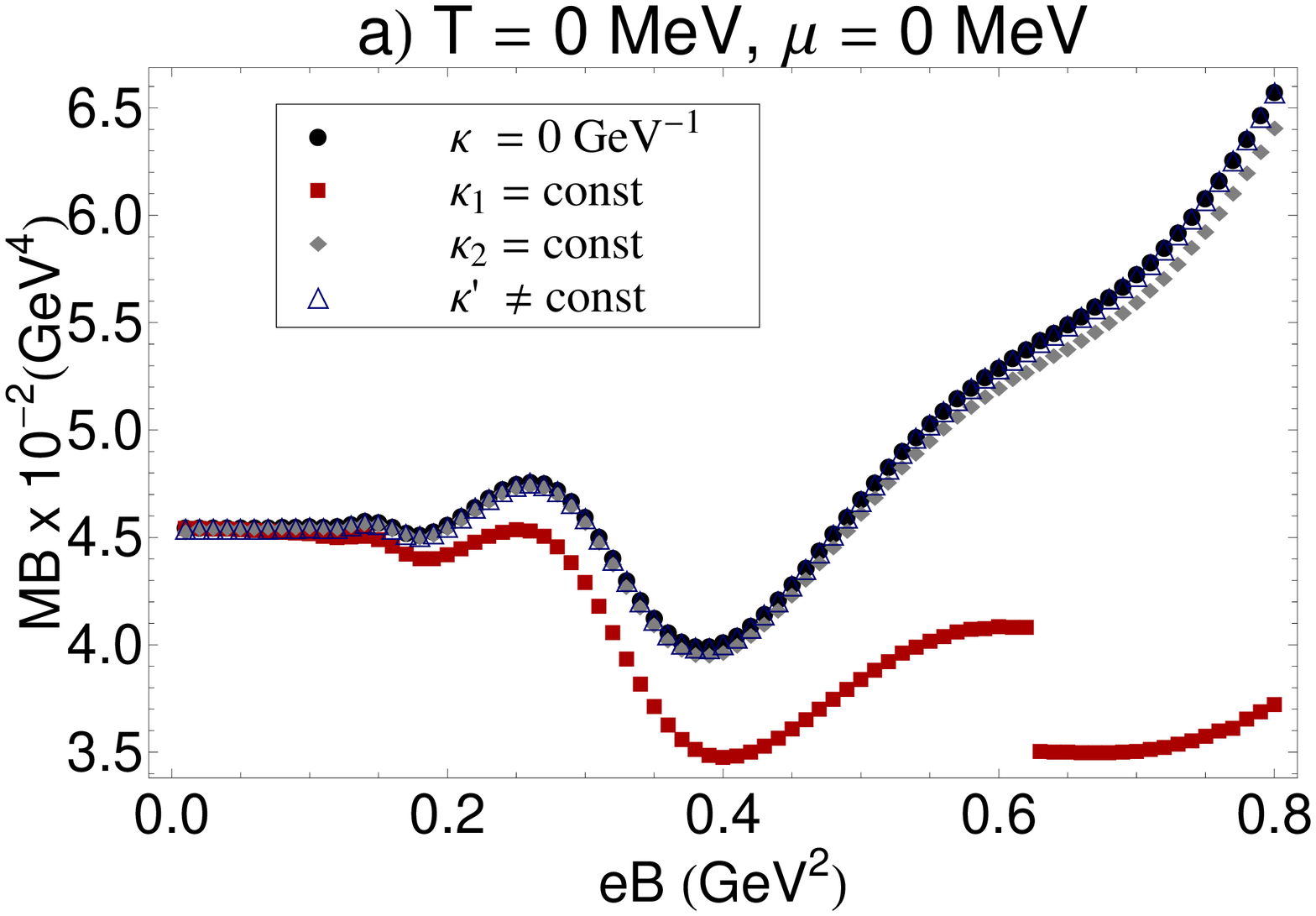}
\includegraphics[width=5.9cm,height=4.4cm]{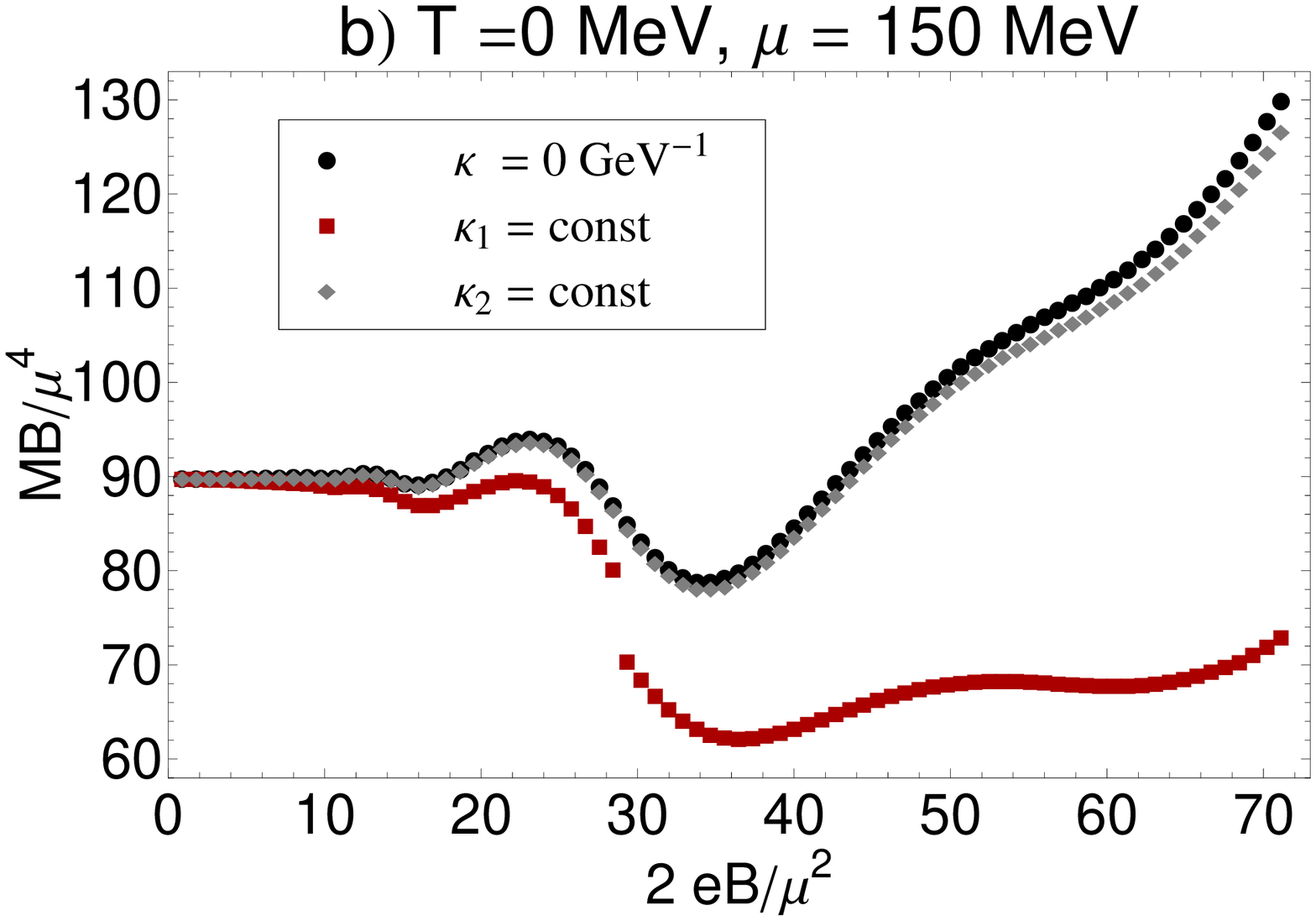}
\includegraphics[width=5.9cm,height=4.4cm]{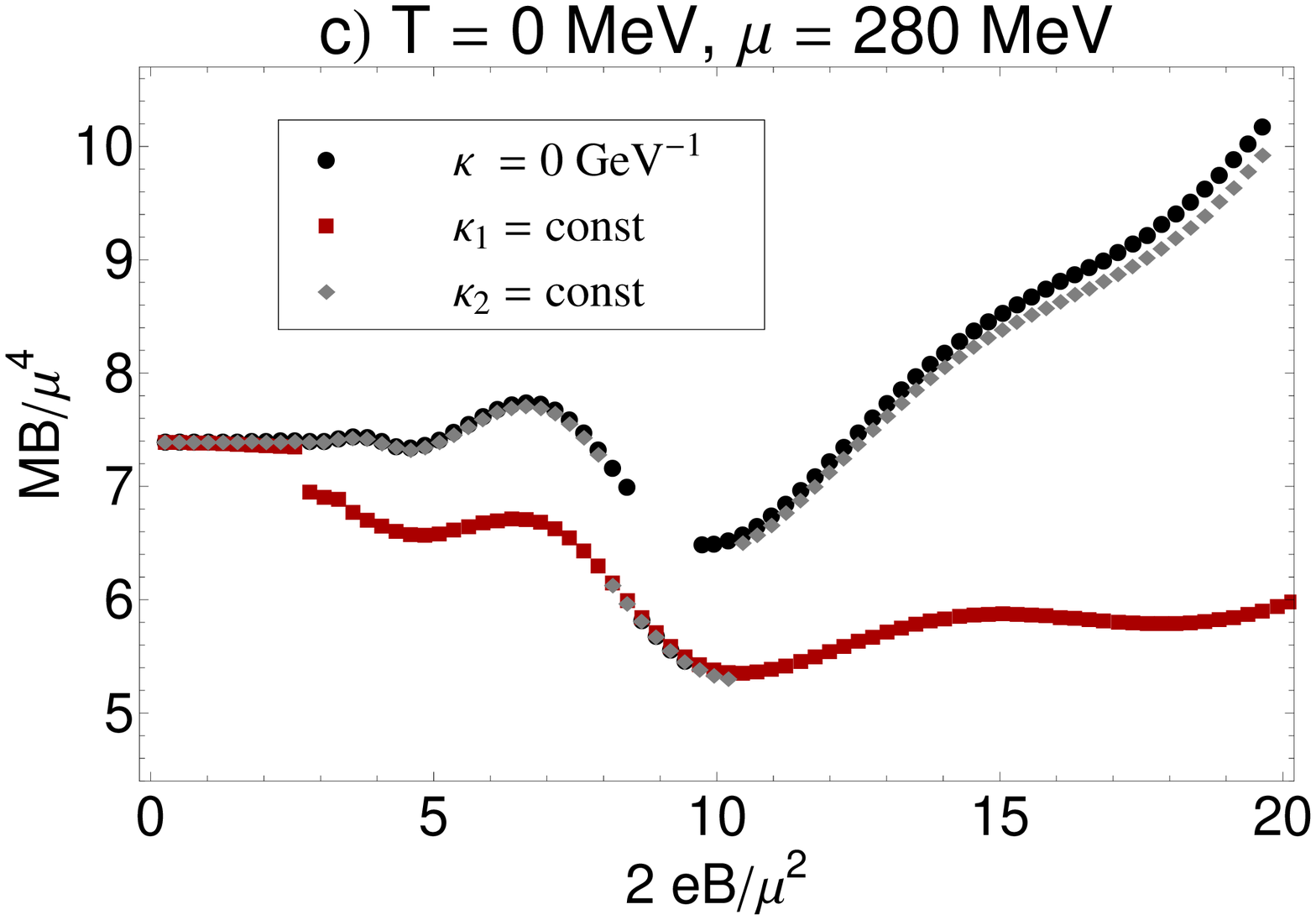}
\caption{(color online).  (a) The product of the magnetization $M$ and the magnetic field strength  $B$ of a two-flavor NJL model is plotted at $T=\mu=0$ MeV and for $\hat{\kappa}=0$ GeV$^{-1}$ (black circles), $\kappa_{1}$ (red squares), $\kappa_{2}$ (gray diamonds) and $\kappa'$ (empty triangles). (b) and (c) The dimensionless quantity $MB/\mu^{4}$ is plotted as a function of 2$eB/\mu^{2}$ at $T=0$ MeV, $\mu=120$ MeV (panel b) and $\mu=280$ MeV (panel c) as well as for $\hat{\kappa}=0$ GeV$^{-1}$ (black circles), $\kappa_{1}$ (red squares), $\kappa_{2}$ (gray diamonds).  }\label{fig14}
\end{figure*}

\par\noindent
In Fig. \ref{fig11}, the $T$--$\mu$ phase diagram of a hot and magnetized two-flavor NJL model is presented for $eB=0.03, 0.28$ GeV$^{2}$ and $\hat{\kappa}=0$ GeV$^{-1}$ (panel a),
$\kappa_{1}$ (panel b) and $\kappa_{2}$ (panel c). The crossover transition lines for $eB=0.03, 0.28$ GeV$^{2}$ are denoted by black and blue solid lines, and the blue dashed line
denotes the first order phase transition for $eB=0.28$ GeV$^{2}$. As expected from the results in Figs. \ref{fig9} and \ref{fig10}, for a fixed $eB$, the critical temperature of the
chiral phase transition decreases with increasing $\mu$, and different values of $\kappa$ do not essentially change this specific feature. However,  for $\kappa_{1}$, in contrast to
the cases of $\hat{\kappa}=0$ GeV$^{-1}$ and $\kappa_{2}$, for fixed $\mu$, the critical temperature decreases with increasing $eB$ (IMC). Comparing the results from Fig. \ref{fig11}(a)
and \ref{fig11}(c) with \ref{fig11}(b), it turns out that the crossover transitions for $eB=0.28$ GeV$^{2}$, $\hat{\kappa}=0$ GeV$^{-1}$ and $\kappa_{2}$ turn into a first order phase transition for $\kappa_{1}$. This result confirms our findings in Figs. \ref{fig8}, \ref{fig9}(b) and \ref{fig10}(b), and can be regarded as one of the main indications of the effect of large $\hat{\kappa}$ on QCD phase diagram.
\subsection{The pressure anisotropy and magnetization of quark matter for nonvanishing  $\hat{\kappa}$}\label{subsec3C}
\par\noindent
Nonvanishing magnetic fields break the Lorentz invariance, and induce certain anisotropies in the pressure of a hot and magnetized quark matter with respect to the direction of the background field. The $eB$ dependence of pressure of a hot and magnetized quark matter for vanishing and nonvanishing $\hat{\kappa}$ has been already demonstrated in \cite{martinez2013, martinez2014-1, martinez2014-2}, where the effect of the quark AMM in the strong magnetic field regime $eB>m_{0}^{2}$ is compared with the weak-field AMM by Schwinger \cite{schwinger1948}. The anisotropy in the pressure of hot and magnetized QCD is also investigated recently in \cite{bali2014} in the framework of lattice gauge theory. Let us denote the longitudinal and transverse pressures with respect to the direction of the magnetic field by $p_{\|}$ and $p_{\perp}$. According to \cite{martinez2014-1, martinez2014-2}, they are defined by
\begin{eqnarray}\label{A5}
p_{\|}&\equiv& -\Omega_{\mbox{\tiny{eff}}}^{\mbox{\tiny{min}}}(m;T,\mu,eB)-\frac{B^{2}}{2},\nonumber\\
p_{\perp}&\equiv&-\Omega_{\mbox{\tiny{eff}}}^{\mbox{\tiny{min}}}(m;T,\mu,eB)-\frac{B^{2}}{2}+BH,
\end{eqnarray}
where $\Omega_{\mbox{\tiny{eff}}}^{\mbox{\tiny{min}}}(m;T,\mu,eB)$ is the value of $\Omega_{\mbox{\tiny{eff}}}$ from (\ref{F11}), evaluated at the minimum of the effective potential and $B^{2}/2$ is the classical magnetic energy. Moreover, $B$ and $H\equiv B-M$ are the external and induced magnetic fields, respectively. Here, $M$ is the magnetization of the quark matter, defined by \cite{fayazbakhsh1}
\begin{eqnarray}\label{A6}
M\equiv -e\frac{\partial\Omega_{\mbox{\tiny{eff}}}(\tilde{m};T,\mu,eB)}{\partial (eB)}\bigg|_{\tilde{m}=m}.
\end{eqnarray}
In what follows, we are, in particular, interested in the effect of different sets of $\hat{\kappa}$ on the $eB$ dependence of $p_{\|}, p_{\perp}$ and on the product $MB$.
\par
In Fig. \ref{fig12}, the longitudinal and transverse pressures, $p_{\|}$ (black solid curves) and $p_{\perp}$ (red dashed curves) are plotted as  functions of $eB$ for $T=120$ MeV, $\mu=0$ MeV and $\kappa_{1}$ (panel a), $\kappa_{2}$ (panel b) as well as $\kappa'$ (panel c).\footnote{In Fig. \ref{fig12}, the anisotropic pressures from (\ref{A5}) are slightly modified by $p_{\|}=-\Omega_{\mbox{\tiny{eff}}}^{\mbox{\tiny{min}}}-e^{-2}b^2/2$ and   $p_{\perp}=-\Omega_{\mbox{\tiny{eff}}}^{\mbox{\tiny{min}}}+e^{-2}b^2/2-\tilde{M}b$ with $b\equiv eB$ and $\tilde{M}\equiv e^{-1}M$. Here, $e^{2}$ is replaced by $e^{2}=4\pi\alpha_{e}$, with $\alpha_{e}=1/137$ the electromagnetic fine structure constant.} Similar to the results presented in \cite{martinez2013}, $p_{\|}$ ($p_{\perp}$) decreases (increases) with increasing $eB$. As it turns out, different choices of $\hat{\kappa}$ do not significantly affect the final results for the $eB$ dependence of anisotropic pressures.
Let us notice at this stage, that there is indeed an ambiguity in determining the longitudinal and transverse pressures in the literature. In \cite{chaichian1999}, for instance, the Maxwell term $B^{2}/2$ is not considered neither in the effective potential, nor in the definitions of $p_{\|}$ and $p_{\perp}$. In \cite{martinez2014-1,martinez2014-2}, however, whereas the Maxwell term is not included in the effective potential $\Omega_{\mbox{\tiny{eff}}}$, it is included in the definitions (\ref{A5}) of anisotropic pressures. To partly overcome these ambiguity, and at the same time to define a measure for the splitting of $p_{\|}$ and $p_{\perp}$ for nonvanishing $eB$, we define the splitting coefficient $\Delta_{s}$ for a fixed $(T^{*},\mu^{*})$ as \cite{martinez2014-2}
\begin{eqnarray}\label{A7}
\Delta_{s}(B;T^{*},\mu^{*})\equiv \bigg|\frac{p_{\|}(B;T^{*},\mu^{*})-p_{\perp}(B;T^{*},\mu^{*})|}{p(0;T^{*},\mu^{*})}\bigg|,\nonumber\\
\end{eqnarray}
where $p(0;T^{*},\mu^{*})$ is the pressure at fixed $(T^{*},\mu^{*})$ and for vanishing magnetic field. In Fig. \ref{fig13}(a), $\Delta_{s}$ is plotted as a function of $eB$ at fixed temperature $T=120$ MeV and for vanishing chemical potential. The results for $\kappa_{1}, \kappa_{2}$ and $\kappa'$ are denoted by red squares, empty diamonds and gray circles, respectively. As it turns out, $\Delta_{s}$ increases with increasing magnetic field $eB$, as expected \cite{martinez2014-2}.
This specific feature is, in particular, not affected by different choices for $\hat{\kappa}$. Small deviations, up to maximum $0.4\%$ occur only in the strong field regime, and increase with increasing $eB$. In comparison with $\hat{\kappa}=0$ GeV$^{-1}$, the same maximum deviation occurs is in the strong-field regime, and as it turns out nonvanishing $\hat{\kappa}$ has a negative effect on $\Delta_{s}$, especially in the strong-field regime. The same observation is also made in \cite{strickland2012}.
\par\noindent
Another useful quantity that quantifies the relation between $p_{\|}$ and $p_{\perp}$ is  $\Delta_{p}$, \cite{strickland2012}
\begin{eqnarray}\label{A8}
\Delta_{p}(B;T^{*},\mu^{*})\equiv\bigg| \frac{p_{\perp}(B;T^{*},\mu^{*})}{p_{\|}(B;T^{*},\mu^{*})}\bigg|,
\end{eqnarray}
where $(T^{*},\mu^{*})$ are fixed temperature and chemical potential. As it turns out from Fig. \ref{fig13}(b), $\Delta_{p}$ has a minimum for a certain $eB_{\mbox{\tiny{min}}}= 0.04$ GeV$^{2}$, and then increases and has a maximum for another $eB_{\mbox{\tiny{max}}}= 0.08$ GeV$^{2}$. It then decreases to values $\Delta_{p}\sim 1$. Almost no differences occurs between different $\hat{\kappa}$s. Let us notice, that $eB_{\mbox{\tiny{min}}}$ and $eB_{\mbox{\tiny{max}}}$ are related to the specific magnetic fields, where  $p_{\perp}$ and $p_{\|}$ almost vanish. As it turns out, $eB_{\mbox{\tiny{min}}}\neq eB_{\mbox{\tiny{max}}}$. This tiny difference is not visible in Fig. \ref{fig12}.
\par
The pressure anisotropy has various effects on astrophysics of dense stellar objects \cite{martinez2014-2, chaichian1999} and the experiments of heavy ion collisions \cite{bali2014}. In \cite{ellia2013, bali2014, simonov2014}, for instance, the magnetization of quark matter $M$ is determined as a function of temperature. In \cite{bali2014} is shown that in the vicinity of the chiral transition point, the magnetization $M$ is positive, and therefore hot and dense QCD at transition point exhibits a paramagnetic response. It is further shown, that the paramagnetic behavior of QCD matter affects the phenomenology of heavy ion collision, and in particular, has ``significant impact on the value of elliptic flow $v_{2}$''.  In Fig. \ref{fig14}(a), the product of the magnetization $M$ and the magnetic field strength $B$ is plotted for our hot and dense NJL model at $T=\mu=0$ and for $\hat{\kappa}=0$ GeV$^{-1}$ (black circles), $\kappa_{1}$ (red squares), $\kappa_{2}$ (gray diamonds) and $\kappa'$ (empty triangles). As it turns out, the results for $\kappa_{1}$ have significant difference with the results corresponding to $\hat{\kappa}=0$ GeV$^{-1}$, $\kappa_{2}$ and $\kappa'$.  As expected, at $T=0$ MeV, there is no difference between the $eB$ dependence of $MB$ for $\hat{\kappa}=0$ GeV$^{-1}$ and $\kappa'$. The difference between the data for $\hat{\kappa}=0$ GeV$^{-1}$ and $\kappa_{2}$ becomes only significant in the LLL dominant regime $eB> 0.5$ GeV$^{2}$. In this regime, as expected, the dHvA oscillations arising in the regime $eB<0.5$ GeV$^{2}$ weaken, and $MB$ monotonically increases with increasing $eB$. These oscillations are also previously observed in \cite{shovkovy2007, fayazbakhsh1}. The discontinuity arising in the data for $\kappa_{1}$ at $eB=0.623$ GeV$^{2}$ (red squares), is related to the first order phase transition from the $\chi$SB into the p$\chi$SR phase, demonstrated in Fig. \ref{fig8}. The latter leads also to a discontinuity in the $eB$ dependence of the constituent quark mass exactly for $eB=0.623$ GeV$^{2}$ [see Fig. \ref{fig1}(a)].
In Figs. \ref{fig14}(b) and (c), the dimensionless quantity $MB/\mu^{4}$ is plotted as a function of $2eB/\mu^{2}$ at $T=0$ MeV and $\mu=150$ MeV [Fig. \ref{fig14}(b)] and $\mu=280$ MeV [Fig. \ref{fig14}(c)] as well as different $\hat{\kappa}$s. Black circles, red squares and gray diamonds denote the results for $\hat{\kappa}=0$ GeV$^{-1}$, $\kappa_{1}$ and $\kappa_{2}$, respectively. This dependence is also studied in \cite{martinez2013} in the regime $2eB/\mu^{2}\in [0,1]$ for $\mu=300$ MeV. Similar to the results of $\mu=0$ MeV, the difference between $\hat{\kappa}=0$ GeV$^{-1}$ and $\kappa_{2}$ increases with increasing $eB$. Except in the regime of weak magnetic field $eB\lesssim 0.17$ GeV$^{2}$ ($2eB/\mu^{2}\lesssim 15$ for $\mu=150$ MeV), and
$eB\lesssim 0.08$ GeV$^{2}$ ($2eB/\mu^{2}\lesssim 2$ for $\mu=280$ MeV),  $MB/\mu^{4}$ for $\kappa_{1}$ is smaller than the data for $\hat{\kappa}=0$ GeV$^{-1}$ and $\kappa_{2}$. According to the $T$--$eB$ phase diagram for $\mu=280$ MeV  in Fig. \ref{fig9}, we expect discontinuities in the $eB$ dependence of  $MB/\mu^{4}$ in the regime $0.34<eB\leq 0.39$ GeV$^{2}$ ($8.7\lesssim 2eB/\mu^{2}\leq 10$)  for $\hat{\kappa}=0$ GeV$^{-1}$,
 $eB\simeq 0.155$ GeV$^{2}$ ($2eB/\mu^{2}\sim 3.95$) for $\kappa_{1}$ and $0.3<eB\leq 0.4$ GeV$^{2}$ ($7.7<2eB/\mu^{2}\leq 10.2$) for
$\kappa_{2}$. We conclude that discontinuities in the $eB$ dependence of $MB$ or $MB/\mu^{4}$ as functions of $eB$ are related, as expected, to first order chiral phase transitions at certain magnetic fields and for fixed values of $T$ and $\mu$.
\section{Concluding remarks}\label{sec4}
\setcounter{equation}{0}
\par\noindent
In recent years, there were a number of attempts to explore the effect of the quark AMM on the phase diagram of QCD at finite temperature, chemical potential and in the presence of uniform magnetic fields \cite{strickland2012, ferrer2013, martinez2013, martinez2014-1}.
Following the method used in \cite{strickland2012}, we have studied, in the present paper, the effects of the (effective) quark AMM on the thermodynamic properties of the constituent quark mass $m$, and on the full phase portrait of a two-flavor magnetized NJL model at finite $T$ and $\mu$. The quark AMM is introduced via an additional minimal coupling term, $\hat{a}\sigma_{\mu\nu}F^{\mu\nu}$, in the Lagrangian density of the NJL model \cite{schwinger1948}. The coefficient $\hat{a}$, defined by $\hat{a}=\hat{Q}\hat{\alpha}\mu_{B}$, includes the nonperturbative (effective) Bohr magneton $\mu_{B}=\frac{e}{2m}$, and is herewith, as function of the constituent quark mass, and receives $(T,\mu,eB)$ corrections. As it turns out, $\hat{a}\sigma_{\mu\nu}F^{\mu\nu}$ leads to an additional term proportional to  $T_{f}=\kappa_{f}q_{f}eB$ in the quark energy dispersion relation (\ref{F9}). In the above expression, $f=u,d$ stands for up ($u$) and down ($d$) quark flavors, and the dimensionful coupling $\kappa_{f}$ is defined by $\kappa_{f}=\frac{\alpha_{f}}{2m}$, with $\alpha_{f}$ related to the deviation of the Land\'e $g$-factor from $2$. Our aim was, in particular, to study the effects of constant as well as $(T,\mu,eB)$ dependent effective coupling $\kappa_{f}$ on the thermodynamic behavior of $m$, as well as on the phase portrait of the hot and magnetized quark matter, described by our model. To this purpose, three different sets for the effective coupling $\kappa_{f}$ are chosen by making use of the method presented in Sec. \ref{sec2} and App. \ref{appA} [see (\ref{F12})-(\ref{F14})].  The dependence of the constituent quark mass $m$ on $T,\mu$ and $eB$ is then determined for each fixed $\kappa_{f}$ in Sec. \ref{subsec3A} (see Figs. \ref{fig1}-\ref{fig7}). Here, we have explicitly described the signatures related to the phenomena of MC and IMC. We have further shown that for large enough $\kappa_{f}$, even in the regime of strong magnetic fields, the $eB$ dependence of $m$ is strongly affected by the phenomenon of IMC.
Then, using the one-loop effective potential $\Omega_{\mbox{\tiny{eff}}}$ from (\ref{F11}) in term of $\kappa_{f}$, we have explored the complete phase portrait of the model in the parameter space $T,\mu,eB$, and $\kappa_{f}$ (see Sec. \ref{subsec3B}). We have shown, that for large enough $\kappa_{f}$ and mainly in the regime of weak magnetic fields $eB<0.5$ GeV$^{2}$, the critical temperature $T_{c}$ and critical chemical potential $\mu_{c}$ decrease with increasing $eB$. This is related to the phenomenon of IMC.
Moreover, it is shown that in certain regimes of the parameter space, the phenomenon of reentrance of chiral symmetry broken phase occurs, mainly as a consequence of dHvA oscillations \cite{alphen1930}. Also, the order of the phase transition turns out to be affected by $\kappa_{f}$ (see Figs. \ref{fig8}-\ref{fig11}). In Sec. \ref{subsec3C}, the pressure anisotropy of the quark matter in the longitudinal and transverse directions with respect to the magnetic field is considered. We have shown that different choices of $\kappa_{f}$ have no specific effect on the pressure anisotropies, demonstrated in Fig. \ref{fig12}, and on $\Delta_{s}$ and $\Delta_{p}$, as quantitative measures for these anisotropies (see Fig. \ref{fig13}). In Fig. \ref{fig14}, the $eB$ dependence of the product of the field strength and the magnetization is demonstrated. We have shown that for large enough $\kappa_{f}$, this product becomes smaller than for the cases of $\hat{\kappa}=0$ GeV$^{-1}$ and small $\kappa_{f}$, but the magnetization is always positive. According to \cite{bali2014}, this is believed to be an indication of paramagnetic behavior of the hot and magnetized quark matter, especially in the vicinity of the phase transition point.
\par
Let us notice at this stage, that the method used in the present paper to introduce the quark AMM is different from the method presented in \cite{ferrer2013}. Here, starting from a one-flavor magnetized NJL model with an appropriate tensor channel, a mechanism for the dynamical generation of the quark AMM in the LLL is presented. It is also shown that the scalar and tensor couplings of the NJL model become anisotropic, and receive longitudinal and transverse components with respect to the direction of the magnetic field. For constant anisotropic couplings, in the subcritical regime, the phenomenon of IMC does not occur. In a subsequent paper \cite{ferrer2014}, however, the authors consider the running of these couplings as a function of $eB$, and show that because of a certain antiscreening effect, induced by quarks that are confined by the magnetic fields to the LLL, the critical temperature of the $\chi$SB decreases with increasing the magnetic field strength. This is believed to be a natural explanation for the phenomenon of IMC, that arises originally in a number of  model calculations \cite{inagaki2003,fayazbakhsh2}, in the framework of gauge/gravity duality \cite{rebhan2011}, and from an ab initio lattice QCD simulation at finite $T$ and $eB$ \cite{bali2011}.
\par
Let us also notice that the linear-in-$B$ ansatz, used in the present paper, is different from the one which is used in \cite{martinez2013,martinez2014-1}. Here, the Schwinger term is defined to be proportional to the Bohr magneton $\mu_{B}^{0}=\frac{e}{2m_{0}}$ in term of the current (bare) quark mass $m_{0}$, in contrast to our approach, described above. Thus far, the inconsistencies from the Schwinger linear-in-$B$ ansatz in the weak-magnetic field approximation, described in \cite{martinez2013, martinez2014-1}, are not expected to occur in our approach. Moreover, in contrast to the previous approaches, we have considered the contributions of all Landau levels, and not restricted ourselves to LLL \cite{ferrer2013}, nor to one-loop approximation in the LLL, as in \cite{martinez2013,martinez2014-1}.
\par
There are many possibilities to improve the approach presented in this paper. As we have argued in Sec. \ref{sec2}, the coefficient $\kappa_{f}$ and the constituent quark mass $m$ are closely entangled. In other words, it is not possible to determine one of them without determining the other one. The reason is indeed formulated in \cite{ferrer2013}, where it is stated that since one and the same symmetry is broken by the quark AMM and the chiral condensate, nothing can guarantee a vanishing AMM, once the chiral symmetry is broken by a nonvanishing chiral condensate. In the present paper, using a method compatible with the constituent quark model, we have fixed $\kappa_{f}$ and determined $m$. This method can gradually be improved. The main idea is to determine $\kappa_{f}$ from the relation $\kappa_{f}=\frac{1}{2m}(\frac{m}{I_{f}}-1)$, where $I_{f}$ for up and down quarks are given in (\ref{appA8}). Instead of fixing $m$ with the phenomenologically reliable $M=420$ MeV and $M=340$ MeV, as it is performed in the present paper, we can replace it, e.g. by $m_{0}(T)\equiv m(T,\mu^{*},eB^{*};\hat{\kappa}=0)$, where $\mu^{*}$ and $eB^{*}$ are fixed values of chemical potential and magnetic field. We then obtain
\begin{eqnarray*}
\kappa_{u}(T,\mu^*,eB^*)=\frac{1}{2m_{0}(T)}\left(\frac{m_{0}(T)}{0.338}-1\right),
\end{eqnarray*}
for up quarks, and
\begin{eqnarray*}
\kappa_{d}(T,\mu^*,eB^*)=\frac{1}{2m_{0}(T)}\left(\frac{m_{0}(T)}{0.322}-1\right),
\end{eqnarray*}
for down quarks. Plugging these relations into (\ref{F9}), and the latter into $\Omega_{\mbox{\tiny{eff}}}(m_{0}(T); T,\mu^{*},eB^{*})$, and eventually looking for the global minima of the resulting expression, a new set of constituent quark mass arises, which replaces the data demonstrated in Figs. \ref{fig6}, for instance. The same procedure may be repeated for the sets $(T^{*},\mu, eB^{*})$ or $(T^{*},\mu^{*},eB)$, where the fixed values of $T,\mu$ and $eB$ are denoted by the superscript ``star''.
It would be interesting to look for the phenomenon of MC and IMC in this framework. We will report about the results of this new approach in the a future publication.
\section*{ACKNOWLEDGMENTS}
\par\noindent
The authors acknowledge F. Ardalan for discussions about the effect of the anomalous magnetic moment of quarks on their spectra, and M. Mohammadi Najafabadi for providing insight into the experimental data for the magnetic moment of protons and neutrons.
\begin{appendix}
\section{Determination of $\hat{\kappa}$ using the constituent quark model}\label{appA}
\setcounter{equation}{0}
\par\noindent
Let us consider a system including up and down quarks in the presence of a uniform magnetic field $\mathbf{B}$. The spin magnetic moment $\hat{\boldsymbol{\mu}}$ of this system is given by
\begin{eqnarray}\label{appA1}
\hat{\boldsymbol{\mu}}=g\hat{Q}\hat{\mu}_{B}{\mathbf{s}}.
\end{eqnarray}
Here, $g=2(1+\hat{\alpha})$ is the Land\'e $g$-factor, with $\hat{\alpha}$ denoting its anomalous contribution. In our two-flavor NJL model $\hat{\alpha}=\mbox{diag}(\alpha_{u},\alpha_{d})$ and $\hat{Q}=\mbox{diag}(q_{u},q_{d})$ are $2\times 2$ diagonal matrices, including the anomalous magnetic moment  $\alpha_{f}, f=u,d$ and electric charge $q_{f}, f=u,d$ of the up and down quarks. Moreover,  ${\mathbf{s}}=\frac{1}{2}{\boldsymbol{\tau}}$ is the quark spin angular momentum of the quarks, and $\hat{\mu}_{B}=\mbox{diag}(\mu_{B}^{u},\mu_{B}^{d})$ with $\mu_{B}^{f}=\frac{e}{2M_{f}}, f=u,d$, the nonperturbative (effective) Bohr magneton in the flavor space, which is given in term of the (bare) electric charge $e$, as well as the up and down quark constituent (effective) masses $M_{u}$ and $M_{d}$, arising in the mass matrix $\hat{M}=\mbox{diag}(M_{u},M_{d})$. Here, ${\boldsymbol{\tau}}$ are the three Pauli matrices. Using (\ref{appA1}), and assuming that the magnetic field $\mathbf{B}$ is directed in the third direction, the $f$-th matrix element of the third component of $\hat{\boldsymbol{\mu}}$, $\hat{\mu}_{3}=\mbox{diag}(\mu_{u},\mu_{d})$, is given by
\begin{eqnarray}\label{appA2}
\mu_{f}=\frac{q_{f}e}{2M_{f}}(1+\alpha_{f})\sigma_{3}.
\end{eqnarray}
Here, $\sigma_{3}=\mbox{diag}(+1,-1)$ is the third Pauli matrix. The eigenvalues of $\mu_{f}$ in the spinor space are therefore given by $\mu_{f}=\frac{q_{f}e}{2M_{f}}(1+\alpha_{f}) s$ with $s=\pm 1$ and $f=u,d$. Using this expression, we can define the following positive ratio
\begin{eqnarray}\label{appA3}
I_{f}\equiv \frac{M_{f}}{1+\alpha_{f}}=\frac{\mu_{N}}{\mu_{f}}q_{f} m_{p},
\end{eqnarray}
which turns out to be a phenomenologically relevant quantity \cite{AMM1998}. Here, $m_{p}\sim 0.938$ GeV the proton mass, and $\mu_{N}\equiv \frac{e}{2m_{p}}$ the nuclear magneton, whose phenomenological values are fixed by experiments.
Using at this stage,
\begin{eqnarray}\label{appA4}
\mu_{p}\sim 2.79~\mu_{N}, \qquad \mu_{n}\sim -1.91~\mu_{N},
\end{eqnarray}
for the magnetic moment of proton (neutron) $\mu_p$ ($\mu_n$), and their relationship to the magnetic moment of up and down quarks, $\mu_{u}$ and $\mu_{d}$,
\begin{eqnarray}\label{appA5}
\mu_{p}=\frac{1}{3}(4\mu_{u}-\mu_{d}),\qquad\mu_{n}=\frac{1}{3}(4\mu_{d}-\mu_{u}),
\end{eqnarray}
that yield
\begin{eqnarray}\label{appA6}
\mu_{u}=\frac{1}{5}(4\mu_{p}+\mu_{n}), \qquad \mu_{d}=\frac{1}{5}(4\mu_{n}+\mu_{p}),
\end{eqnarray}
we obtain
\begin{eqnarray}\label{appA7}
\mu_{u}\sim 1.852~\mu_{N},\qquad\mu_{d}\sim -0.972~ \mu_{N}.
\end{eqnarray}
Using these data, the ratio $I_{f}=\frac{\mu_{N}}{\mu_{f}}q_{f}m_{p}$ for $f=u,d$ is fixed to be
\begin{eqnarray}\label{appA8}
I_{u}\sim 0.338~\mbox{GeV},\qquad I_{d}\sim 0.322~\mbox{GeV}.
\end{eqnarray}
Same results are also reported in \cite{AMM1998}. The above phenomenological values for $I_{f}, f=u,d$ can be used to determine phenomenological values for $\alpha_{f}, f=u,d$ through $I_{f}=\frac{M_{f}}{1+\alpha_{f}}$ from (\ref{appA3}).  In order to have a sizable quark AMM, we choose $M_{u}=M_{d}=0.420$ GeV for the quark (effective) constituent mass $M_{f}, f=u,d$ \cite{AMM1998}.
 We arrive at
\begin{eqnarray}\label{appA9}
\alpha_{u}\sim 0.242,\qquad \alpha_{d}\sim 0.304,
\end{eqnarray}
which satisfy the condition $\alpha_{u}-\alpha_{d}\simeq 0.05$. The latter guarantees the isospin symmetry \cite{AMM1998}. Plugging (\ref{appA9}) into the relation $\kappa_{f}=\frac{\alpha_{f}}{2M_{f}}$ from (\ref{F9}), and choosing $M_{u}=M_{d}=0.420$ GeV, we obtain
\begin{eqnarray}\label{appA10}
\kappa_{u}\sim 0.290~\mbox{GeV}^{-1},\qquad\kappa_{d}\sim 0.360~\mbox{GeV}^{-1},
\end{eqnarray}
[see also (\ref{F12})]. Choosing, on the other hand, $M_{u}=M_{d}=0.340$ GeV, and following the same steps as above, we obtain
\begin{eqnarray}\label{appA11}
\alpha_{u}\sim 0.006,\qquad\alpha_{d}\sim 0.056,
\end{eqnarray}
which lead to
\begin{eqnarray}\label{appA12}
\kappa_{u}\sim 0.009~\mbox{GeV}^{-1},\qquad\kappa_{d}\sim 0.080~\mbox{GeV}^{-1},
\end{eqnarray}
[see also (\ref{F13})]. As described in Sec. \ref{sec2}, in the present work, $\hat{\kappa}$, appearing explicitly in the quark energy dispersion relation (\ref{F9}) is fixed by  (\ref{appA10}) and (\ref{appA12}) [see \ref{F12}) and (\ref{F13})]. The $(T,\mu,eB)$ dependent constituent quark mass are then determined by plugging this dispersion relation into the thermodynamic potential (\ref{F11}), and minimizing it appropriately. We have shown that for $\hat{\kappa}_{1}$ from (\ref{appA10}), leading to large $\hat{\alpha}$ from (\ref{appA9}), the phenomenon of IMC occurs.
\end{appendix}

\end{document}